\pdfoutput=1
\documentclass[11pt,a4paper,american]{extarticle}
\usepackage{a4wide}
\usepackage[active]{srcltx}
\usepackage[export]{adjustbox}
\usepackage{amssymb}
\usepackage{amsmath}
\usepackage{amsfonts}
\usepackage{amsthm}
\usepackage{array,multirow,bigdelim}
\usepackage{authblk}
\usepackage{babel}
\usepackage{bbm}
\usepackage{bm}
\usepackage{braket}
\usepackage{cite}
\usepackage{color}
\usepackage{comment}
\usepackage{enumerate}
\usepackage{esint}
\usepackage{etoolbox}
\usepackage{float}
\usepackage[T1]{fontenc}
\usepackage{graphicx}
\usepackage{hyperref}
\usepackage[latin1]{inputenc}
\usepackage{latexsym}
\usepackage{mathrsfs}
\usepackage{mathtools}
\usepackage{relsize}
\usepackage{setspace}
\usepackage{slashed}
\usepackage{soul}
\usepackage{subcaption}
\usepackage{todonotes}
\usepackage{xcolor}

\newcommand{\xdownarrow}[1]{%
	{\left\downarrow\vbox to #1{}\right.\kern-\nulldelimiterspace}
}

\apptocmd{\sloppy}{\hbadness 10000\relax}{}{}

\def\a{\alpha}
\def\dif{\text{d}}
\newcommand{\s}{\sigma}
\newcommand{\al}{\alpha}

\graphicspath{ {Images/} }

\newcommand{\Tr}{ \mathrm{Tr}}

\newcommand{\be}{\begin{equation}\label}
\newcommand{\ee}{\end{equation}}
\newcommand{\bea}{\begin{eqnarray}\label}
\newcommand{\eea}{\end{eqnarray}}

\makeatletter
\newcommand*{\textoverline}[1]{$\overline{\hbox{#1}}\m@th$}
\newcommand*\bigcdot{\mathpalette\bigcdot@{.65}}
\newcommand*\bigcdot@[2]{\mathbin{\vcenter{\hbox{\scalebox{#2}{$\m@th#1\bullet$}}}}}
\makeatother

\newcommand{\eq}[1]{\begin{equation}#1\end{equation}}
\newcommand{\eqs}[1]{\begin{equation}\begin{split}#1\end{split}\end{equation}}

\newcommand{\pf}{\mathrm{Pf}}

\date{}

\makeatletter

\allowdisplaybreaks
\numberwithin{equation}{section}
\author[1]{C. Armstrong\thanks{connor.armstrong@durham.ac.uk}}
\author[1,2]{H. Gomez\thanks{humberto.gomez@durham.ac.uk}}
\author[3]{R. Lipinski Jusinskas\thanks{renannlj@fzu.cz}}
\author[1]{A. Lipstein\thanks{arthur.lipstein@durham.ac.uk}}
\author[1]{J. Mei\thanks{jiajie.mei@durham.ac.uk}}

\affil[1]{Department of Mathematical Sciences, Durham University,
\authorcr  Stockton Road, DH1 3LE, Durham, United Kingdom}
\affil[2]{Facultad de Ciencias Basicas,  Universidad Santiago de Cali,
\authorcr  Calle 5 $N^\circ$  62-00 Barrio Pampalinda, Cali, Valle, Colombia}
\affil[3]{Institute of Physics of the Czech Academy of Sciences \& CEICO,
\authorcr  Na Slovance 2, 182 21, Prague, Czech Republic}

\makeatother

\begin{document}

\title{Effective Field Theories and Cosmological Scattering Equations}

\maketitle

\begin{abstract}
We propose worldsheet formulae for wavefunction coefficients of the massive non-linear sigma model (NLSM), scalar Dirac-Born-Infeld (DBI), and special Galileon (sGal) theories in de Sitter momentum space in terms of the recently proposed cosmological scattering equations constructed from conformal generators in the future boundary. The four-point integrands are assembled from simple building blocks and we identify a double copy prescription mapping the NLSM wavefunction coefficient to the DBI and sGal wavefunction coefficients, including mass deformations and curvature corrections. Finally, we compute the soft limits of these wavefunction coefficients and find that they can be written in terms of boundary conformal generators acting on contact diagrams.
\end{abstract}

\pagebreak

\tableofcontents

\section{Introduction}

In recent years, there has been exciting progress in adapting methods for computing scattering amplitudes, which probe physics at its shortest distances, to cosmological observables, which probe physics at its largest distances. In the latter, we have in mind boundary correlators in four dimensional de Sitter space (dS$_4$), which provides an approximate description of the early Universe according to the inflationary paradigm \cite{Guth:1980zm,Linde:1981mu,Albrecht:1982wi,Mukhanov:1981xt}. The cosmological observables we focus on are known as coefficients of the wavefunction of the universe \cite{Hartle:1983ai}. They are computed via a Wick rotation of Witten diagrams in Anti de Sitter (AdS) and can be treated like conformal field theory (CFT) correlators in the future boundary of dS \cite{Strominger:2001gp,Maldacena:2002vr,McFadden:2009fg,McFadden:2010vh,Maldacena:2011nz,Raju:2011mp,Ghosh:2014kba}. In-in correlators \cite{Weinberg:2005vy} can then be computed by squaring of the wavefunction and computing expectation values \cite{Maldacena:2002vr}.  Recent developments in this direction include geometric approaches \cite{Arkani-Hamed:2017fdk,Bzowski:2020kfw}, methods based on factorisation \cite{Arkani-Hamed:2015bza,Arkani-Hamed:2018kmz,Baumann:2019oyu,Baumann:2020dch,Baumann:2021fxj} and unitarity \cite{Meltzer:2021zin,Hillman:2021bnk,Alday:2017vkk,Meltzer:2020qbr,Goodhew:2020hob,Melville:2021lst,Jazayeri:2021fvk,Goodhew:2021oqg}, Mellin-Barnes representations \cite{Sleight:2019hfp,Sleight:2021plv}, color/kinematics duality \cite{Armstrong:2020woi,Albayrak:2020fyp,Alday:2021odx,Diwakar:2021juk,Sivaramakrishnan:2021srm,Cheung:2022pdk,Herderschee:2022ntr,Drummond:2022dxd}, the double copy \cite{Farrow:2018yni,Lipstein:2019mpu,Jain:2021qcl,Zhou:2021gnu}, and the cosmological scattering equations (CSE) \cite{Gomez:2021qfd,Gomez:2021ujt}. See also \cite{Fichet:2021xfn,Herderschee:2021jbi,Gadde:2022ghy,Heckelbacher:2022hbq} for other recent work adapting  amplitude ideas to (A)dS correlators. 

Here we focus on the double copy and the CSE. In flat space, the double copy relates graviton amplitudes to the square of gluon amplitudes \cite{Bern:2008qj,Bern:2010ue}, and the scattering equations of Cachazo, He, and Yuan (CHY)  \cite{fairlie,Gross:1987kza} provide a universal framework for describing scattering amplitudes in terms of worldsheet integrals \cite{Cachazo:2013hca,Cachazo:2013iea,Mason:2013sva}. They uncover a vast web of relations among quantum field theories such as the non-linear sigma model (NLSM), Dirac-Born-Infeld (BDI), and special Galileon (sGal) theories, extending the scope of the double copy \cite{Cachazo:2014xea,Casali:2015vta,Cheung:2017ems}. The CSE represent an extension of the flat space scattering equations to de Sitter momentum space. They were initially proposed to describe massive $\phi^4$ theory as a toy model of inflation, building on earlier work on bi-adjoint scalar scalar theory in AdS position space \cite{Eberhardt:2020ewh,Roehrig:2020kck}. One of the key properties of this proposal is an operatorial integrand built out of conformal generators in the future boundary of dS which acts on a contact diagram. An important open question was how to extend this structure to more general scalar theories, and here we take the first steps to address it.

Towards this goal, we will extend the CSE framework to the NLSM, scalar DBI, and sGal theories in dS space. When lifting derivative interactions to curved background, new subtleties arise. First of all, the spacetime Lagrangians receive curvature corrections whose coefficients cannot be fixed by the flat space limit. At four points, we find that such corrections can be described using simple building blocks at the level of the worldsheet integrand. We can also deform the integrand to describe bulk scalar fields with arbitrary mass. The other class of subtleties is related to more technical aspects of the problem. Becasue of derivative interactions in the curved background, the worldsheet formulae may have ordering ambiguities involving differential operators with nontrivial commutators. Such ambiguities can arise in the NLSM at six points and the DBI and sGal theories at four points, although we use a simple prescription for avoiding them at four points. 

Our worldsheet formulae give rise to four-point wavefunction coefficients in terms of boundary conformal generators acting on contact diagrams. We also identify a simple prescription for mapping the NLSM wavefunction coefficient into the DBI and sGal wavefunction coefficients at the level of the worldsheet integrand, which we refer to as a generalised double copy. After specifying a prescription for evaluating the worldsheet integrals to avoid ordering ambiguities, this provides a systematic way to implement the double copy of four-point wavefunction coefficients in these theories. Finally, we study the soft limits of the four-point wavefunction coefficients and find that they can once again be written in terms of boundary conformal generators acting on certain three-point  contact diagrams. In the flat space limit, the scattering amplitudes of these theories are well-known to exhibit vanishing soft limits \cite{Adler:1964um,Cheung:2014dqa} as a result of underlying hidden symmetries \cite{Low:2014nga,Hinterbichler:2015pqa,Padilla:2016mno}. In our treatment, the coefficients of curvature corrections are left unfixed and the soft limits we derive are completely general. Lagrangians for the DBI and sGal theories with certain hidden symmetries in dS were recently proposed in \cite{Bonifacio:2021mrf}. It would be interesting to compute their four-point wavefunction coefficients in terms of the building blocks we derive in this paper and check if their soft limits exhibit any additional simplicity. Soft limits of flat space wavefunction coefficients in these theories were also recently analysed in \cite{Bittermann:2022nfh}.

This paper is organised as follows. In section \ref{sec:EFTFlatSpace} we review the NLSM, DBI, and sGal theories in flat space and the CHY representation for their amplitudes. In section \ref{sec:EFTdS}, we then discuss how to lift these theories to dS space up to four-point interactions, and compute their four-point wavefunction coefficients in terms of Witten diagrams. In section \ref{worldsheet}, we then propose worldsheet formulae for these wavefunction coefficients in terms of CSE, and a generalised double copy which maps the NLSM wavefunction coefficient into the DBI and sGal wavefunction coefficients, including curvature corrections and mass deformations. In section \ref{soft}, we analyse the soft limits of these four-point wavefunction coefficients and show that they can be written in terms of boundary conformal generators acting on certain three-point contact diagrams. We present our conclusions in section \ref{sec:conclusion}. We also have a number of Appendices in which we compute wavefunction coefficients for the six-point NLSM in dS using Witten diagrams and worldsheet methods, and provide more details on four-point wavefunction coefficients.

\section{Effective Field Theory Amplitudes}
\label{sec:EFTFlatSpace}

In this work we focus on certain scalar effective field theories (EFT), namely the non-linear sigma model (NLSM), scalar Dirac-Born-Infeld theory (DBI), and special Galileon theory (sGal). Their scattering amplitudes take a particularly simple form in the CHY approach \cite{Cachazo:2014xea}. 

\subsection{Lagrangians} \label{lagrangians}

We start with a quick review of the scalar EFT Lagrangians.

The NLSM Lagrangian is given by
\eqs{
\mathcal{L}_{\mathrm{NLSM}} &= \frac{1}{8 \lambda^2} \Tr(\partial_\mu U^\dagger \partial^\mu U), \label{eq:nlsmlagrangian}\\
&=-\Tr\left[\tfrac{1}{2}\partial_{\mu}\Phi\partial^{\mu}\Phi+\lambda^{2}\Phi^{2}\partial_{\mu}\Phi\partial^{\mu}\Phi+\lambda^{4}\left(\Phi^{4}\partial_{\mu}\Phi\partial^{\mu}\Phi+\tfrac{1}{2}\Phi^{2}\partial_{\mu}\Phi\Phi^{2}\partial^{\mu}\Phi\right)+\ldots\right], 
}
where $U =(\mathbb{I}+\lambda\Phi)(\mathbb{I}-\lambda\Phi)^{-1}$, $\Phi$ is in the adjoint representation of SU$(N)$, and the ellipsis denotes higher-point interactions. For the DBI theory we have
\eqs{
\mathcal{L}_{\mathrm{DBI}}&=\frac{1}{\lambda}\left(\sqrt{1-\lambda\left(\partial\phi\right)^{2}}-1\right), \label{eq:dbilagrangian} \\ &=-\frac{1}{2}\partial\phi\cdot\partial\phi-\frac{\lambda}{8}\left(\partial\phi\cdot\partial\phi\right)^{2}+\ldots.
}
Finally, in $d=4$ the special Galileon theory is given by \cite{Hinterbichler:2015pqa}
\eq{
\mathcal{L}_{\mathrm{sGal}}=-\frac{1}{2}\left(\partial\phi\right)^{2}-\frac{\lambda}{8}\left(\partial_{\mu}\partial_{\nu}\phi\right)^{2}\left(\partial\phi\right)^{2}.
\label{eq:sgalagrangian}
}
Note that the four-point interactions in the DBI and sGal theories contain four derivative and six-derivatives, respectively. They are unique up to integration by parts and equations of motion. Their lift to curved backgrounds, however, is not unique because covariant derivatives no longer commute and there are curvature corrections, as we describe later on.

\subsection{Worldsheet formulae} \label{flatspaceintegrands}

The CHY formulae express tree-level scattering amplitudes as integrals over the Riemann sphere, which localise onto solutions of the scattering equations (SE),
\begin{equation}\label{eq:SE-flat}
S_a=\sum_{a\neq b}\frac{2\, k_{a}\cdot k_b}{\sigma_{ab}}=0, \qquad \sigma_{ab} \equiv \sigma_{a}-\sigma_{b},
\end{equation}
where $\sigma_a$ is the holomorphic coordinate of the $a$-th puncture, and the punctures are in one-to-one correspondence with the external legs of the amplitude. The details of the theory appear only in the worldsheet integrand, and a generic $n$-point tree level amplitude takes the form
\begin{equation}\label{eq:CHY-amp}
\mathcal{A}_n=\int_\gamma  \prod_{a=1 \atop a\neq b,c,d}^n  \dif\sigma_a \, (S_a)^{-1} \,  (\sigma_{bc}\sigma_{cd}\sigma_{db})^2 \, \mathcal{I}_n,
\end{equation}
where $\left\{ b,c,d\right\} $ are fixed punctures. The integration contour is defined by the intersection $\gamma= \bigcap_{a\neq b,c,d} \gamma_{S_a}$, where $\gamma_{S_a}$ encircles the poles of $S_a$. 

The CHY integrands for the scalar EFTs of interest are defined using simple building blocks, with various operations to transform between different  theories \cite{Cachazo:2014xea}. They are summarised in Table \ref{table:EFTIntegrands} and we will briefly review their construction.
\begin{table}[h]
\centering
\begin{tabular}{ll}
Theory	&Integrand\\
\hline
NLSM			&$\mathrm{PT} (\mathrm{Pf}'A)^2$\\
DBI				&$\mathrm{Pf} X (\mathrm{Pf}'A)^3$\\
sGal	&$(\mathrm{Pf}'A)^4$\\
\hline
\end{tabular}
\caption{A summary of CHY integrands for a selection of scalar EFTs.}\label{table:EFTIntegrands}
\end{table}

The integrand for color-ordered amplitudes of the NLSM is
\eq{
\mathcal{I}^{\mathrm{NLSM}} (\alpha_1,\alpha_2,\ldots,\alpha_n) =\rm{PT}(\alpha_1,\alpha_2,\ldots,\alpha_n) (\mathrm{Pf}'A)^2,
\label{eqn:NLSMIntegrand}
}
where $\a=(\alpha_1,\alpha_2,\ldots,\alpha_n)$ denotes a specific ordering, and
$\rm{PT}$ is the Parke-Taylor factor
\begin{equation}
{\rm PT}(\alpha_1,\alpha_2,\ldots,\alpha_n)=\left(\sigma_{\a_1\a_2} \sigma_{\a_2\a_3}  \cdots \sigma_{\a_n\a_1}\right)^{-1}.
\end{equation} 
In practice we will use the canonical ordering, {\it i.e.} ${\rm PT}(1,2,\ldots, n)$, so for simplicity we introduce the notation
\begin{equation}
{\rm PT}\equiv {\rm PT}(1,2,\ldots, n).
\end{equation}
The reduced Pfaffian ${\rm Pf}^\prime A$ is given by 
\begin{eqnarray}
{\rm Pf}^\prime A&=&\frac{(-1)^{c+d}}{\sigma_{cd}}  {\rm Pf}A^{cd}_{cd}, \label{eq:PfA} \\
{\rm Pf}A^{cd}_{cd} &=&
\frac{ \epsilon^{r_1 s_1\ldots r_{p-1} s_{p-1}} (A^{cd}_{cd})_{r_1 s_1} \cdots (A^{cd}_{cd})_{r_{p-1} s_{p-1}}   }{2^{p-1}(p-1)!},
\end{eqnarray}
where the matrix $A^{cd}_{cd}$ is obtained from the $n \times n$ matrix
\begin{align}
A_{rs} & =
\begin{cases} 
\displaystyle \frac{2 k_r \cdot k_s }{\sigma_{rs}},  & r\neq s,\\
\displaystyle  ~~ 0 , & r=s,
\end{cases}
\end{align}
by removing any pair of rows and columns $\{c,d\}$, with $n=2p$. Alternatively the Pfaffian can be computed from the square root of the determinant.

For the scalar DBI theory, the worldsheet integrand is given by
\eq{
\mathcal{I}^{\mathrm{DBI}} = \pf X(\pf'A)^3,
\label{eqn:DBIIntegrand}
}
where 
\begin{align}
X_{rs} & =
\begin{cases} 
\displaystyle \frac{ 1}{\sigma_{rs}},  & r\neq s,\\
\displaystyle  ~~ 0 , & r=s.
\end{cases}
\label{xmatrix}
\end{align}
The $X$-matrix arises from dimensional reduction of the CHY formula for Yang-Mills amplitudes \cite{Cachazo:2014xea}. These reason for this will be further explained below. Finally, the integrand for the sGal theory is given by
\eq{
\mathcal{I}^{\mathrm{sGal}} = (\pf'A)^4.
\label{sGalintegrand}
}

Observe that the integrands of both DBI and sGal can be obtained from NLSM via the following substitutions, respectively:
\begin{eqnarray}
{\rm PT} &\rightarrow&  \pf X(\pf'A) \label{DBIdouble}, \\
{\rm PT} &\rightarrow& \left({\rm Pf'}A\right)^{2} \label{sGaldouble},
\end{eqnarray}
which encode the double copy structure of these theories. Roughly speaking, $\rm{sGal}=\rm{NLSM}^2$ and $\rm{DBI}=\rm{NLSM}\times \rm{YM}$, where $\rm{YM}$ corresponds to the dimensional reduction of Yang-Mills theory. More precisely, $\pf X(\pf'A) $ can be written as a linear combination of $(n-2)!$ Parke-Taylor factors, where the coefficients are the Yang-Mills BCJ master numerators after the identification $\epsilon_{i}\cdot\epsilon_{j}=1$ and $\epsilon_{i}\cdot k_{j}=0$. 
\cite{Cachazo:2013iea,Cachazo:2014xea,Bjerrum-Bohr:2016axv,Fu:2017uzt,Edison:2020ehu,Cheung:2017ems,Bjerrum-Bohr:2020syg,Low:2020ubn}.

\section{de Sitter wavefunction coefficients from Effective Actions}
\label{sec:EFTdS}

In this section we review the computation of field theory observables in de Sitter. For convenience, we use the Poincar\'e patch with radius set to one,
\begin{equation}
ds^{2}=\frac{1}{\eta^{2}}(d\vec{x}^{2}-d\eta^{2}), \label{eq:dSmetric}
\end{equation}
where $-\infty<\eta<0$ is the conformal time, and $\vec{x}$ denotes the boundary coordinate, with individual components $x^i$, $i=1,..,d$. In practice, we set the dimension of the boundary $d=3$.

In-in correlators \cite{Weinberg:2005vy} can be computed from a cosmological wavefunction as follows \cite{Maldacena:2002vr}:
\begin{equation}
\langle \phi(\vec{k}_{1})\ldots\phi(\vec{k}_{n})\rangle =\frac{\int\mathcal{D}\phi \, \phi(\vec{k}_{1})...\phi(\vec{k}_{n})\left|\Psi[\phi]\right|^{2}}{\int\mathcal{D}\phi\left|\Psi[\phi]\right|^{2}}.
\end{equation}
The scalars $\phi$ are taken to be in the future boundary, Fourier transformed to momentum space. The functional $\Psi[\phi]$ is the cosmological wavefunction, which can be perturbatively expanded as
\begin{equation}
\ln\Psi[\phi]=-\sum_{n=2}^{\infty}\frac{1}{n!}\int\prod_{i=1}^{n}\frac{{\rm d}^{d}k_{i}}{(2\pi)^{d}}\Psi_{n}(\vec{k}_{1},\ldots,\vec{k}_{n})\phi(\vec{k}_{1})\ldots\phi(\vec{k}_{n}).
\end{equation}

The wavefunction coefficients $\Psi_n$ can be treated as $n$-point CFT wavefunction coefficients in the future boundary. In momentum space, they can be expressed as
\begin{equation}
\Psi_{n}=\delta^{d}(\vec{k}_T)\langle\langle \mathcal{O}(\vec{k}_{1})...\mathcal{O}(\vec{k}_{n})\rangle\rangle ,
\label{psid}
\end{equation}
where $\vec{k}_T=\vec{k}_{1}+\ldots+\vec{k}_{n}$, and the double brackets denote a CFT correlator on the boundary. The scalar operators $\mathcal{O}$ have scaling dimension $\Delta$, and are dual to scalar fields $\phi$ in the bulk with mass
\begin{equation}
m^{2}=\Delta(d-\Delta).
\label{eq:massdelta}
\end{equation}
Note that $\Delta=d$ describes minimally coupled scalars while $\Delta=(d+1)/2$ describes conformally coupled scalars.

The wavefunction coefficients $\Psi_n$ satisfy conformal Ward identities (CWIs), which are a consequence of the de Sitter isometries. The conformal generators are $D$ (dilatation), $P_i$ (translations), $K_i$ (special conformal transformations),  and $M_{ij}$ (rotations). The CWIs can be cast as
\begin{equation}
\sum_{a=1}^{n}P_{a}^{i}\Psi_{n}=\sum_{a=1}^{n}D_{a}\Psi_{n}=\sum_{a=1}^{n}K_{a}^{i}\Psi_{n}=\sum_{a=1}^{n}M_{a}^{ij}\Psi_{n}=0,
\label{eq:cwi}
\end{equation}
where $a,b,...$ are particle labels and
\eqs{
P^{i} & =  k^{i}, \\
D& =  k^{i}\partial_{i}+(d-\Delta), \label{eq:CGG-boundary}\\
K_{i} & =  k_{i}\partial^{j}\partial_{j}-2k^{j}\partial_{j}\partial_{i}-2(d-\Delta)\partial_{i}, \\
M_{ij} &= \left(k_i\partial_j-k_j\partial_i\right),
}
with $\partial_i=\tfrac{\partial}{\partial k^i}$. Observe here that the conformal dimension of the scalars appear explicitly in generators. Boundary vector indices will be freely raised and lowered here using a flat metric.

\subsection{Witten Diagrams}

From a more traditional field theory perspective, the wavefunction coefficients can be computed through Witten diagrams \cite{Maldacena:2002vr,Raju:2011mp,Maldacena:2011nz}.

The scalar equation of motion is given by
\begin{equation}
[\eta^{2}\partial_{\eta}^{2}+(1-d)\eta\partial_{\eta}-\eta^{2}\partial^i \partial_i+m^{2}] \phi =0.
\end{equation}
We will consider momentum eigenstates, such that $\phi = \mathcal{K}_{\nu}(k,\eta) e^{i \vec{k}\cdot \vec{x}}$, with $\mathcal{K}_{\nu}(k,\eta) $ denoting the bulk-to-boundary propagator:
\begin{equation}
\mathcal{K}_{\nu}(k,\eta)=\mathcal{N}k^{\nu}\eta^{d/2}H_{\nu}(-k \eta).
\label{bulktoboundaryprop}
\end{equation}
Here, $\nu=\Delta-d/2$, $k=|\vec{k}|$, $H_\nu$ is a Hankel function of the second kind, and we will leave the normalisation $\mathcal{N}$ unspecified for now. 
$\mathcal{K}_{\nu}(k,\eta)$ satisfies
$ (\mathcal{D}_k^{2}+m^{2})\mathcal{K}_{\nu} =0$, with
\begin{equation}
\mathcal{D}_k^{2} \equiv \eta^{2}\partial_{\eta}^{2}+(1-d)\eta\partial_{\eta}+\eta^{2}k^{2}. \label{eq:D2def}
\end{equation}

Contact diagrams are given by a product of bulk-to-boundary propagators integrated over the bulk, expressed as
\begin{eqnarray}
\mathcal{C}^{\Delta}_n & \equiv &  \int  \frac{d\eta}{\eta^{d+1}} U_{1,n}(\eta), \\
U_{m,n}(\eta) & = &  \prod_{a=m}^{n}\mathcal{K}_{\nu}(k_{a},\eta). \label{eq:prodK}
\end{eqnarray}
More general Witten diagrams involve also bulk-to-bulk propagators, $G_{\nu}(k,\eta,\tilde{\eta})$, satisfying
\begin{equation}\label{eq:eom-bulk-to-bulk}
(\mathcal{D}_{k}^{2}+m^{2})G_{\nu}=\eta^{d+1}\delta(\eta-\tilde{\eta}).
\end{equation} As it turns out, however, any tree level diagram can be obtained from contact diagrams through certain differential operations. 

In order to see this, let us first consider the action of the boundary generators in \eqref{eq:CGG-boundary} on bulk-to-boundary propagators. It can be expressed in terms of derivatives with respect to conformal time
\begin{equation}\label{eqn:ConfTimeGenerators}
\begin{array}{ccc}
D\mathcal{K}_{\nu} =  \eta\tfrac{\partial}{\partial\eta}\mathcal{K}_{\nu},  & &  P^{i}\mathcal{K}_{\nu} =  k^{i}\mathcal{K}_{\nu}, \\
 K_{i}\mathcal{K}_{\nu} =  \eta^{2}k_{i}\mathcal{K}_{\nu}, & & M_{ij}\mathcal{K}_{\nu} = 0.
\end{array}
\end{equation}
Now consider the following operator
\begin{equation}\label{eq:DaDb}
\mathcal{D}_{a}\cdot\mathcal{D}_{b}= \frac{1}{2}(P_{a}^{i}K_{bi}+K_{ai}P_{b}^{i} - M_{a, ij}M_b^{ij})+D_{a}D_{b}.
\end{equation}
Using \eqref{eqn:ConfTimeGenerators} one finds that
\begin{equation}
(\mathcal{D}_{a}\cdot\mathcal{D}_{b})\mathcal{K}_{\nu}^{a}\mathcal{K}_{\nu}^{b}=\eta^{2}[\partial_{\eta}\mathcal{K}_{\nu}^{a}\partial_{\eta}\mathcal{K}_{\nu}^{b}+(\vec{k}_{a}\cdot\vec{k}_{b})\mathcal{K}_{\nu}^{a}\mathcal{K}_{\nu}^{b}],\label{eq:DaDb}
\end{equation}
with shorthand notation $\mathcal{K}_{\nu}^{a}=\mathcal{K}_{\nu}(k_{a},\eta)$. This observation can then be used to show that
\begin{equation}
(\mathcal{D}_{1\ldots p}^{2}U_{1,p})U_{p{+}1, n}= (\mathcal{D}_{1}+\ldots + \mathcal{D}_{p})^2 U_{1,n}. \label{eq:bulkvsboundary-props}
\end{equation}
where in the left hand side $\mathcal{D}_{1\ldots p}^{2}$ is defined in \eqref{eq:D2def} with $k=|\vec{k}_{1}+\ldots +\vec{k}_{p}| \equiv k_{1\ldots p}$ and $p<n$,
and the right hand side is built using the boundary conformal generators in momentum space (\ref{eq:CGG-boundary}), satisfying $\mathcal{D}_a \cdot \mathcal{D}_a  = -m^2$. In particular, we can derive the following identity,
\begin{equation} \label{eq:exchnage=difxcontact}
[(\mathcal{D}_1+\ldots +\mathcal{D}_p)^2+m^2]^{-1} \mathcal{C}^{\Delta}_n =  \int  \frac{d\eta}{\eta^{d+1}}  \frac{d\tilde{\eta}}{\tilde{\eta}^{d+1}}   U_{p+1,n}(\eta) G_{\nu}(k_{1...p},\eta,\tilde{\eta})U_{1,p} (\tilde{\eta}),
\end{equation}
which can be easily demonstrated using \eqref{eq:bulkvsboundary-props} and the equation of motion of the bulk-to-bulk propagator \eqref{eq:eom-bulk-to-bulk}. 
Therefore, any tree level exchange diagram can be recast as a differential operator given in terms of the boundary momenta acting on a contact diagram.

Finally, observe that acting with $\mathcal{D}_{a}\cdot\mathcal{D}_{b}$ on a pair of bulk-to-boundary propagators is equivalent to acting with $\nabla_a \cdot \nabla_b$, where $\nabla_a$ is a bulk covariant derivative acting on leg $a$. This identity will be very useful for computing Witten diagrams for derivative interactions containing terms of the form $\nabla \phi \cdot \nabla \phi$. Another useful identity is
\begin{align} \label{eq:commute}
[(\mathcal{D}_{a}\cdot\mathcal{D}_{b}),(\mathcal{D}_{b}\cdot\mathcal{D}_{c})]\mathcal{C}_n^{\Delta} &= 2(K_a\cdot P_c-P_a\cdot K_c)D_b \mathcal{C}_n + \mathrm{cyc}(abc), \nonumber \\
& = 0.
\end{align}
The commutator is not zero but vanishes when acting on a contact diagram. This was first derived in the embedding space formalism \cite{Diwakar:2021juk}, and its generalization to momentum space is straightforward. Using (\ref{eqn:ConfTimeGenerators}) we can easily show that the right hand side vanishes.

\subsection{Four-point wavefunction coefficients} \label{4ptwitten}

Our strategy in this section will be to lift the effective actions in section \ref{lagrangians} to de Sitter space up to four point interactions, use them to compute the four-point wavefunction coefficients using Witten diagrams, and express the result in terms of boundary conformal generators acting on a contact term. It will be convenient to define the following operators:
\begin{equation} \label{eq:GenMandelstam}
\begin{array}{ccccc}
\hat{s}=\mathcal{D}_{1}\cdot\mathcal{D}_{2}, & & \hat{t}=\mathcal{D}_{1}\cdot\mathcal{D}_{4}, & & \hat{u}=\mathcal{D}_{1}\cdot\mathcal{D}_{3},
\end{array}
\end{equation}
which satisfy
\begin{equation}
\hat{s}+\hat{t}+\hat{u}=m^{2},
\label{eq:mandelstamsum}
\end{equation}
when acting on contact diagrams. This can be seen using the CWIs in \eqref{eq:cwi}.

In de Sitter space, the expansion of the NLSM action up to four points can be cast as
\begin{equation}
S_{4}^{NLSM}= -\int d^{4}x\sqrt{-g}\Tr\{\tfrac{1}{2}\nabla\Phi\cdot\nabla\Phi+\tfrac{1}{2}m^{2}\Phi^{2}+\lambda^{2}\Phi^{2}\nabla\Phi\cdot\nabla\Phi+\tfrac{1}{4}C\Phi^{4}\},
\label{nlsmlift}
\end{equation}
where we have included a mass term and a possible quartic interaction coming from a curvature correction (recall we have set the dS radius to one). The mass is also proportional to the curvature and vanishes in the flat space limit. Using the identity \eqref{eq:DaDb}, it is straightforward to show that the four-point wavefunction coefficient obtained from Witten diagrams is given by
\begin{equation}
\Psi_{4}^{NLSM}=\delta^{3}(\vec{k}_{T})\left[2\lambda^{2}\left(\hat{s}+\hat{t}\right)-C\right]\mathcal{C}_{4}^{\Delta}=-\delta^{3}(\vec{k}_{T})\left(2\lambda^{2}\hat{u}+C\right)\mathcal{C}_{4}^{\Delta}.
\label{nlsmcorr}
\end{equation}

Up to quartic vertices and six-derivative interactions, the most general effective action for a scalar field in (A)dS is given by \cite{Heemskerk:2009pn}
\begin{equation}
S^{(6)}_4= -\int d^{4}x\sqrt{-g}\{\tfrac{1}{2}\nabla\phi\cdot\nabla\phi+\tfrac{1}{2}m^{2}\phi^{2}+\tfrac{1}{8}A(\nabla_{\mu}\nabla_{\nu}\phi)^{2}\nabla\phi\cdot\nabla\phi+\tfrac{1}{8}B(\nabla\phi\cdot\nabla\phi)^{2}+\tfrac{1}{4!}C\phi^{4}\},\label{effectiveaction}
\end{equation}
where $A$, $B$, and $C$ are undetermined numerical coefficients. Other possible interactions are related by integration by parts or the free equation of motion $\nabla^{2}\phi=-m^{2}\phi.$ For the sGal theory, the 6-derivative interaction is the naive uplift of the one in \eqref{eq:sgalagrangian} while the lower-derivative interactions correspond to curvature corrections and a mass term. In the flat space limit, these are subleading and the action reduces to \eqref{eq:sgalagrangian} for $A=\lambda$. Hence, \eqref{effectiveaction} represents the uplift of the special Galileon theory to a curved background, with unfixed coefficients corresponding to curvature corrections. Additional data must be specified in order to fix them, such as soft limits, and we will explore this section \ref{soft}. For the DBI theory, we set $A=0,B=\lambda$, and there is a single curvature correction with unfixed coefficient $C$. In the flat space limit, the action reduces to \eqref{eq:dbilagrangian} up to quartic interactions. 

The four-point wavefunction coefficient obtained from \eqref{effectiveaction} is
\begin{equation}
\Psi^{(6)}_{4}=\delta^{3}(\vec{k}_{T})[A(\hat{s}^{3}+\hat{t}^{3}+\hat{u}^{3})+(d A-B)(\hat{s}^{2}+\hat{t}^{2}+\hat{u}^{2})-C]\mathcal{C}_{4}^{\Delta}.
\label{4pteffective}
\end{equation}
We can illustrate this derivation using Witten diagrams. For example, let's consider the the six-derivative interaction term
\begin{eqnarray}
\left(\nabla_{\mu}\nabla_{\nu}\phi\right)^{2}\nabla\phi\cdot\nabla\phi &=& \eta^6 \eta^{\mu\nu}\eta^{\rho\sigma}\eta^{\kappa\lambda}(\nabla_\mu\phi)(\nabla_\nu\phi)(\nabla_\rho\nabla_\kappa\phi)(\nabla_\sigma\nabla_\lambda\phi), \nonumber\\
&=&\eta^6(\partial_\mu\phi\partial^\mu\phi)\eta^{\rho\sigma}\eta^{\kappa\lambda}(\partial_\rho\partial_\kappa - \Gamma^\alpha_{\rho\kappa}\partial_\alpha\phi)(\partial_\sigma\partial_\lambda\phi - \Gamma^\beta_{\sigma\lambda}\partial_\beta\phi).
\end{eqnarray}
The corresponding tree-level Witten diagrams are given by
\eqs{
&\int_{-\infty}^{0}\frac{d\eta}{\eta^{4}} \eta^6\bigg[(\vec{k}_1\cdot\vec{k}_2)^2\mathcal{K}_1\mathcal{K}_2 + 2\vec{k}_1\cdot\vec{k}_2\dot{\mathcal{K}}_1\dot{\mathcal{K}}_2 + \ddot{\mathcal{K}}_1\ddot{\mathcal{K}}_2\\
&\qquad + \frac{1}{\eta}\left(2\vec{k}_1\cdot\vec{k}_2(\mathcal{K}_1\dot{\mathcal{K}}_2+\dot{\mathcal{K}}_1\mathcal{K}_2)-k_1^2\mathcal{K}_1\dot{\mathcal{K}}_2-k_2^2\dot{\mathcal{K}}_1\mathcal{K}_2+\dot{\mathcal{K}}_1\ddot{\mathcal{K}}_2 + \ddot{\mathcal{K}}_1\dot{\mathcal{K}}_2\right)\\
&\qquad + \frac{2}{\eta^2}\left(\vec{k}_1\cdot\vec{k}_2\mathcal{K}_1\mathcal{K}_2 + 2\dot{\mathcal{K}}_1\dot{\mathcal{K}}_2\right)\bigg]\left[\vec{k}_3\cdot\vec{k}_4\mathcal{K}_3\mathcal{K}_4 + \dot{\mathcal{K}}_3\dot{\mathcal{K}}_4\right] +...,
\label{eqn:s3sGalVertex}
}
where $\mathcal{K}_a$ are bulk-to-boundary propagators, $\dot{\mathcal{K}}_a$ denotes a derivative with respect to conformal time, and the ellipsis denotes the $t$ and $u$ channels. If we recast this result in terms of differential operators acting on a contact diagram we see that
\begin{equation}
-\frac{1}{8}\left(\nabla_{\mu}\nabla_{\nu}\phi\right)^{2}\nabla\phi\cdot\nabla\phi\rightarrow\left[\hat{s}^{3}+\hat{t}^{3}+\hat{u}^{3}+d\left(\hat{s}^{2}+\hat{t}^{2}+\hat{u}^{2}\right)\right]\mathcal{C}_{4}^{\Delta}.
\label{mainidentity}
\end{equation}
The other terms in \eqref{4pteffective} can be directly derived from \eqref{effectiveaction} using \eqref{eq:DaDb}. In Appendix \ref{app:6ptNLSM} we evaluate a 6-point NLSM wavefunction coefficient using Witten diagrams. 

\section{de Sitter wavefunction coefficients from the Worldsheet} \label{worldsheet}

In this section we will use the cosmological scattering equations introduced in \cite{Gomez:2021qfd} to compute the four-point wavefunction coefficients obtained from Witten diagrams in the previous section.

\subsection{Cosmological Scattering Equations}

In order to lift the scattering equations to de Sitter space, we replace $k_a \cdot k_b$ in \eqref{eq:SE-flat} with differential operators acting in the future boundary, given by \eqref{eq:DaDb}, and introduce a mass deformation. The scattering equations then have an operatorial character,
\begin{align}\label{eq:20R}
S_a = \sum_{b=1 \atop b\neq a}^{n} \frac{2\,  ({\cal D}_a \cdot {\cal D}_b )+\mu_{ab} }{\sigma_{ab}} \equiv \sum_{b=1 \atop b\neq a}^{n} \frac{\a_{ab} }{\sigma_{ab}} \, \, ,
\end{align}
where  $\mu_{a\,a\pm 1} =-m^2$ modulo $n$ and zero otherwise. This mass deformation is analogous to the flat space one in \cite{Dolan:2013isa} and assumes canonical ordering of the external legs $\mathbb{I}_n = (1,2,\ldots ,n)$. 
These are referred to as cosmological scattering equations (CSE). The CWIs in \eqref{eq:cwi} can be recast as
\begin{equation}
\sum_{b\neq a}\alpha_{ab}\Psi_n=0,
\end{equation}
where $a$ is any external leg and we sum over $b$. This implies an underlying ${\rm SL}(2,\mathbb{C})$ symmetry in the scattering equations, just like in flat space. This symmetry can be used to fix the location of three punctures. For more details, see \cite{Gomez:2021ujt}.

The worldsheet formula in \eqref{eq:CHY-amp} can then be lifted to de Sitter space as follows:
\begin{equation}
\Psi_{n}=\delta^{d}(\vec{k}_{T})\int_{\gamma}\prod_{a\neq b,c,d}^{n}\dif\sigma_{a}\,S_{a}^{-1}(\s_{bc}\s_{cd}\s_{db})^{2}\,\,{\cal I}_{n}{\cal C}_{n}^{\Delta},
\end{equation}
where the integrand may also contain boundary conformal generators. For theories with $\phi^n$ interactions, we are free to shuffle the CSE with other terms in the integrand  $\mathcal{I}_n$ \cite{Gomez:2021ujt}. On the other hand, this may not be the case for derivative interactions, and we believe this issue calls for a more systematic investigation in the future. In this work we fix the ordering with CSE appearing to the left of the integrand.

The flat space integrands were constructed from Pfaffians defined in \eqref{eq:PfA}. Their obvious uplift  to de Sitter is through the matrix
\begin{align}
A_{rs} & =
\begin{cases} 
\displaystyle \frac{ \a_{rs} }{\sigma_{rs}},  & r\neq s,\\
\displaystyle  ~~ 0 , & r=s.
\end{cases}
\end{align}
As discussed previously, theories with higher-derivative interactions can have curvature corrections that are absent in flat space. Therefore, we cannot simply lift the flat space integrands in section \ref{flatspaceintegrands} to dS by replacing kinematic invariants with differential operators. This procedure has to be supplemented by curvature corrections and mass deformations, which we will describe in the following subsections.

\subsection{Building Blocks}

There are four basic building blocks for the four-point integrands. Here we will describe each of them, explaining how to evaluate the corresponding worldsheet integrals. 

\begin{figure}[h]
\centering
\parbox[c]{10.5em}{\includegraphics[scale=0.33]{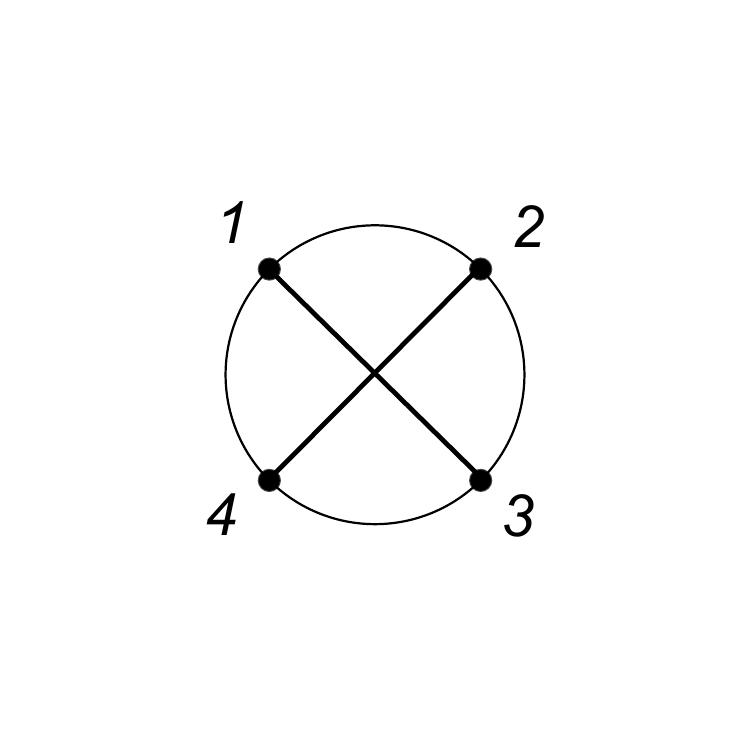}} 
\vspace{-0.35cm}
\caption{Graphic representation of ${\rm PT}\, {\rm Pf} X|_{\rm conn}$. The circle refers to the ${\rm PT}$ factor while the intersecting lines correspond to ${\rm Pf} X|_{\rm conn}$.}\label{PTPfXconn} 
\end{figure}

The simplest building block is
\begin{equation} \label{eq:phi4-integrand}
\mathcal{I}^{\phi^4}={\rm PT}\left.{\rm Pf}X\right|_{{\rm {conn}}}{\rm Pf'}A,
\end{equation}
where the matrix $X$ is defined in \eqref{xmatrix} and 
\begin{eqnarray}
{\rm Pf}\,X &=&\frac{1}{\sigma_{12}\sigma_{34}}-\frac{1}{\sigma_{13}\sigma_{24}}+\frac{1}{\sigma_{23}\sigma_{14}},\label{pfX} \\
\left.{\rm Pf}\,X\right|_{\rm{conn}}&=&-\frac{1}{\sigma_{13}\sigma_{24}}.
\end{eqnarray}
The integrand \eqref{eq:phi4-integrand} describes a contact diagram for a $\phi^4$ interaction \cite{Cachazo:2014xea,Gomez:2021qfd}. Note that at four-points, $\rm{Pf}\,X$ can be written as a sum of three terms which correspond to perfect matchings and $\left.{\rm Pf}X\right|_{\rm{conn}}$ refers to the connected perfect matching with respect to the ordering of the Parke-Taylor factor 
(see Figure \ref{PTPfXconn}).
\vspace{-0.45cm}
Fixing legs $\left\{ 1,2,4\right\} $ and deleting legs $\left\{ 2,4\right\} $ from the $A$-matrix in the Pfaffian leads to the wavefunction coefficient
\begin{equation}
\Psi_{4}^{\phi^4}=-\delta^{3}(\vec{k}_{T})\int_{\gamma_{3}}d\sigma_{3}\left(\sigma_{41}\sigma_{12}\sigma_{24}\right)^{2}\left[\frac{1}{\sigma_{12}\sigma_{23}\sigma_{34}\sigma_{41}}\frac{\sigma_{13}\sigma_{23}\sigma_{43}}{\hat{S}_{3}}\frac{\alpha_{13}}{\left(\sigma_{13}\sigma_{24}\right)^{2}}\right]\mathcal{C}_{4}^{\Delta},
\end{equation}
where the contour $\gamma_3$ encircles the pole arising from $S_3$. The first term in the integrand is the standard Jacobian associated with the ${\rm SL}(2,\mathbb{C})$ fixing, and we have rescaled the third scattering equation to 
\begin{equation}
\hat{S}_{3}=\alpha_{13}\sigma_{23}\sigma_{43}+\alpha_{23}\sigma_{13}\sigma_{43}+\alpha_{43}\sigma_{13}\sigma_{23}.
\end{equation}
After some cancellations we see there is only  a simple pole at $\sigma_{13}=0$ so we wrap the contour around this pole to obtain
\begin{equation}
\Psi_{4}^{\phi^4}=\delta^{3}(\vec{k}_{T})\int_{\gamma_{3}}d\sigma_{3}\left[\frac{\sigma_{41}\sigma_{12}}{\sigma_{13}}\frac{1}{\hat{S}_{3}}\alpha_{13}\right]\mathcal{C}_{4}=-\delta^{3}(\vec{k}_{T})\int_{\sigma_{13}}d\sigma_{3}\left[\frac{\sigma_{41}\sigma_{12}}{\sigma_{13}}\frac{1}{\hat{S}_{3}}\alpha_{13}\right]\mathcal{C}_{4}^{\Delta}.
\end{equation}
Computing the residue then gives
\begin{equation}
\Psi_{4}^{\phi^4}=-\delta^{3}(\vec{k}_{T})\left[\left.\frac{\sigma_{12}\sigma_{41}}{\hat{S}_{3}}\right|_{\sigma_{13}=0}\alpha_{13}\right]\mathcal{C}_{4}^{\Delta}=\delta^{3}(\vec{k}_{T})\,\mathcal{C}_{4}^{\Delta}.
\label{phi4answer}
\end{equation}

Next we consider the naive uplifts of the flat space integrands of section \ref{flatspaceintegrands}. Note that these uplifts will not describe the full four-point wavefunction coefficients in dS because they will be missing mass deformations and curvature corrections. We will explain how to encode these additional terms in the worldsheet integrands in the next subsection. The naive uplift of the NLSM integrand in \eqref{eqn:NLSMIntegrand} is given by
\begin{equation}
\mathcal{I}^{NLSM}= {\rm PT}\left({\rm Pf}'A\right)^{2}.
\end{equation}
As before, we will fix legs $\left\{ 1,2,4\right\} $ and delete legs $\left\{ 2,4\right\} $ from the $A$-matrix to obtain
\begin{equation}
\Psi_{4}^{NLSM}=\delta^{3}(\vec{k}_{T})\int_{\gamma_{3}}d\sigma_{3}\left(\sigma_{41}\sigma_{12}\sigma_{24}\right)^{2}\left[\frac{1}{\sigma_{12}\sigma_{23}\sigma_{34}\sigma_{41}}\frac{\sigma_{13}\sigma_{23}\sigma_{43}}{\hat{S}_{3}}\left(\frac{\alpha_{13}}{\sigma_{13}\sigma_{24}}\right)^{2}\right]\mathcal{C}_{4}^{\Delta}.
\end{equation}
After simplifying the integrand there is once again a simple pole at $\sigma_{13}=0$, so we wrap the contour around this pole:
\begin{equation}
\Psi_{4}^{NLSM}=-\delta^{3}(\vec{k}_{T})\int_{\gamma_{3}}d\sigma_{3}\left[\frac{\sigma_{12}\sigma_{41}}{\sigma_{13}}\frac{1}{\hat{S}_{3}}\alpha_{13}^{2}\right]\mathcal{C}_{4}^{\Delta}=\delta^{3}(\vec{k}_{T})\int_{\sigma_{13}=0}d\sigma_{3}\left[\frac{\sigma_{12}\sigma_{41}}{\sigma_{13}}\frac{1}{\hat{S}_{3}}\alpha_{13}^{2}\right]\mathcal{C}_{4}^{\Delta}.
\end{equation} 
Evaluating the residue of this pole finally gives
\begin{equation}
\Psi_{4}^{NLSM}=\delta^{3}(\vec{k}_{T})\left[\left.\frac{\sigma_{12}\sigma_{41}}{\hat{S}_{3}}\right|_{\sigma_{13}=0}\alpha_{13}^{2}\right]\mathcal{C}_{4}^{\Delta}=-\delta^{3}(\vec{k}_{T})\alpha_{13}\,\mathcal{C}_{4}^{\Delta}.
\label{nlsmanswer}
\end{equation}
In Appendix \ref{sec:6-pt-worldsheet}, we generlise this computation to 6-points. This will illustrate a number subtleties that can arise for worldsheet descriptions of theories with derivative interactions, such as the presence of higher-order poles and potential ordering ambiguities.  

Now we analyse the naive uplift of the DBI integrand,
\begin{equation} \label{eq:DBIintegrand}
\mathcal{I}^{DBI}={\rm Pf}X\left({\rm Pf}'A\right)^{3}.
\end{equation}
Fixing legs $\left\{ 1,2,4\right\} $ as above, we find
\begin{equation}
\Psi_{4}^{DBI}=\delta^{3}(\vec{k}_{T})\int_{\gamma_{3}}d\sigma_{3}(\sigma_{14}\sigma_{12}\sigma_{24})^{2}\left[{\rm Pf}X\frac{1}{S_{3}}\left({\rm Pf}'A\right)^{3}\right]\mathcal{C}_{4}^{\Delta}.
\label{4ptdbi}
\end{equation}
Recall from \eqref{pfX} that $\rm{Pf}X$ has three terms. To evaluate the worldsheet integral containing the first term,
it is convenient to choose $({\rm Pf}'A)^{3}=({\rm Pf}A_{23}^{23})^{2}({\rm Pf}A_{24}^{24})$ such that
\begin{equation}
\Psi_{4}^{12,34}=\delta^{3}(\vec{k}_{T})\int_{\gamma_{3}}d\sigma_{3}\frac{\left(\sigma_{14}\sigma_{12}\sigma_{24}\right)^{2}}{\sigma_{12}\sigma_{34}}\left[\frac{\sigma_{13}\sigma_{23}\sigma_{43}}{\hat{S}_{3}}\left(\frac{\alpha_{14}}{\sigma_{14}\sigma_{23}}\right)^{2}\frac{\alpha_{13}}{\sigma_{13}\sigma_{24}}\right]\,\mathcal{C}_{4}^{\Delta},
\end{equation}
where the superscript on $\Psi_{4}$ denotes the contribution from the first term in \eqref{pfX}. The contour integral can be evaluated as above to obtain 
\begin{equation}
\Psi_{4}^{12,34}=\delta^{3}(\vec{k}_{T})\left[\left.\frac{\sigma_{12}\sigma_{24}}{\hat{S}_{3}}\right|_{\sigma_{23}=0}\alpha_{14}^{2}\alpha_{13}\right]\mathcal{C}_{4}^{\Delta}=-\delta^{3}(\vec{k}_{T})\alpha_{14}\alpha_{13}\mathcal{C}_{4}^{\Delta},
\end{equation}
where we wrapped the contour around the pole $\sigma_{23}=0$, evaluated the residue, and used the CWI to cancel $\alpha_{23}$ in the denominator with $\alpha_{14}$ in the numerator. Note that the ordering of $\alpha_{14}$ and $\alpha_{13}$ in the final result is not important due to \eqref{eq:commute}. For the remaining two terms in \eqref{4ptdbi} we choose $({\rm Pf}'A)^{3}$ to be $({\rm Pf}A_{14}^{14})^{2}({\rm Pf}A_{12}^{12})$ and $({\rm Pf}A_{13}^{13})^{2}({\rm Pf}A_{12}^{12})$, respectively. Using similar manipulations, we finally obtain
\begin{equation}
\Psi_{4}^{DBI}=-\delta^{3}(\vec{k}_{T})\left(\alpha_{12}\alpha_{14}+\alpha_{14}\alpha_{13}+\alpha_{13}\alpha_{12}\right)\mathcal{C}_{4}^{\Delta}.
\label{dbianswer}
\end{equation}
This choice of Pfaffians avoids higher-order poles in the worldsheet coordinates which are more subtle to evaluate. We describe such an example in  Appendix \ref{4point-DP}. It will also be useful to consider the following integrand, whose corresponding wavefunction coefficient follows trivially from the above calculation:
\begin{equation}
({\rm Pf}'A)^{3}\left.{\rm Pf}X\right|_{{\rm conn}}\rightarrow-\delta^{3}(\vec{k}_{T})\alpha_{12}\alpha_{14}\mathcal{C}_{4}^{\Delta}.
\label{dbi2}
\end{equation}

Finally, let us consider the naive uplift of the special Galileon integrand
\begin{equation}
\mathcal{I}^{sGal}=\left({\rm Pf}'A\right)^{4}.
\end{equation}
In this case, the four-point wavefunction coefficient is given by
\begin{equation}
\Psi_4^{sGal}=\delta^{3}(\vec{k}_{T})\int_{\gamma_{3}}d\sigma_{3}\left(\sigma_{14}\sigma_{12}\sigma_{24}\right)^{2}\left[\frac{1}{S_{3}}\left({\rm Pf}'A\right)^{4}\right]\mathcal{C}_{4}^{\Delta}.
\end{equation}
As in the DBI case, there are Pfaffian choices exclusively leading to simple poles. The following choice leads to a permutation invariant result:
\begin{equation}
({\rm Pf}'A)^{4}=\frac{1}{3}\left\{ \frac{1}{\s^2_{34}} ({\rm Pf}A_{34}^{34})^{2} \, \frac{(-1)}{\s_{23}} ({\rm Pf}A_{23}^{23}) \frac{1}{\s_{24}} \,  ({\rm Pf}A_{24}^{24})+{\rm cyclic}(2,3,4)\right\}.
\end{equation}
Note that other choices can give different results due to non-trivial commutators which only vanish in the flat space limit. Hence we must specify a choice of Pfaffian. Following closely the computations above, we obtain
\begin{equation}
\Psi_{4}^{sGal}=\frac{1}{3}\delta^{3}(\vec{k}_{T})\left(\alpha_{12}\alpha_{14}\alpha_{13}+\alpha_{14}\alpha_{13}\alpha_{12}+\alpha_{13}\alpha_{14}\alpha_{12}\right)\mathcal{C}_{4}^{\Delta}.
\label{sgalanswer}
\end{equation}
Using \eqref{eq:commute}, it is not difficult to see that the above expression is permutation invariant.

\subsection{Generalised Double Copy} \label{gendoublecopy}

Using the building blocks in the previous subsection (in particular \eqref{phi4answer} and \eqref{nlsmanswer}), we see that the four-point NLSM wavefunction coefficient (including a curvature correction) in \eqref{nlsmcorr} follows from the following integrand:
\begin{equation}
\mathcal{I}_{4}^{NLSM}=\lambda^{2}{\rm PT}\left({\rm Pf}'A\right)^{2}+c{\rm PT}\left.{\rm Pf}X\right|_{\rm{conn}}{\rm Pf'}A,
\label{nlsmdsintegrand}
\end{equation}
if we identify the unfixed parameter $c$ with $-C$. Now consider the following shift of the first term in \eqref{nlsmdsintegrand}:
\begin{equation}
\lambda^2 {\rm PT}\rightarrow a{\rm Pf'}A\left({\rm Pf'}A+m^{2}\left.{\rm Pf}X\right|_{{\rm conn}}\right)+b\left({\rm Pf'}A{\rm Pf}X+m^{2}{\rm PT}\right),
\label{eq:doublecopy}
\end{equation}
where $a$ and $b$ are also free coefficients. We then obtain the following integrand:
\begin{multline}
\mathcal{I}_{4}^{(6)}=a({\rm Pf}'A)^{3}({\rm Pf'}A+m^{2}\left.{\rm Pf}X\right|_{{\rm conn}}) +b({\rm Pf}'A)^{2}({\rm Pf'}A{\rm Pf}X+m^{2}{\rm PT})+c {\rm PT}\left.{\rm Pf}X\right|_{\rm{conn}}{\rm Pf'}A.
\label{6derivintegrand}
\end{multline}
Using the prescription for evaluating worldsheet integrals described the previous subsection (in particular \eqref{phi4answer}, \eqref{nlsmanswer}, \eqref{dbianswer},  \eqref{dbi2}, and \eqref{sgalanswer}), we see that the corresponding wavefunction coefficient is
\begin{equation}
\Psi_{4}^{(6)}=\delta^{3}(\vec{k}_{T})\left[\tfrac{8}{3}a(\tilde{s}\tilde{t}\tilde{u}+\tilde{t}\tilde{u}\tilde{s}+\tilde{u}\tilde{s}\tilde{t})-4b(\tilde{s}\tilde{t}+\tilde{t}\tilde{u}+\tilde{u}\tilde{s})+c\right]\mathcal{C}_{4}^{\Delta},
\label{6dervwavefunction coefficient}
\end{equation}
where $\tilde{s}$, $\tilde{t}$, and $\tilde{u}$ are related to their hatted counterparts \eqref{eq:GenMandelstam} as
\begin{equation}
\begin{array}{ccccc}
\tilde{s}=\hat{s}-\tfrac{1}{2}m^{2}, & & \tilde{t}=\hat{t}-\tfrac{1}{2}m^{2}, & & \tilde{u}=\hat{u}-\tfrac{1}{2}m^{2}.
\end{array}
\label{stilde}
\end{equation}

It is then straightforward to match \eqref{6dervwavefunction coefficient} with \eqref{4pteffective} via the following identification of unfixed coefficients:  
\begin{equation}
\begin{array}{ccccc}
A=\tfrac{8}{3}a,\,\,\,B=2a\left(m^{2}+\frac{4}{3}d\right)-2b,\,\,\,C=-\tfrac{1}{3}am^{6}+bm^{4}-c.
\end{array}
\end{equation}
Hence, the worldsheet integrand \eqref{6derivintegrand} encodes the effective action in \eqref{effectiveaction}. Moreover, \eqref{eq:doublecopy} can be thought of as a double copy procedure encoding mass deformations and curvature corrections. In particular, it reduces to \eqref{sGaldouble} and \eqref{DBIdouble} in the flat space limit for $a \neq 0$ and $a=0$, respectively (recalling that the mass is measured in units of the inverse dS radius). We therefore refer to this as a generalised double copy. Note that this prescription leaves the curvature corrections and mass deformations unfixed. To fix these coefficients, we must specify additional data beyond the flat space limit. We will suggest a strategy for doing this in the next section.

\section{Soft Limits} \label{soft}

In the previous section, we found the building blocks of four-point wavefunction coefficients of scalar EFTs in dS to be:
\begin{equation}\label{eq:4pt-list}
\mathcal{C}_{4}^{\Delta},\,\,\,\hat{u}\mathcal{C}_{4}^{\Delta},\,\,\,\left(\hat{s}^{2}+\hat{t}^{2}+\hat{u}^{2}\right)\mathcal{C}_{4}^{\Delta},\,\,\,\left(\hat{s}^{3}+\hat{t}^{3}+\hat{u}^{3}\right)\mathcal{C}_{4}^{\Delta}.
\end{equation} 
For notational simplicity, we will leave out the delta function imposing momentum conservation along the boundary. The first one is a $\phi^4$ contact diagram, while the remaining building blocks arise from the action on this contact diagram with the differential operators defined in \eqref{eq:GenMandelstam}. The first two terms contribute when lifting NLSM to dS, while the last two arise when lifting the DBI and sGal theories. In this section, we will compute these objects explicitly in the cases $\Delta=2,3$ and analyse their soft limits. In practice, all the integrals we encounter will be of the form $\int^0_{-\infty} \eta^\alpha e^{-iE \eta}d\eta$. To evaluate them, we rotate the contour clockwise onto the complex plane so that the integrand is exponentially damped as $\eta \to -\infty$ \cite{Maldacena:2011nz}. To simplify our expressions we will set the normalisation of the bulk-to-boundary propagators in \eqref{bulktoboundaryprop} to be $\mathcal{N}=\left(-1\right)^{\nu-\frac{1}{2}}\sqrt{2/\pi}$.

We will show that the soft limits of the building blocks in \eqref{eq:4pt-list} are given in terms of boundary conformal generators acting on certain three-point contact diagrams. Soft limits play an important role in cosmology where they appear in constraints relating higher-point functions to symmetry transformations of lower-point functions \cite{Maldacena:2002vr,Creminelli:2012ed,Hinterbichler:2013dpa,Kundu:2014gxa} and allow one to deduce 3-point inflationary correlators from four-point dS correlators \cite{Creminelli:2003iq,Assassi:2012zq,Arkani-Hamed:2015bza,Kundu:2015xta,Shukla:2016bnu}. Soft limits of DBI and sGal wavefunction coefficients in flat space were recently analysed in \cite{Bittermann:2022nfh}. Note that the soft limit of the four-point wavefunction coefficients in \eqref{nlsmcorr} and \eqref{4pteffective} can be obtained from linear combinations of soft limits we derive in this section. By appropriately choosing the coefficients of curvature corrections in the corresponding EFTs in \eqref{nlsmlift} and \eqref{effectiveaction} we can set many of these terms to zero, which may signal the existence of hidden symmetries.

\subsection{Contact diagram}

Let us first consider the four-point contact diagram. For conformally coupled scalars, we find
\begin{equation}
\mathcal{C}_{4}^{\Delta=2}=\int\frac{d\eta}{\eta^{4}}\left(\mathop{\prod} \limits_{i=1}^{4}\mathcal{K}_{1/2}^{i}\right)=\frac{1}{E}.
\end{equation}
It is trivial to see that the soft limit of this quantity corresponds to a three-point contact diagram for conformally coupled scalars with a time-dependent interaction:
\begin{equation}
\lim_{\vec{k}_{1}\rightarrow0}C_{4}^{\Delta=2}=\mathcal{C}_{3,\eta}^{\Delta=2},
\end{equation}
where 
\begin{equation}
\mathcal{C}_{3,\eta}^{\Delta=2}=\int\frac{d \eta}{\eta^{4}} \left(\eta \, \mathop{\prod} \limits_{i=2}^{4}\mathcal{K}_{1/2}^{i}\right)=\frac{1}{E}.
\label{3ptconf}
\end{equation}
This three-point contact diagram will also arise in the soft limit of more complicated four-point wavefunction coefficients of conformally coupled scalars. In particular, we will obtain conformal generators acting on \eqref{3ptconf}.

For minimally coupled scalars, the integrals need to be regulated. We will use the prescription $d\rightarrow d+2\epsilon$, $\Delta\rightarrow\Delta+\epsilon$, which leaves the spectral parameter $\nu=\Delta-d/2$ unchanged \cite{Bzowski:2015pba,Bzowski:2017poo,Bzowski:2018fql,paul}, such that
\begin{eqnarray}
\mathcal{C}_{4}^{\Delta=3+\epsilon}&=&\int\frac{d\eta}{\eta^{4+2\epsilon}}\mathop{\prod} \limits_{i=1}^{4}\mathcal{K}_{3/2}^{i}, \nonumber \\
& = &\frac{1}{E^{2\epsilon}}\big\{E^{3} [\Gamma(-2+2\epsilon)+\Gamma(-3+2\epsilon)]+\Gamma(2\epsilon)(k_{1}k_{2}k_{3}+\ldots)\\
& & +\Gamma(-1+2\epsilon)E\left(k_{1}k_{2}+\ldots\right)+\frac{k_{1}k_{2}k_{3}k_{4}}{E}\Gamma(1+2\epsilon)\big\}\nonumber,
\end{eqnarray}
where the ellipsis inside each parenthesis denotes all permutations involving the four legs. We then find the following soft limit:
\begin{align}
\lim_{\vec{k}_{1}\rightarrow0}\mathcal{C}_{4}^{\Delta=3+\epsilon} = &\frac{1}{E^{2\epsilon}}\{ E^3[\Gamma(-2+2\epsilon) +\Gamma(-3+2\epsilon)] +\Gamma(2\epsilon)k_{2}k_{3}k_{4} \nonumber\\ & +\Gamma(-1+2\epsilon)E(k_{2}k_{3}+k_{2}k_{4} +k_{3}k_{4}) \}.
\label{softmincontact}
\end{align}

We can now compare this with the three-point contact diagram of minimally coupled scalars:
\eqs{
\mathcal{C}_{3}^{\Delta=3+\epsilon}=\int\frac{d\eta}{\eta^{4+2\epsilon}}\mathop{\prod} \limits_{i=2}^{4}\mathcal{K}_{3/2}^{i}.
\label{min3pt}
}
Unlike the contact diagram in \eqref{3ptconf}, this contact diagram does not contain a time-dependent interaction. After performing the integral in \eqref{min3pt} and changing $\epsilon\rightarrow 2 \epsilon$, we finally  obtain \eqref{softmincontact}:
\begin{equation}
\lim_{\vec{k}_{1}\rightarrow0}C_{4}^{\Delta=3+\epsilon}=\mathcal{C}_{3}^{\Delta=3+2\epsilon},
\end{equation}
which, for clarity, can be easily expanded in $\epsilon$:
\begin{equation}
\mathcal{C}_{3}^{\Delta=3+\epsilon}=\frac{k_2^3+k_3^3+k_4^3}{3}\left(\frac{1}{\epsilon}-\gamma_E-\ln E+ 1\right)+\frac{1}{9} E^3 - k_2k_3k_4 + \mathcal{O}(\epsilon),
\label{3ptminexpanded}
\end{equation}
where $\gamma_E$ is the Euler-Mascheroni constant. As we will see, \eqref{min3pt} will continue to play a role in the soft limit of more complicated four-point wavefunction coefficients of minimally coupled scalars. 

\subsection{Two Derivatives}

Next, we will analyse the soft limit of  $\hat{u} \mathcal{C}_4^{\Delta}$. In the conformally coupled case, it can be cast as
\eq{
\hat{u}\mathcal{C}_{4}^{\Delta=2}=-\frac{2}{E^{3}}[\vec{k}_1\cdot\vec{k}_3 - k_1k_3-\tfrac{E}{2}(k_{2}+k_{4})],
\label{uhatcontact}
}
with soft limit
\eq{
\lim_{\vec{k}_{1}\to0}\hat{u}\mathcal{C}_{4}^{\Delta=2}=\frac{1}{E}-\frac{k_{3}}{E^{2}}.
}
We recognise this expression as the dilatation operation:
\eq{
D_3 \left(\frac{1}{E}\right) = \frac{1}{E} - \frac{k_3}{E^2}.
\label{eqn:NLSMconfSoft}
}
Hence, we find that the soft limit of \eqref{uhatcontact} can be obtained by acting with a dilatation on the three-point contact diagram of \eqref{3ptconf}:
\begin{equation}
\lim_{\vec{k}_{1}\to0}\hat{u}\mathcal{C}_{4}^{\Delta=2}= D_{3} \, \mathcal{C}_{3,\eta}^{\Delta=2}.
\label{2dervsoftconf}
\end{equation}

In the minimally coupled case, we find
\eqs{
\hat{u}\mathcal{C}_{4}^{\Delta=3} &=\frac{\vec{k}_1\cdot\vec{k}_3}{E}\left[-E^2 + \sum_{i<j}k_ik_j+\frac{k_1k_2k_3k_4}{E}\left(\frac{1}{k_1} + \frac{1}{k_2} +\frac{1}{k_3} +\frac{1}{k_4} +\frac{2}{E} \right)\right]\\
&\qquad -2\frac{k_1^2k_3^2}{E}\left(1 + \frac{k_2+k_4}{E} + \frac{2k_2k_4}{E^2}\right),
\label{eqn:NLSM4ptMinCHY}
}
where requires dimensional regularisation in intermediate steps but has a finite output. Taking the soft limit then gives
\eq{
\lim_{\vec{k}_{1}\to0}\hat{u}\mathcal{C}_{4}^{\Delta=3}=\frac{\vec{k}_1\cdot\vec{k}_3}{E}\left(k_2^2+k_3^2+k_4^2 + k_2k_3+k_3k_4+k_4k_2-\frac{k_2k_3k_4}{E}\right),
}
which can be obtained by acting with a conformal boost on the three-point contact diagram in \eqref{min3pt}: 
\begin{equation}
\lim_{\vec{k}_{1}\to0}\hat{u}\mathcal{C}_{4}^{\Delta=3}=(\vec{k}_{1}\cdot K_{3})\, \mathcal{C}_{3}^{\Delta=3}.
\label{2dervsoftmin}
\end{equation}
Note that the divergences in \eqref{3ptminexpanded} are removed by the action of conformal generators so we set $\epsilon=0$. This will continue to hold for higher-derivative wavefunction coefficients so we will set $\epsilon=0$ in those cases as well.

\subsection{Four Derivatives}

We now consider the term $(\hat{s}^{2}+\hat{t}^{2}+\hat{u}^{2})\mathcal{C}_{4}^{\Delta}$. For conformally coupled scalars, we obtain
\eq{
(\hat{s}^{2}+\hat{t}^{2}+\hat{u}^{2})\mathcal{C}^{\Delta=2}_{4} = \frac{24}{E^5}[(\vec{k}_1\cdot\vec{k}_2-k_1k_2)(\vec{k}_3\cdot\vec{k}_4-k_3k_4) + \mathrm{cyc}(234)] - \frac{8}{E^3}\sum_{i<j}k_ik_j + \frac{4}{E}.
}
In this case the soft limit is simply
\eq{
\lim_{{\vec{k}_1}\to0}(\hat{s}^{2}+\hat{t}^{2}+\hat{u}^{2})\mathcal{C}^{\Delta=2}_{4}  = \frac{4}{E^3}(k_2^2+k_3^2+k_4^2),
}
which can be recast as a second order combination of dilatation operators acting on the three-point contact diagram of \eqref{3ptconf}:
\eq{
\lim_{{\vec{k}_{1}}\to0}(\hat{s}^{2}+\hat{t}^{2}+\hat{u}^{2})\mathcal{C}_{4}^{\Delta=2}=2(D_{2}^{2}+D_{3}^{2}+D_{4}^{2})\mathcal{C}_{3,\eta}^{\Delta=2}.
\label{soft4derconf}
}

In the minimally coupled case, we obtain
\begin{multline}
(\hat{s}^{2}+\hat{t}^{2}+\hat{u}^{2})\mathcal{C}^{\Delta=3}_{4} = \frac{24k_1k_2k_3k_4}{E^5}(\vec{k}_1\cdot\vec{k}_2 - k_1k_2)(\vec{k}_3\cdot\vec{k}_4 - k_3k_4)\\ 
 + \frac{2}{E^3}(\vec{k}_1\cdot\vec{k}_2)(\vec{k}_3\cdot\vec{k}_4)\big(E^2 + E\sum_{i<j}k_ik_j + \sum_{i<j<l}k_ik_jk_l\big)\\
 - \frac{2}{E^4}[(\vec{k}_1\cdot\vec{k}_2)k_3^2k_4^2(E+2(k_1+k_2) + (\vec{k}_3\cdot\vec{k}_4)k_1^2k_2^2(E+2(k_3+k_4)) + \mathrm{cyc}(234)].
\end{multline}
We then find that the soft limit is given by 
\eqs{
\lim_{{\vec{k}_{1}}\to0}(\hat{s}^{2}+\hat{t}^{2}+\hat{u}^{2})\mathcal{C}_{4}^{\Delta=3} &= \frac{\vec{k}_1\cdot \vec{k}_2}{E^3}\big[4k_2^2 k_3k_4+k_2 E(k_3 k_4+2k_2(k_3+k_4))\\
&\qquad\qquad+E^2(2k_2^2+k_2k_3+k_3k_4+k_4k_2)-E^4)\big] +  \mathrm{cyc}(234),
} which corresponds to applying the following quadratic combination of conformal generators to a three-point contact diagram: 
\begin{equation}
\lim_{{\vec{k}_{1}}\to0}(\hat{s}^{2}+\hat{t}^{2}+\hat{u}^{2})\mathcal{C}_{4}^{\Delta=3}=-2\big[D_{2}(\vec{k}_{1}\cdot K_{2})+\mathrm{cyc}(234)\big]\mathcal{C}_{3}^{\Delta=3}.
\label{soft4dermin}
\end{equation}

In practice, the soft limits in \eqref{soft4derconf} and \eqref{soft4dermin} are most easily derived at the level of the integrand. In particular, this requires taking the soft limit of bulk-to-boundary propagators and their derivatives, and then using equations of motion to remove derivatives acting on the bulk-to-boundary propagator for the soft leg as well as factors of $k_a^2$. For example, in the conformally coupled case we have
\eqs{
\lim_{{\vec{k}_{1}}\to0}(\hat{s}^{2}+\hat{t}^{2}+\hat{u}^{2})\mathcal{C}_{4}^{\Delta=2} &= \int d\eta\left(\frac{1}{\eta}(-k_2^2\dot{\mathcal{K}}_1\mathcal{K}_2 + \dot{\mathcal{K}}_1\ddot{\mathcal{K}}_2) - \frac{1}{\eta^2}\dot{\mathcal{K}}_1\dot{\mathcal{K}}_2\right)\mathcal{K}_3\mathcal{K}_4 + \mathrm{cyc}(234),\\
&= \int \frac{d\eta}{\eta^3}[2\eta^2 \ddot{\mathcal{K}}_2 - \eta\dot{\mathcal{K}}_2 - 2\mathcal{K}_2]\mathcal{K}_3\mathcal{K}_4 + \mathrm{cyc}(234),\\
&= 2\left(D_2^2 + D_3^2+D_4^2\right)\int\frac{d\eta}{\eta^3}\mathcal{K}_2\mathcal{K}_3\mathcal{K}_4.
\label{softintegrand}
}
The analogous construction in the minimally coupled case is given by
\eqs{
\lim_{{\vec{k}_{1}}\to0}\left(\hat{s}^{2}+\hat{t}^{2}+\hat{u}^{2}\right)\mathcal{C}_{4}^{\Delta=3} &=(\vec{k}_1\cdot\vec{k}_2) \int d\eta \left(\frac{2}{\eta}\mathcal{K}_1\dot{\mathcal{K}}_2 + \frac{1}{\eta^2}\mathcal{K}_1\mathcal{K}_2\right)\mathcal{K}_3\mathcal{K}_4 + \mathrm{cyc}(234),\\
&= -2[D_2(\vec{k}_1\cdot K_2) +\mathrm{cyc}(234)]\int\frac{d\eta}{\eta^4}\mathcal{K}_2\mathcal{K}_3\mathcal{K}_4.
}
Using this method, one can also derive the soft limits in \eqref{2dervsoftconf} and \eqref{2dervsoftmin}.

\subsection{Six Derivatives} 

Finally we consider the six derivative interaction $(\hat{s}^3 + \hat{t}^3+\hat{u}^3)\mathcal{C}_4^{\Delta}$. In the conformally coupled case we obtain
\begin{multline}
(\hat{s}^3 + \hat{t}^3+\hat{u}^3)\mathcal{C}_4^{\Delta=2} = \Big\{\frac{90}{E^7}[(\vec{k}_1\cdot \vec{k}_2)+(\vec{k}_3\cdot \vec{k}_4)-k_1k_2 - k_3k_4]^3  + \frac{156}{E^5}[(\vec{k}_1\cdot \vec{k}_2)-k_1k_2][(\vec{k}_3\cdot \vec{k}_4)-k_3k_4]\\
\qquad\qquad - \frac{54}{E^5}(k_1k_2 + k_3k_4)[(\vec{k}_1\cdot \vec{k}_2)+(\vec{k}_3\cdot \vec{k}_4)-k_1k_2 - k_3k_4] - \frac{79}{E^3}(k_1k_2 + k_3k_4) + \mathrm{cyc}(234)\Big\}+\frac{93}{4E},
\end{multline}
with soft limit
\eq{
\lim_{{\vec{k}_{1}}\to0}(\hat{s}^{3}+\hat{t}^{3}+\hat{u}^{3})\mathcal{C}_{4}^{\Delta=2}  = -\frac{108}{E^4}k_2k_3k_4 + \frac{52}{E^3}(k_2k_3+k_3k_4+k_4k_2)-\frac{12}{E}.
}
Using the methods described above, we can obtain this by acting with the following combination of conformal generators on a three-point contact diagram:
\eq{
\lim_{\vec{k}_{1}\to0}(\hat{s}^{3}+\hat{t}^{3}+\hat{u}^{3})\mathcal{C}_{4}^{\Delta=2}=\left(6(D_{2}^{3}+D_{3}^{3}+D_{4}^{3})-22(D_{2}^{2}+D_{3}^{2}+D_{4}^{2})+20\right)\mathcal{C}_{3,\eta}^{\Delta=2}.
\label{eqn:4ptSoftsGal}
}

In the minimally coupled case, the expression for the integrated wavefunction coefficient can be found in appendix \ref{app:sGalMin}. The soft limit is given by
\begin{multline}
\lim_{\vec{k}_{1}\to0}(\hat{s}^{3}+\hat{t}^{3}+\hat{u}^{3})\mathcal{C}_{4}^{\Delta=3}= \frac{(\vec{k}_1\cdot\vec{k}_2)}{E^4}\big[3 k_2^5 + 12 k_2^4 (k_3 + k_4) + k_2^3 (k_3^2 + 36 k_3 k_4 + k_4^2)\\
 - k_2^2 (k_3 + k_4) (11 k_3^2 + 21 k_3 k_4 + 11 k_4^2) -  4 k_2 (k_3 + k_4)^2 (k_3^2 + k_3 k_4 + k_4^2)\\
 - (k_3 + k_4)^3 (k_3^2 + k_3 k_4 +  k_4^2)\big] + \mathrm{cyc}(234).
\end{multline}
This expression can also be recast in terms of conformal generators acting on a three-point contact diagram. For example, we can write
\eqs{
\lim_{\vec{k}_{1}\to0}(\hat{s}^{3}+\hat{t}^{3}+\hat{u}^{3})\mathcal{C}_{4}^{\Delta=3}=[2(3D_{2}^{2}-11D_{2})(\vec{k}_{1}\cdot K_{2})+\mathrm{cyc}(234)]\mathcal{C}_{3}^{\Delta=3}.
\label{eqn:sGalMinSoft}
}
Observe that these expressions are not unique. For example, the operators $K_2$ and $D_2$ do not commute and so we could choose to order them differently and pick up different coefficients. Note also that unlike \eqref{eqn:4ptSoftsGal}, the term in brackets in \eqref{eqn:sGalMinSoft} contains no constant term. This comes from the different behavior of the soft limits for different values of $\Delta$. The constant term in \eqref{eqn:4ptSoftsGal}  arises from $D_2+D_3+D_4$ and the equivalent term in the case of massless scalars would be a sum over $\vec{k}_1\cdot K_i$. We can rewrite this using CWI to get a contribution of order $k_1^2$ which we then drop as it is subleading.

Let us close this section with some general comments. Overall, we find that soft limits require both special conformal and dilatation operators acting on a contact diagram. In the conformally coupled case, it is always possible to express them exclusively in terms of dilatation operators but at the level of the integrand this is not the whole story. When considering how the integrand behaves in the soft limit we find that both types of conformal generators appear. Another interesting point is that we can obtain particularly simple soft limits by tuning coefficients in the effective action in \eqref{effectiveaction}. For example, if we choose the mass to be that of a conformally coupled scalar and set the coefficients $\left\{ A,B,C\right\} =\lambda\left\{ 1,-8,-20\right\} $, then the soft limit of the corresponding four-point wavefunction coefficient in \eqref{4pteffective} is simply given by   
\eq{
\lim_{\vec{k}_{1}\to0}\Psi_4^{(6)}=6\lambda(D_{2}^{3}+D_{3}^{3}+D_{4}^{3})\mathcal{C}_{3,\eta}^{\Delta=2}.
}
It would be interesting to look for possible hidden symmetries in the corresponding EFT. We could perform a similar set of steps for $\Delta=3$, but in this case there is an ambiguity in how to fix the coefficients since the commutators between the operators $D_i$ and $K_i$ are given by lower order contributions so different orderings  might lead to different preferred choices for $\left\{ A,B,C\right\} $.

\section{Conclusion}
\label{sec:conclusion}

The study of correlation functions in (Anti) de Sitter space using curved-space analogues of the scattering equations is still an ongoing endeavour. Prior to this paper, they have only been formulated for scalar theories with polynomial interactions, notably $\phi^3$ \cite{Eberhardt:2020ewh,Roehrig:2020kck} and $\phi^4$ \cite{Gomez:2021qfd,Gomez:2021ujt}. We have now initiated the study of scalar theories with derivative interactions using this framework. In particular, we study the wave function coefficients of the NLSM, DBI, and sGal theories in dS using the cosmological scattering equations. These effective scalar theories are of particular interest in flat space since their scattering amplitudes have a very elegant description and are related to each other in a simple way in terms of CHY formulae \cite{Cachazo:2014xea}.

In this paper we have proposed new formulae for four-point wavefunction coefficients of different scalar EFTs in the form of worldsheet integrals which encode both curvature corrections and mass deformations. We showed that the DBI and sGal integrands can be obtained from the NLSM integrand by replacing a Parke-Taylor factor with a linear combination of simple building blocks involving Pfaffians of certain operatorial matrices. Because the integrands are constructed from differential operators which do not generally commute, this leads to potential ordering ambiguities that are absent in theories with polynomial interactions. Such ambiguities can occur in the DBI and sGal theories at four points and the NLSM at six points. At four points we introduced a simple prescription for defining the worldsheet integrals such that the final results are permutation-invariant. We have also studied the soft limits of the resulting four-point wavefunction coefficients and derived formulae in the form of differential operators acting on three-point contact diagrams. For conformally coupled scalars, the three-point contact diagrams involve a time-dependent interaction, while for minimally coupled scalars the contact diagram is divergent and needs to be regulated. However, all divergences cancel out after acting with the appropriate combinations of boundary conformal generators.

There are a number of directions for future investigation. For example, it would be of interest to generalise our formulae to any number of points. In order to do so, there are several technical questions that need to be addressed. First of all, we need to develop a systematic classification of higher-derivative corrections to scalar EFTs analogous to the one in \cite{Heemskerk:2009pn} beyond four-points, which to our knowledge has not been carried out yet. It would then be interesting to see if there is a simple generalisation of the double copy prescription in \eqref{eq:doublecopy} which encodes such corrections at higher points. Secondly, we need a systematic understanding of ordering ambiguities that could arise in the corresponding worldsheet integrals and a prescription for removing them. A natural starting point along these lines would be to analyse potential ordering ambiguities in the six-point NLSM calculation in Appendix \ref{sec:6-pt-worldsheet}. 

It would also be of great interest to apply our approach for computing wavefunction coefficients to the EFTs recently constructed in de Sitter space based on hidden symmetries \cite{Bonifacio:2021mrf}. Note that the effective actions we consider in this paper have general massess and curvature corrections up to four-points, so the actions derived in \cite{Bonifacio:2021mrf} should correspond to a particular choice of these parameters. It would then be interesting to compute the corresponding four-point wavefunction coefficients and their soft limits in terms of the building blocks derived in this paper and investigate the interplay of hidden symmetries with soft limits. Note that our double copy prescription does not fix massess or coefficients of curvature corrections, but these can be constrained by imposing simplicity of the soft limits, as we illustrate in the end of section \ref{soft}. It would also be interesting to relate our results to those recently obtained in flat space \cite{Bittermann:2022nfh} using the methods recently developed in \cite{Hillman:2021bnk}. 

Finally, it is important to extend the techniques of this paper to more realistic models by considering different worldsheet integrands encoding more general interactions and the breaking of conformal symmetry by inflation \cite{Cheung:2007st,Green:2020ebl,Pajer:2020wxk}. Moreover inflationary three-point correlators can be obtained by giving a small mass to one of the legs of a four-point function in de Sitter space (proportional to the slow-roll parameter) and then taking a soft limit \cite{Creminelli:2003iq,Assassi:2012zq,Arkani-Hamed:2015bza,Kundu:2015xta,Shukla:2016bnu}. Our results in section \ref{soft} should be useful in this regard.

\begin{center}
\textbf{Acknowledgements}
\end{center}
We thank Paul McFadden, Ashish Shukla, Allic Sivaramakrishnan, and Fei Teng for useful discussions. CA, HG, and AL are supported by the Royal Society via a PhD studentship, PDRA grant, and a University Research Fellowship, respectively. JM is supported by a Durham-CSC Scholarship. RLJ acknowledges the Czech Science Foundation - GA\v{C}R for financial support under the grant 19-06342Y.

\appendix

\section{Six-point NLSM wavefunction coefficients from Witten diagrams} \label{app:6ptNLSM}

In this appendix we extend the calculations in section \ref{4ptwitten} to six-points for the NLSM. At this level, we have the following Lagrangian,
\eqs{
\mathcal{L}_{\mathrm{NLSM}}^{\textrm{6-pt}}={\rm Tr}\left[-\frac{1}{2}\partial\Phi\cdot\partial\Phi-\frac{1}{2}m^{2}\Phi^{2}-\lambda^{2}\Phi^{2}\partial_{\mu}\Phi\partial^{\mu}\Phi-\frac{1}{4}C\Phi^{4}\right. \\
\quad \left.-\lambda^{4}\left(\Phi^{4}\partial_{\mu}\Phi\partial^{\mu}\Phi+\frac{1}{2}\Phi^{2}\partial_{\mu}\Phi\Phi^{2}\partial^{\mu}\Phi\right)-\frac{1}{6}F\Phi^{6}\right],
\label{6ptnlsmlagrangian}
}
where $m$ is the mass, and $C$ and $F$ are free coefficients coming from possible curvature corrections. For simplicity, we will set $m=C=F=0$. We will write the result in terms of the boundary differential operators in \eqref{eq:DaDb} acting on a contact diagram, which we will compare against the results of a worldsheet calculation in Appendix \ref{sec:6-pt-worldsheet}. The final result will be free of ambiguities since all the differential operators which appear commute. A similar formula was previously derived in \cite{Diwakar:2021juk} using AdS embedding coordinates.

\begin{figure}[b]
\centering
\includegraphics[width=12cm]{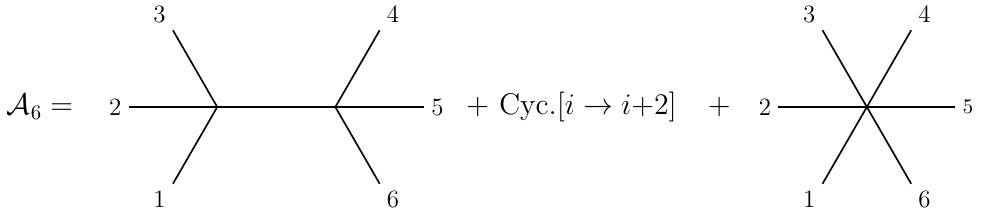}
\caption{Witten diagrams for the six-point NLSM amplitude. The exchange diagram is summed over the three inequivalent cyclic permutations.}
\label{fig:6ptDiagram}
\end{figure}

The wavefunction coefficient is given by a sum of exchange and contact diagrams shown in Figure \ref{fig:6ptDiagram}. 
 To obtain the desired form, we must write the four-point vertices in the exchange diagrams in such a way that derivatives only act on bulk-to-boundary propagators. 
\begin{figure}[b]
\centering
\includegraphics[width=4.5cm]{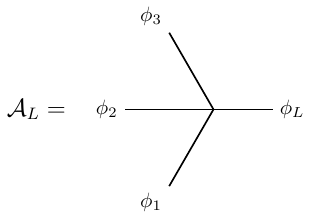}
\caption{Four-point vertex appearing on the left-hand-side of a 6-point NLSM exchange diagram.}
\label{fig:6ptIBP}
\end{figure}
Let us consider the four-point vertex appearing on the left-hand-side of an exchange diagram, illustrated in Figure \ref{fig:6ptIBP}. Using the Feynman rules derived from \eqref{6ptnlsmlagrangian} and the identity in \eqref{eq:DaDb} we find
\eqs{
\mathcal{A}_L &= \int \frac{d\eta}{\eta^{d+1}}\eta^2\big\{\mathcal{K}_1\mathcal{K}_2\mathcal{K}_3\mathcal{K}_L[(\vec{k}_1\cdot\vec{k}_2) + (\vec{k}_2\cdot\vec{k}_3) + (\vec{k}_3\cdot\vec{k}_L) + (\vec{k}_L\cdot\vec{k}_1)]\\
&\qquad\qquad + \dot{\mathcal{K}}_1\dot{\mathcal{K}}_2\mathcal{K}_3\mathcal{K}_L + \mathcal{K}_1\dot{\mathcal{K}}_2\dot{\mathcal{K}}_3\mathcal{K}_L + \mathcal{K}_1\mathcal{K}_2\dot{\mathcal{K}}_3\dot{\mathcal{K}}_L + \dot{\mathcal{K}}_1\mathcal{K}_2\mathcal{K}_3\dot{\mathcal{K}}_L\big\},
}
where $\mathcal{K}_L$ is associated to the bulk-to-bulk propagator and $\dot{\mathcal{K}}=\partial_\eta \mathcal{K}$. Using integration by parts to remove the derivatives acting on $\mathcal{K}_L$ gives
\eqs{
\mathcal{A}_L &= \int \frac{d\eta}{\eta^{d-1}}\mathcal{K}_L\Big[(-2\vec{k}_1\cdot\vec{k}_3\mathcal{K}_1\mathcal{K}_2\mathcal{K}_3 - 2\dot{\mathcal{K}}_1\mathcal{K}_2\dot{\mathcal{K}}_3 - k_1^2\mathcal{K}_1\mathcal{K}_2\mathcal{K}_3 - k_3^2\mathcal{K}_1\mathcal{K}_2\mathcal{K}_3\\
&\qquad\qquad -\mathcal{K}_1\mathcal{K}_2\ddot{\mathcal{K}}_3 - \ddot{\mathcal{K}}_1\mathcal{K}_2\mathcal{K}_3 + \frac{1-d}{\eta}\left(\dot{\mathcal{K}}_1\mathcal{K}_2\mathcal{K}_3 + \mathcal{K}_1\mathcal{K}_2\dot{\mathcal{K}}_3\right)\Big],
}
where we have also used momentum conservation at the vertex, such that $\vec{k}_L = -(\vec{k}_1+\vec{k}_2+\vec{k}_3)$. Using the equations of motion for $\mathcal{K}_1$ and $\mathcal{K}_3$, we are left with 
\eq{
\mathcal{A}_L= -2\int \frac{dz}{z^{d+1}}\left[\mathcal{K}_2\mathcal{K}_L\eta^2\left(\vec{k}_1\cdot\vec{k}_3\mathcal{K}_1\mathcal{K}_2\mathcal{K}_3 + \dot{\mathcal{K}}_1\mathcal{K}_2\dot{\mathcal{K}}_3\right)\right].
}

Performing similar manipulations on the other four-point vertices, we finally obtain
\eq{
\Psi_{6}^{NLSM}=\lambda^{4}\delta^{3}(\vec{k}_{T})[(\alpha_{13}\alpha_{123}^{-1}\alpha_{46}-\alpha_{24})+\mathrm{cyc}(i\to i{+}2)]\mathcal{C}_{6}^{\Delta}
\label{nlsmaction6pt}
}
where  $\alpha_{ijk} = \left(\mathcal{D}_i+\mathcal{D}_j+\mathcal{D}_k\right)^2$. Note that the first term comes from the exchange diagrams and the second term is from the contact diagram. These can be seen from the Witten diagrams in Figure \ref{fig:6ptDiagram}. This is manifestly free of any ordering ambiguities since the products of operators that appear can be written in any order. Note that this result can be obtained from flat space Feynman diagrams simply by replacing $2 (k_i \cdot k_j)$ with $\alpha_{ij}$. Similarly, the flat space limit of \eqref{nlsmaction6pt} is equivalent to the result given in \cite{Kampf:2013vha}.

\section{Six-point NLSM from the worldsheet}\label{sec:6-pt-worldsheet}

In this Appendix we will compute the six-point wavefunction coefficient of the NLSM in dS using the CSE. For simplicity, we will set the mass and curvature corrections to zero as we did in the previous Appendix. The worldsheet calculation turns out to be very intricate and we obtain an expression involving boundary differential operators which do not commute, in contrast to the result of the previous Appendix. It is easy to match the two expressions if these commutators are ignored. We therefore expect that the additional terms coming from commutators should cancel out although we have not verified this due to the complexity of the expression. We leave a systematic study of such commutators and potential ordering ambiguities to future work.

By lifting the flat space formula in 
\eqref{eqn:NLSMIntegrand} to dS, as explained in section \ref{worldsheet}, the six-point NLSM wavefunction coefficient in the CSE framework can be written as $\Psi_{6}^{NLSM}=\delta^{3}(\vec{k}_{T})A_{{\rm NLSM}}{\cal C}_{6}^{\Delta}$ where 
\begin{align}
A_{{\rm NLSM}}=\int_{\gamma}\,\prod_{i=2}^{4}{\rm d}\s_{i}\,(S_{i})^{-1}\left(\sigma_{15}\sigma_{56}\sigma_{61}\right)^{2}{\rm PT}\,\frac{1}{\s_{15}^{2}}\,({\rm Pf}A_{15}^{15})^{2}.
\end{align}
Here we have fixed legs $1,5,6$ and removed the rows and columns $\{1,5\}$ from the $A$-matrix. The contour encircles the poles corresponding to the CSE for legs $2,3,4$, i.e. $\gamma \equiv \gamma_{S_2}\cap\gamma_{S_3}\cap \gamma_{S_4}$.

In the following we will focus our attention on $A_{{\rm NLSM}}$, which is understood to act on the contact diagram, and impose momentum conservation along the boundary. In order to evaluate the worldsheet integral, we expand $ ({\rm Pf} A^{15}_{15} )^2 $ as
\begin{align}\label{PfA-six}
 ({\rm Pf} A^{15}_{15} )^2  = 
\frac{\al_a}{\s_{23}^2 \s_{46}^2} + \frac{\al_b}{\s_{24}^2 \s_{36}^2} + \frac{\al_c}{\s_{26}^2 \s_{34}^2   }
-  {\rm PT} (2,3,6,4)\, \al_d -  {\rm PT} (2,3,4,6)\, \al_e -  {\rm PT} (2,4,3,6)\, \al_f ,
\end{align} 
where\begin{align}
&
\al_a\equiv \al_{23}^2\, \al_{46}^2\, , \quad \al_b\equiv \al_{24}^2\, \al_{36}^2\, , \quad \al_c\equiv \al_{26}^2\, \al_{34}^2\, , \quad
\al_d \equiv  \al_{23}\, \al_{46} \, \al_{24} \, \al_{36} +  \al_{24} \, \al_{36} \, \al_{23}\, \al_{46}\, , \nonumber \\
& \al_e \equiv  \al_{23}\, \al_{46} \, \al_{26} \, \al_{34} +  \al_{26} \, \al_{34} \, \al_{23}\, \al_{46}\, , \quad
\al_f \equiv  \al_{24}\, \al_{36} \, \al_{26} \, \al_{34} +  \al_{26} \, \al_{34} \, \al_{24}\, \al_{36}\,,
\end{align}
and evaluate the integral for each term separately.

Therefore, $A_{\rm NLSM}$ is recast as
\begin{align}\label{calA}
A_{\rm NLSM} = {\cal A}^{(a)}\a_a+ {\cal A}^{(b)}\al_b+ {\cal A}^{(c)} \al_c - {\cal A}^{(d)}\al_d - {\cal A}^{(e)}\al_e - {\cal A}^{(f)}\al_f,
\end{align}
where we have denoted
\begin{equation}
{\cal A}^{(i)} =\int_\gamma \, \prod_{i=2}^4 {\rm d}\s_i \, (S_i)^{-1}  \left(\sigma_{15}\sigma_{56}\sigma_{61}\right)^{2} {\cal I}^{(i)}, \qquad 
i\in \{a,b,c,d,e,f\},
\end{equation}
and defined the integrands 
\begin{align}
&{\cal I}^{(a)} = {\rm PT} \, \frac{1}{\s_{15}^2 \,\s_{23}^2 \, \s_{46}^2} , \quad
{\cal I}^{(b)} = {\rm PT} \, \frac{1}{\s_{15}^2 \,\s_{24}^2 \, \s_{36}^2}, \quad
{\cal I}^{(c)} = {\rm PT} \, \frac{1}{\s_{15}^2 \,\s_{26}^2 \, \s_{34}^2}, \nonumber\\
&{\cal I}^{(d)} =   {\rm PT}\, \frac{ {\rm PT}(2,3,6,4)  }{\s_{15}^2 }, \quad
{\cal I}^{(e)} =   {\rm PT} \, \frac{ {\rm PT}(2,3,4,6)  }{\s_{15}^2 }, \quad
{\cal I}^{(f)} =   {\rm PT}\, \frac{ {\rm PT}(2,4,3,6)  }{\s_{15}^2}.
\end{align}

\begin{figure}[h]
\centering
\parbox[c]{5.1em}{\includegraphics[scale=0.2]{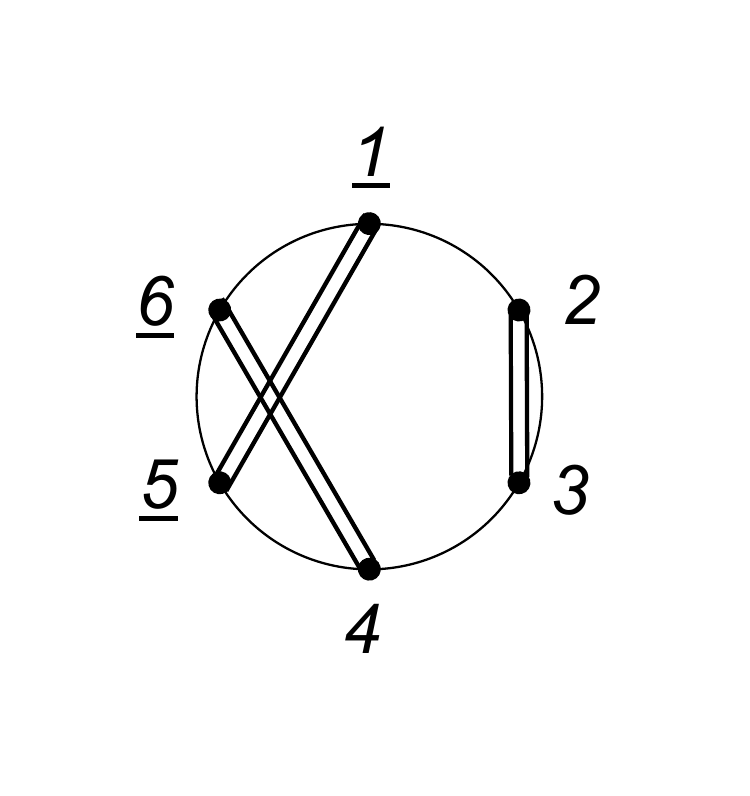}} $\,\al_a \,+\!\!\!\!\!\!\!$ 
\parbox[c]{5.1em}{\includegraphics[scale=0.2]{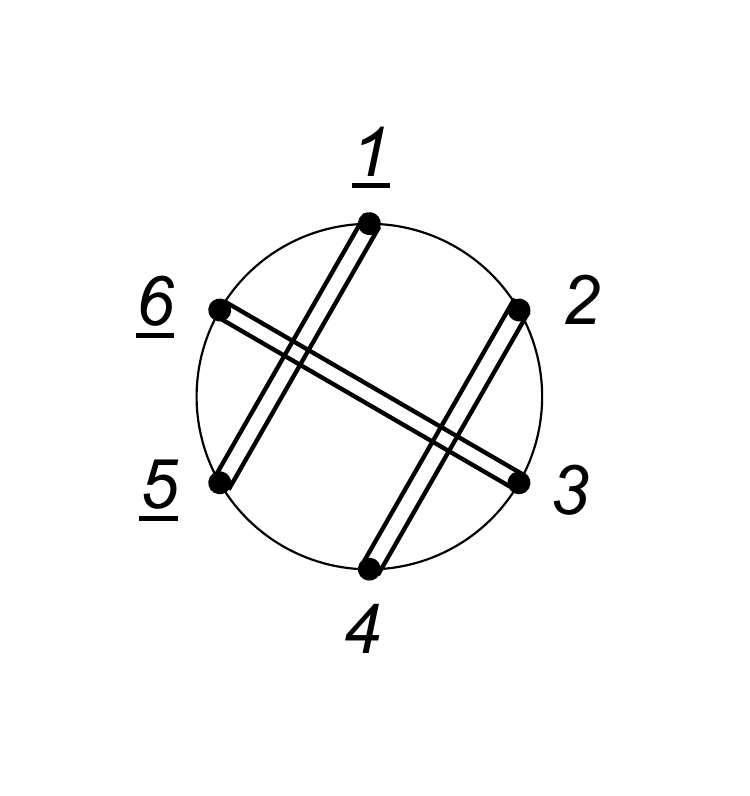}} $\, \al_b \,+\!\!\!\!\!\!\!$
\parbox[c]{5.1em}{\includegraphics[scale=0.2]{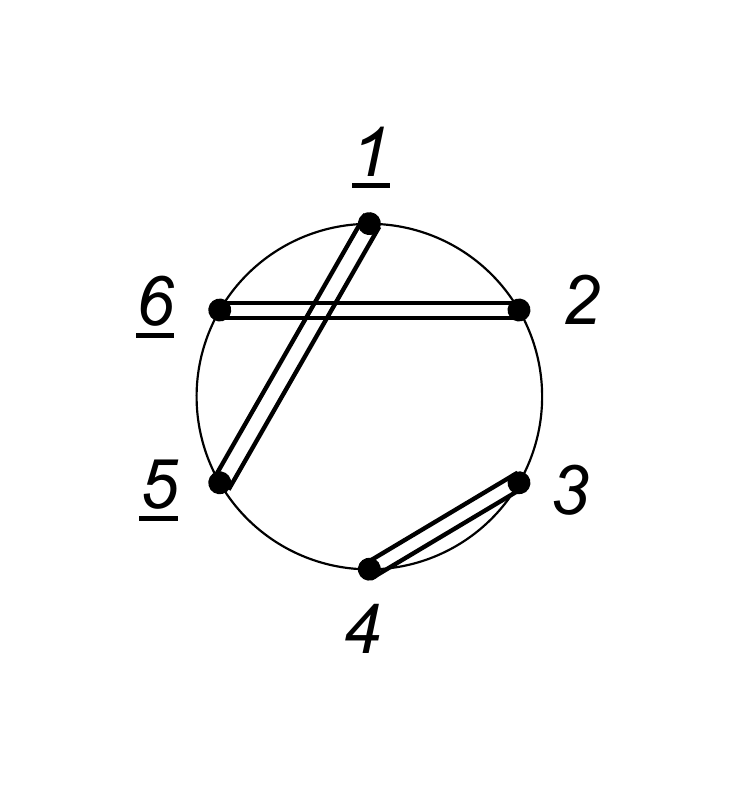}} $\, \al_c \,-\!\!\!\!\!\!\!$
\parbox[c]{5.1em}{\includegraphics[scale=0.2]{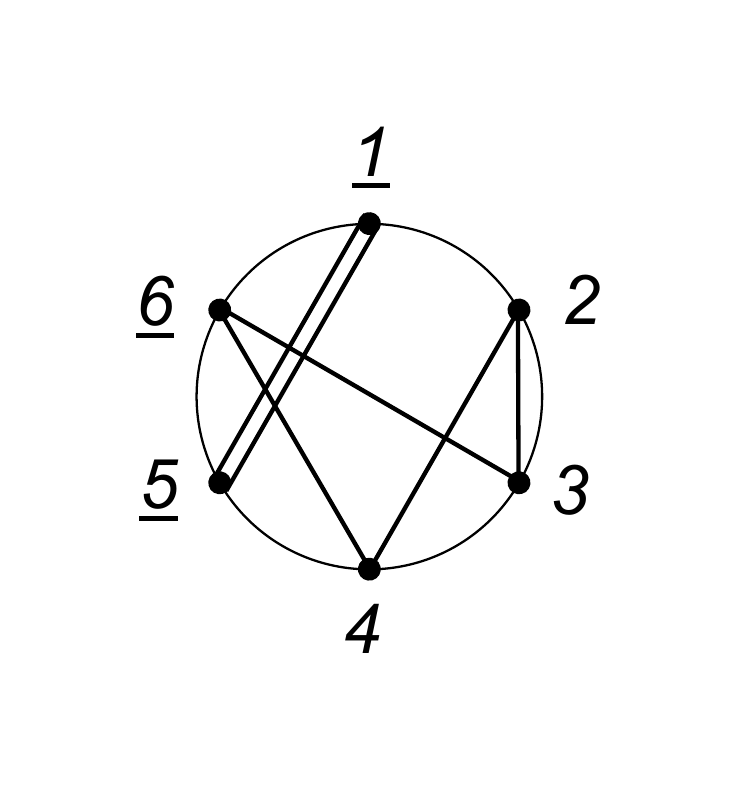}}  $\, \al_d \,-\!\!\!\!\!\!\!$
\parbox[c]{5.1em}{\includegraphics[scale=0.2]{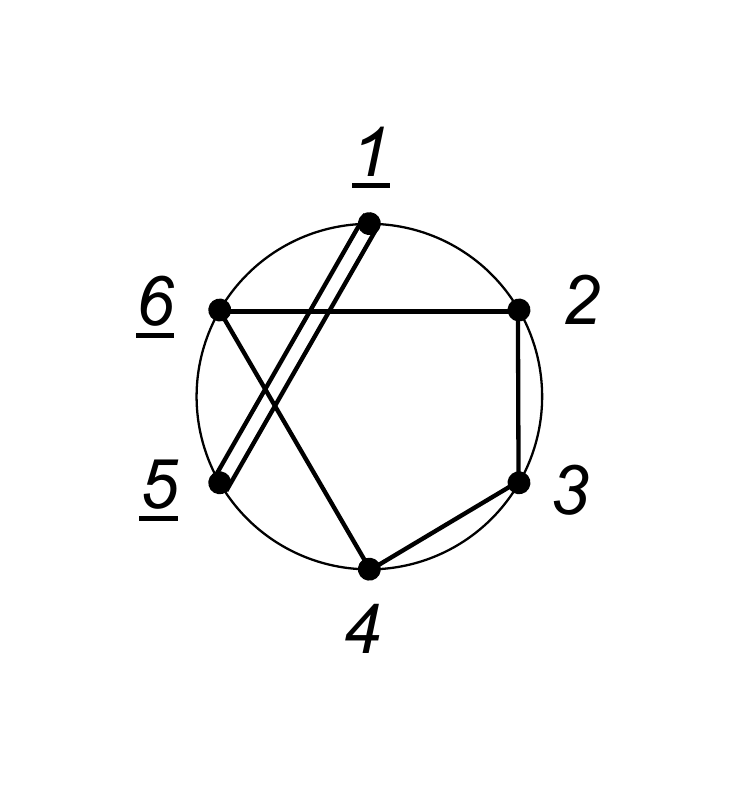}}  $\, \al_e \,-\!\!\!\!\!\!\!$
\parbox[c]{5.1em}{\includegraphics[scale=0.2]{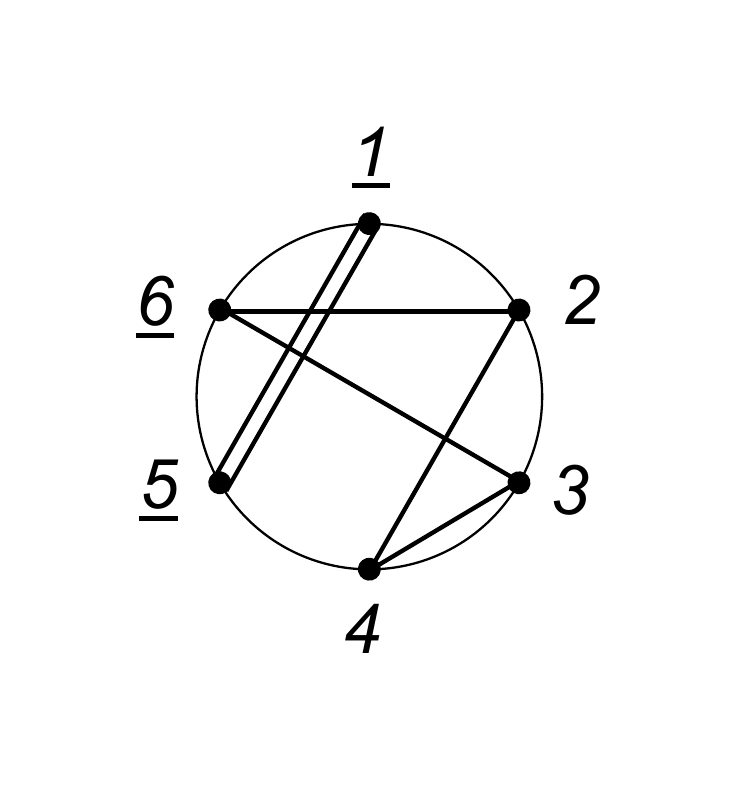}}  $\, \al_f$
\caption{All diagrams for $A_{\rm NLSM}$.}\label{Fig6-point} 
\end{figure}

We will evaluate the worldsheet integrals using the graph representation and integration rules introduced in \cite{Gomez:2021qfd,Gomez:2021ujt}. The graphs corresponding to each term in $A_{\rm NLSM}$  are shown in Fig. \ref{Fig6-point}. The chords in each circle encode the poles in worldsheet coordinates not contained in the Parke-Taylor factor. For example, ${\cal A}^{(a)}$ contains double poles $\sigma_{15}$, $\sigma_{23}$, and $\sigma_{46}$ so we draw double lines between these points on the circumference. When evaluating contour integrals, the worldsheet will factorise when a subset of punctures collide. We then draw closed loops around the punctures that collided and refer to them as factorisation cuts. There is a simple set of rules for determining which factorisation cuts contribute:
\begin{itemize}
	\item {\bf Rule I.}  All  factorization cuts with fewer  than two fixed
		marked-points (labels with underlines) vanish.
\item {\bf Rule II.} If the factorization cuts more than four lines in the corresponding graph then, this contribution vanishes.
\end{itemize}
In practice, these rules make calculations much more efficient.

\begin{figure}[h]
\centering
$(I)$\!\!\!\!\!\!
\parbox[c]{11.1em}{\includegraphics[scale=0.23]{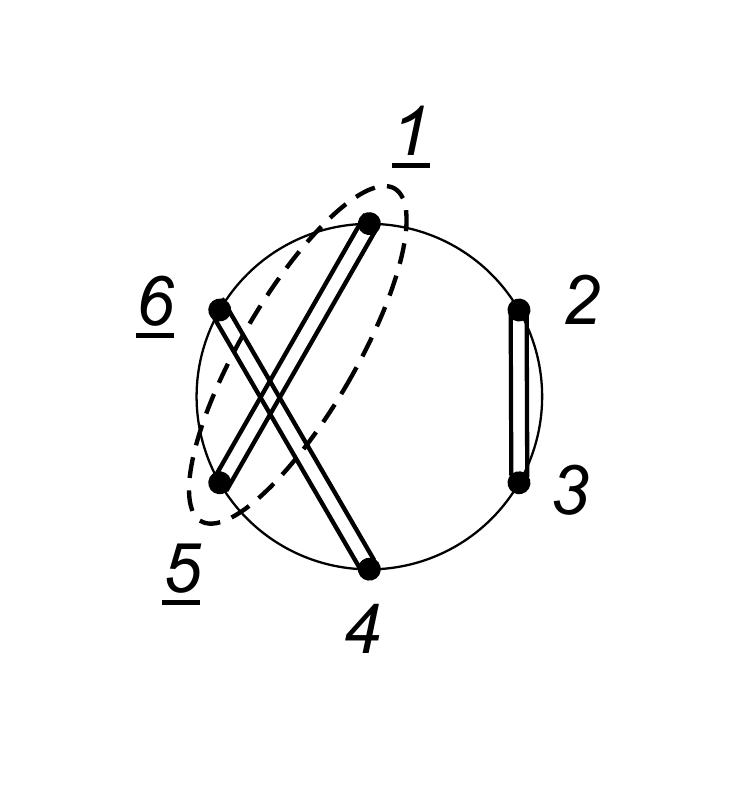}} 
$(II)$\!\!\!\!\!\!
\parbox[c]{5.1em}{\includegraphics[scale=0.23]{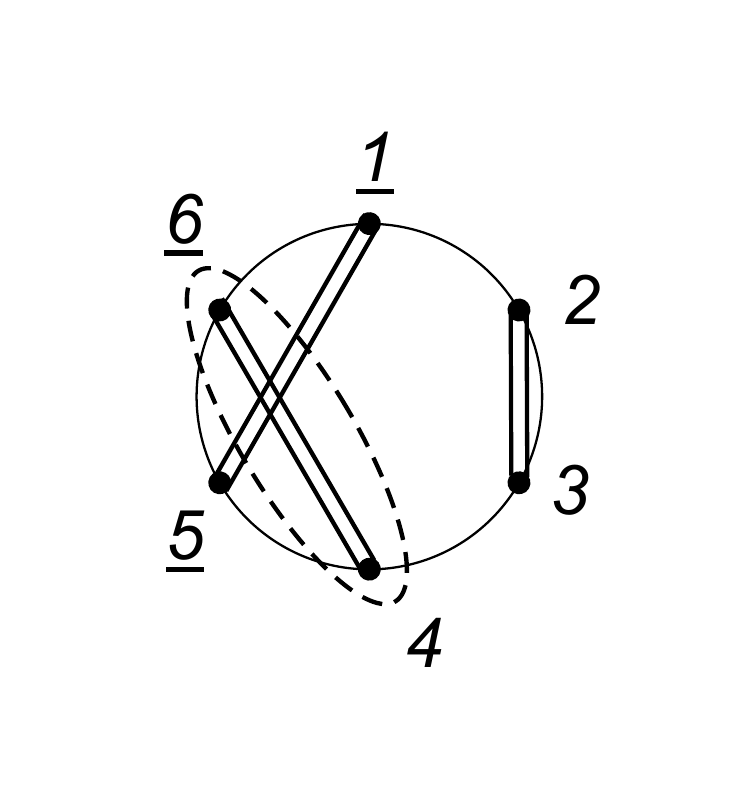}} 
\caption{Factorization contributions for ${\cal A}^{(a)}$.}\label{Figgraph-a} 
\end{figure}

Let us illustrate in detail how to evaluate the first diagram in Figure \ref{Fig6-point}. From the integration rules, it is simple to see that  there are two factorisation contributions for this diagram, which are illustrated in Fig. \ref{Figgraph-a}.
The contribution $({ I})$, is given when $\s_2\rightarrow \s_3 \rightarrow \s_4 \rightarrow \s_6$. To perform this computation, we use the parametrisation, $\sigma_a=\epsilon x_a+\sigma_6$,  
with $a=2,3,4,6$, $x_4=\text{constant}$, $x_6=0$, $\sigma_6=\sigma_L$,
and expand around $\epsilon=0$, 
\begin{align}\label{eq:measure}
&
\dif\sigma_2\wedge\dif\sigma_3\wedge \dif\sigma_4= \epsilon^2 \, x_{46}\,\dif x_2\wedge\dif x_3\wedge \dif\epsilon
\nonumber \\
&
(\sigma_{15} \sigma_{56} \sigma_{61})^2 \, 
 {\rm PT} \, \frac{1}{\s_{15}^2 \,\s_{23}^2 \, \s_{46}^2}
=\frac{1}{\epsilon^6}\, \frac{(\sigma_{15} \sigma_{5L} \sigma_{L1})^2 }{ ( \sigma_{L1}^2\, \sigma_{5 L }^2\, \sigma_{15}^2 ) }\,
\frac{1 }{(x_{23}^2\, x_{34} \, x_{46}^2 )} + {\cal O}(\epsilon^{-5}), 
\end{align}
and
\begin{align}\label{eq:SE234}
& 
S_2 =  \frac{1}{\epsilon}    \left[\hat S_2 + {\cal O}(\epsilon) \right] , \qquad  
S_3 = \frac{1}{\epsilon}   \left[ \hat S_3  + {\cal O}(\epsilon) \right] , \qquad
S_4 = \frac{1}{\epsilon}   \left[ \hat S_4 + {\cal O}(\epsilon) \right] , \nonumber\\
&
\hat S_2=\frac{\a_{23}}{x_{23}}  +\frac{\a_{24}}{x_{24}} +\frac{\a_{26}}{x_{26}} + \frac{\a_{2R}}{x_{2R}} ,
  \,\,  \hat S_3=\frac{\a_{32}}{x_{32}} + \frac{\a_{34}}{x_{34}}  +\frac{\a_{36}}{x_{36}} + \frac{\a_{3R}}{x_{3R}} , 
 \,\,  \hat S_4=\frac{\a_{42}}{x_{42}} + \frac{\a_{43}}{x_{43}}  +\frac{\a_{46}}{x_{46}} + \frac{\a_{4R}}{x_{4R}} , 
\end{align}
where $x_{R}=\infty$, and $\a_{a R}= \a_{a1}+\a_{a5} $, $a=2,3,4$.
Now we can deform the contour 
 $\gamma=\gamma_{ \hat S_2  + {\cal O}(\epsilon) }\cap \gamma_{\hat S_3  + {\cal O}(\epsilon)} \cap \gamma_{\hat S_4  + {\cal O}(\epsilon)}$  into $\tilde \gamma= \gamma_{\epsilon}\cap \gamma_{\hat S_2  + {\cal O}(\epsilon)} \cap \gamma_{\hat S_3  + {\cal O}(\epsilon)}$, where $\gamma_{\epsilon} = \{  |\epsilon| = \delta \}$.

After integrating over $\gamma_\epsilon$, the full integral is split into two parts: one with $\{\sigma_1,\sigma_5,\sigma_L\}$ and the other with $\{x_2,x_3,x_4,x_6,x_{R}\}$. Moreover, using the identity
\begin{equation}
{\rm PT}(4,6,R )\,\hat S_4 + {\rm PT}(3,6,R )\, \hat S_3+ {\rm PT}(2,6,R )\, \hat S_2  =  {\rm PT}(6,R )\,(\a_{23}+\a_{24}+ \a_{26} + \a_{34} +\a_{36} + \a_{46} ),
\end{equation}
we find that on the support of $\gamma_{\hat S_2} \cap \gamma_{\hat S_3}$, $\hat S_4$ reduces to
\begin{equation}
\hat S_4\Big|_{
\gamma_{\hat S_2} \cap \gamma_{\hat S_3  }
} = \frac{{\rm PT}(6,R) }{{\rm PT}(4,6,R)} \, (\a_{15}),
\end{equation}
where we have used the CWI. Putting everything together, the contribution $(I)$ for ${\cal A}^{(a)}$ is given by
\begin{align}\label{eq:Aa(I)}
{\cal A}^{(a)}\Big|_{(I)} &= \frac{(\sigma_{15} \sigma_{5L} \sigma_{L1})^2 }{ ( \sigma_{L1}^2\sigma_{5 L }^2 \sigma_{15}^2 ) } \,  (\a_{15})^{-1} \int_{\hat\gamma}
\prod_{i=2}^3 \dif x_{i} (\hat S_i)^{-1} (x_{46} x_{6R} x_{R4})^2 \frac{{\rm PT}(2,3,4,6,R) }{x^2_{23} \, x_{46} \,x_{6R}\,x_{R4} } \nonumber \\
& = 
 (\a_{15})^{-1} \int_{\hat\gamma}
\prod_{i=2}^3 \dif x_{i} (\hat S_i)^{-1} (x_{46} x_{6R} x_{R4})^2\, \frac{{\rm PT}(2,3,4,6,R) }{x^2_{23} \, x_{46} \,x_{6R}\,x_{R4} },
\end{align}
with $\hat \gamma = \gamma_{\hat S_2} \cap \gamma_{\hat S_3}$. 
\begin{figure}[h]
\centering
\parbox[c]{7.1em}{\includegraphics[scale=0.23]{graph-a-cut1.pdf}} 
= \,\,$ (\a_{15})^{-1}$
\!\!\!\!\!\!\!
\!\!\!\!\!
\parbox[c]{5.1em}{\includegraphics[scale=0.24]{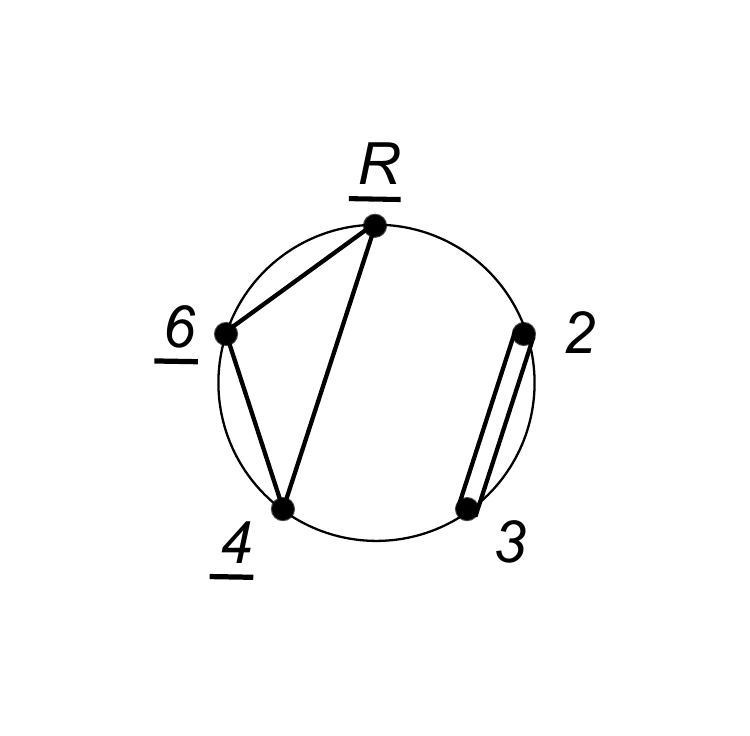}} 
\caption{Diagrammatic representation of ${\cal A}^{(a)}\Big|_{(I)} $.}\label{cut1-5p} 
\end{figure}

\begin{figure}[h]
\centering
$(I\,A)$\!\!\!\!\!\!
\parbox[c]{11.1em}{\includegraphics[scale=0.23]{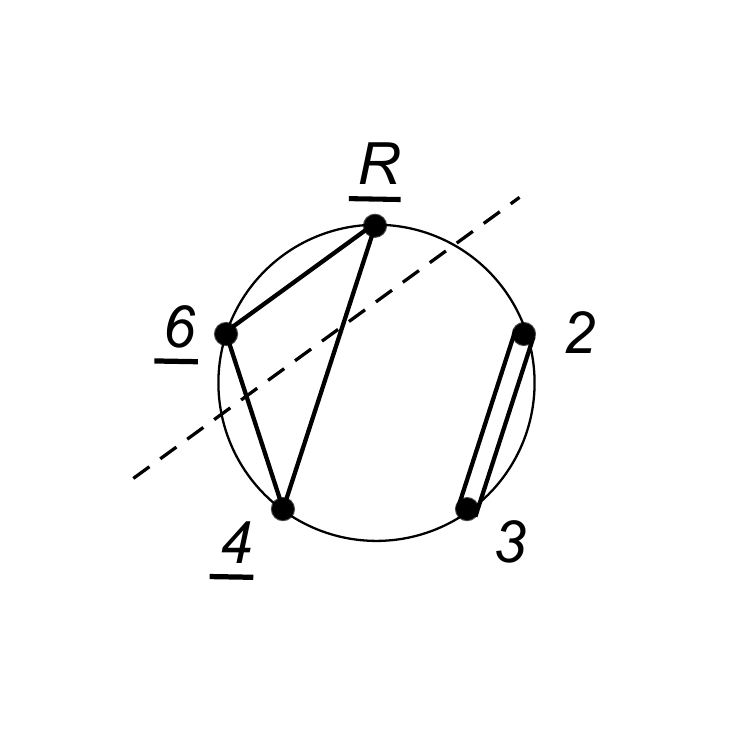}} 
$(I\,B)$\!\!\!\!\!\!
\parbox[c]{5.1em}{\includegraphics[scale=0.23]{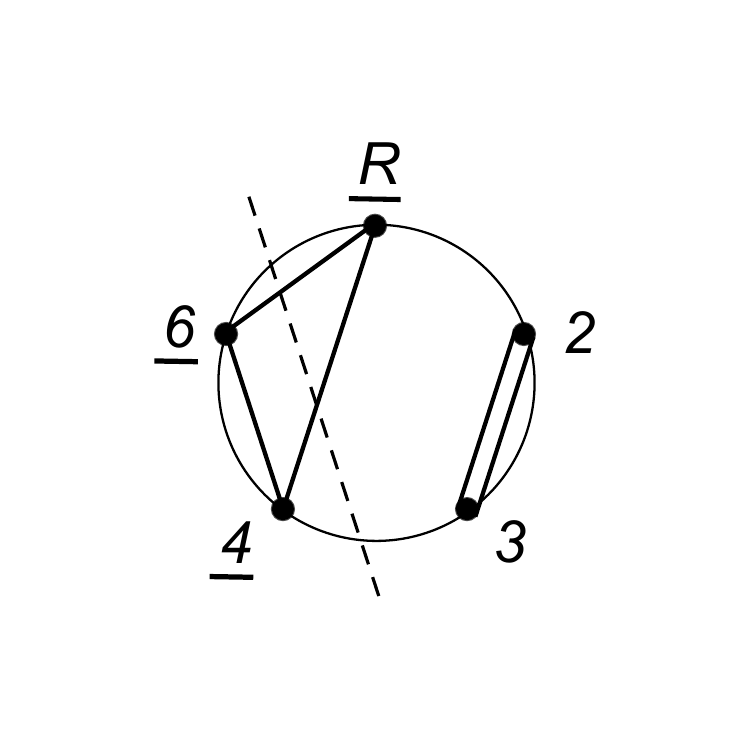}} 
\caption{Factorization contributions for the five-point diagram obtained in ${\cal A}^{(a)}\Big|_{(I)} $.}\label{5point-cut12} 
\end{figure}

The expression in \eqref{eq:Aa(I)} is depicted in Fig. \ref{cut1-5p}. This diagram has two factorisation contributions which are illustrated in Fig. \ref{5point-cut12}.
The contribution $({ I\, A})$ arises when $x_2\rightarrow x_3\rightarrow x_4$. In this case we use the parametrisation  
$x_a = \epsilon\, y_a + x_4$, $a=2,3,4$, $y_2=$constant, $y_4=0$ and $x_4=x_L$. We then have
\begin{align}\label{eq:}
&
\dif x_2\wedge\dif x_3= \epsilon \, y_{24}\,\dif\epsilon\wedge\dif y_3
\nonumber \\
&
(x_{46} x_{6R} x_{R4})^2\, \frac{{\rm PT}(2,3,4,6,R) }{x^2_{23} \, x_{46} \,x_{6R} }
=\frac{1}{\epsilon^4}\, \frac{ (x_{L6} x_{6R} x_{RL})^2 }{ (x_{L6} x_{6R} x_{RL})^2 }\,
\frac{1 }{(y_{23}^3\, y_{34}  )} + {\cal O}(\epsilon^{-3}), 
\end{align}
and
\begin{align}\label{eq}
& 
\hat S_2 =  \frac{1}{\epsilon}    \left[\tilde S_2 + {\cal O}(\epsilon) \right] , \qquad  
\hat S_3 = \frac{1}{\epsilon}   \left[ \tilde S_3  + {\cal O}(\epsilon) \right] , \nonumber\\
&
\tilde S_2=\frac{\a_{23}}{y_{23}}  +\frac{\a_{24}}{y_{24}} + \frac{\a_{2\hat R}}{y_{2 \hat R}} ,
\qquad  
\tilde S_3=\frac{\a_{32}}{y_{32}} + \frac{\a_{34}}{y_{34}}  + \frac{\a_{3\hat R}}{y_{3\hat R}} , 
\end{align}
where $y_{\hat R}=\infty$, and $\a_{a \hat R}= \a_{a1}+\a_{a5}+\a_{a6} $, $a=2,3$.
By the GRT, we deform the contour $\hat \gamma=\gamma_{ \tilde S_2  + {\cal O}(\epsilon) }\cap \gamma_{\tilde S_3  + {\cal O}(\epsilon)} $ into $\gamma^\prime =- \gamma_{\epsilon}\cap \gamma_{\tilde S_3  + {\cal O}(\epsilon)} $.
Thus, integrating around $\epsilon=0$, the five-point integral breaks into two parts: one with $\{x_6, x_R,x_L\}$ and the other with $\{y_2,y_3,y_4,y_{\hat R}\}$.
Using the identity
\begin{equation}
{\rm PT}(2,4,\hat R )\,\tilde S_2 + {\rm PT}(3,4,\hat R )\, \tilde S_3  =  {\rm PT}(4,\hat R )\, \left[ ( {\cal D}_2+{\cal D}_3+ {\cal D}_4 )^2 + m^2 \right],
\end{equation}
we find that on the support of $ \gamma_{\tilde S_3}$, $\tilde S_2$ reduces to
\begin{equation}
\tilde S_2\Big|_{
 \gamma_{\tilde S_3  }
} = \frac{{\rm PT}(4,\hat R) }{{\rm PT}(2,4,\hat R)} \, \left[ ( {\cal D}_2+{\cal D}_3+ {\cal D}_4 )^2 + m^2 \right] .
\end{equation}
The contribution $({ I\, A})$ is therefore given by
\begin{align}
{ (I \, A)} &=  \frac{ (x_{L6} x_{6R} x_{RL})^2 }{ (x_{L6} x_{6R} x_{RL})^2 }\, \left[ ( {\cal D}_2+{\cal D}_3+ {\cal D}_4 )^2 + m^2 \right]^{-1} \!\!\int_{\gamma_{\tilde S_3}} \!\!\! \dif y_3 (\tilde S_3)^{-1} ( y_{2 4} y_{4 \hat R}  y_{\hat R 2}  )^{2}\, \frac{{\rm PT}(2,3,4,\hat R)}{ y_{23}^2\, y_{4 \hat R}^2} \nonumber\\
&=\left[ ( {\cal D}_2+{\cal D}_3+ {\cal D}_4 )^2 + m^2 \right]^{-1} \!\! \int_{\gamma_{\tilde S_3}} \!\!\! \dif y_3 (\tilde S_3)^{-1} ( y_{2 4}  y_{4 \hat R}  y_{\hat R 2}  )^{2}\, \frac{{\rm PT}(2,3,4,\hat R)}{ y_{23}^2\, y_{4 \hat R}^2}\, .
\label{IAterms}
\end{align}

\begin{figure}[h]
\centering
\parbox[c]{6.4em}{\includegraphics[scale=0.23]{5point-c1.pdf}} 
$= \left[ ( {\cal D}_2+{\cal D}_3+ {\cal D}_4 )^2 + m^2 \right]^{-1} $
\!\!\!\!\!\!\!\!\!
\!\!\!\!
\parbox[c]{5.1em}{\includegraphics[scale=0.23]{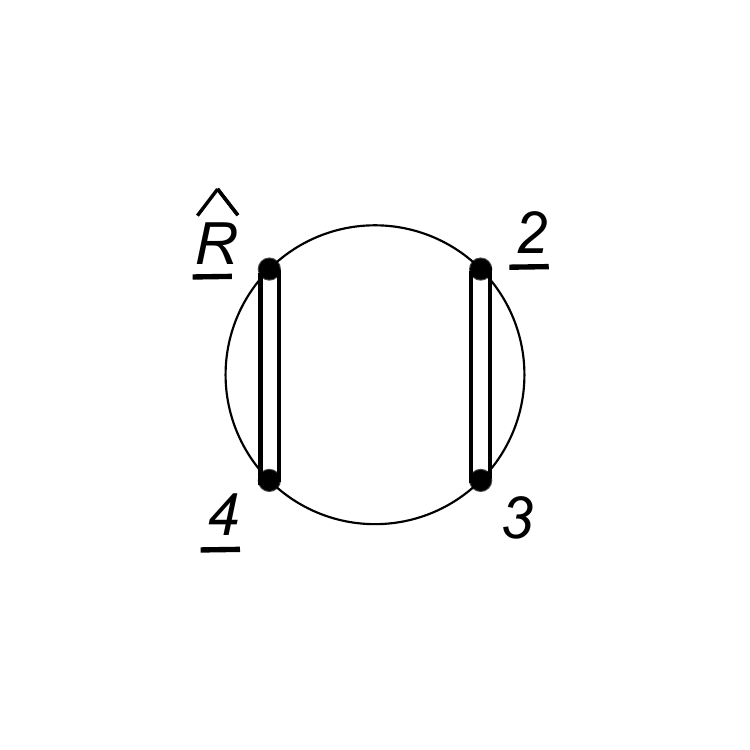}} 
\caption{Diagramatic representationn for the contribution $({ I\, A})$.}\label{5point-IA} 
\end{figure}

The contribution in \eqref{IAterms} is represented in Fig. \ref{5point-IA}.
Since the four-point diagram in Fig. \ref{5point-IA} has a double pole, it can potentially give rise to ambiguities associated with the ordering of non-commuting differential operators. In appendix \ref{4point-DP}, we compute this contribution in detail and show that it is well defined. In particular, we find that
\begin{align}
\int_{\gamma_{\tilde S_3}} \!\!\! \dif y_3 (\tilde S_3)^{-1} ( y_{2 4}  y_{4 \hat R}  y_{\hat R 2}  )^{2}\, \frac{{\rm PT}(2,3,4,\hat R)}{ y_{23}^2\, y_{4 \hat R}^2} = \left(\a_{23} \right)^{-1} \left( \a_{23} + \a_{34}\right) \left(\a_{23} \right)^{-1}.
\end{align}
Plugging this into \eqref{IAterms} then gives
\begin{align}
{ (I \, A)} =  \left[ ( {\cal D}_2+{\cal D}_3+ {\cal D}_4 )^2 + m^2 \right]^{-1}  \left(\a_{23} \right)^{-1} \left( \a_{23} + \a_{34}\right)\left(\a_{23} \right)^{-1}.
\end{align}
Similarly, we find the $({\rm I \, B})$ contribution in Figure \ref{5point-cut12} to be  
\begin{align}
{ (I \, B)} =  ( \a_{1235}- \a_{15} )^{-1}  \left(\a_{23} \right)^{-1} \left( \a_{23} + \a_{34}+\a_{36}\right)\left(\a_{23} \right)^{-1}
\end{align}
where $\a_{1235} \equiv \a_{12}+\a_{13}+\a_{15} + \a_{23} + \a_{25} +\a_{35} $.
Hence, the factorization contribution $({\rm I})$ in Fig. \ref{Figgraph-a} is given by
\begin{align}\label{}
{\cal A}^{(a)}\Big|_{({ I})} &= 
(\a_{15})^{-1} \left\{  \left[ ( {\cal D}_2+{\cal D}_3+ {\cal D}_4 )^2 + m^2 \right]^{-1}  \left(\a_{23} \right)^{-1} \left( \a_{23} + \a_{34}\right) \left(\a_{23} \right)^{-1}
 \right. \nonumber\\
 &+ \left.
( \a_{46} - \a_{15} )^{-1}  \left(\a_{23} \right)^{-1} \left( \a_{23} + \a_{34}+\a_{36}\right) \left(\a_{23} \right)^{-1}
 \right\}.
\end{align}
The computation for the contribution $({\rm II})$ in Fig. \ref{Figgraph-a} is completely analogous and we obtain
\begin{align}\label{}
{\cal A}^{(a)}\Big|_{({ II})} \! &= 
 \! (\a_{46})^{-1} \! \left\{  \left[ ( {\cal D}_1+{\cal D}_2+ {\cal D}_3 )^3 + m^2 \right]^{-1} \! \left(\a_{23} \right)^{-1} \!\left( -\a_{13} \right) \left(\a_{23} \right)^{-1}
 \right. \nonumber\\
 &+ \left.
  ( \a_{15} - \a_{46 } )^{-1}  \left(\a_{23} \right)^{-1} \left( \a_{23} + \a_{34}+\a_{36}\right) \left(\a_{23} \right)^{-1}
 \right\}.
\end{align}
Putting it all together, we finally obtain the following formula for the first diagram in Figure \ref{Fig6-point}:
\begin{align}\label{figa}
{\cal A}^{(a)} &= {\cal A}^{(a)}\Big|_{({ I})} +{\cal A}^{(a)}\Big|_{({ II})} 
\nonumber \\
 &= 
 \left[ ( {\cal D}_1+{\cal D}_2+ {\cal D}_3 )^2 + m^2 \right]^{-1} (\a_{46})^{-1} \left(\a_{23} \right)^{-1} \left( - \a_{13} \right) \left(\a_{23} \right)^{-1}
 \nonumber\\
 &+  \left[ ( {\cal D}_2+{\cal D}_3+ {\cal D}_4 )^2 + m^2 \right]^{-1} (\a_{15})^{-1} \left(\a_{23} \right)^{-1} \left(  \a_{23} + \a_{34} \right) \left(\a_{23} \right)^{-1}
  \nonumber\\
 &+  (\a_{15})^{-1} (\a_{46})^{-1} \left(\a_{23} \right)^{-1} \left(  \a_{23} + \a_{34}+\a_{36} \right) \left(\a_{23} \right)^{-1}.
\end{align}

\begin{figure}[h]
\centering
(a)\qquad
\parbox[c]{6.3em}{\includegraphics[scale=0.22]{graph-b.pdf}} 
$ \rightarrow$
\!\!\!\!\!
\parbox[c]{6.3em}{\includegraphics[scale=0.22]{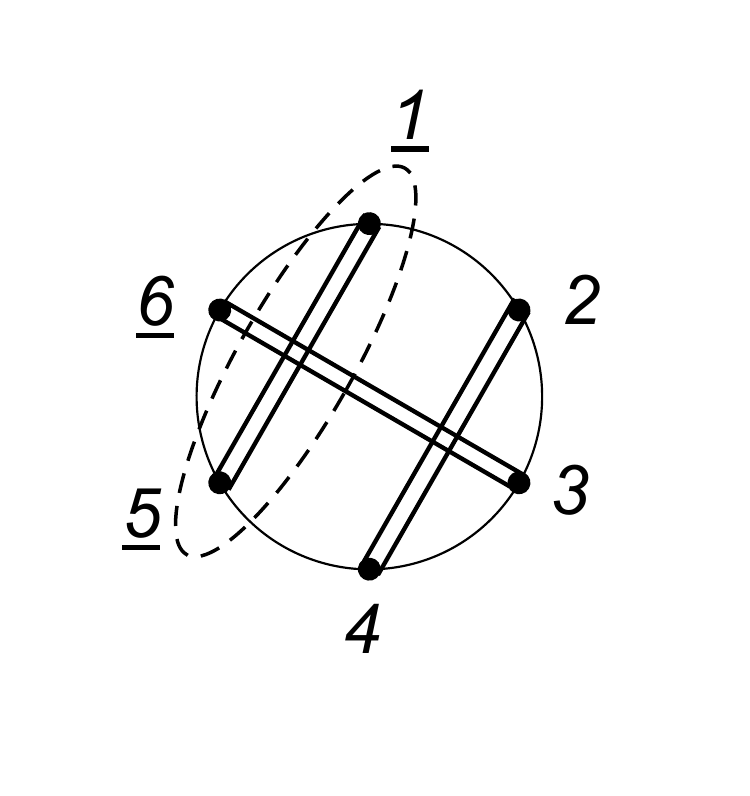}} 
= \,$ (\a_{15})^{-1} $
\!\!\!\!\!\!\!
\!\!\!
\parbox[c]{5.1em}{\includegraphics[scale=0.22]{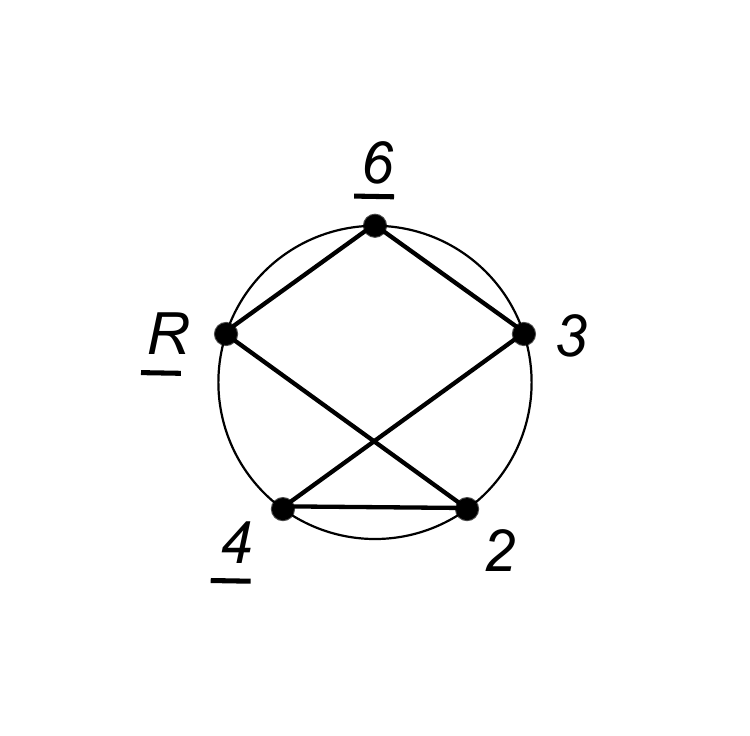}} \\
(b)\qquad
\parbox[c]{6.3em}{\includegraphics[scale=0.22]{graph-c.pdf}} 
$ \rightarrow$
\!\!\!\!\!
\parbox[c]{6.3em}{\includegraphics[scale=0.22]{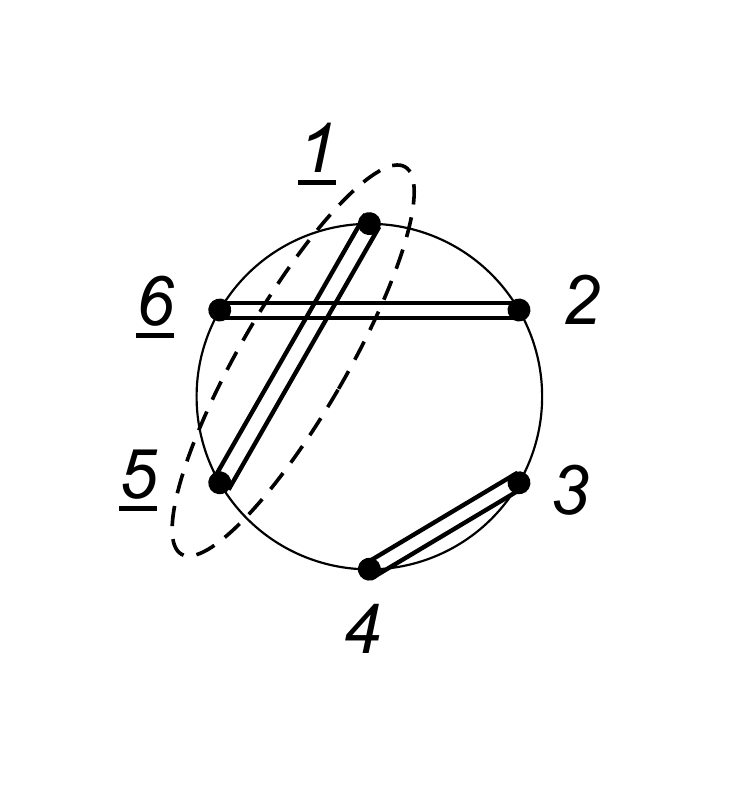}} 
= \,$ (\a_{15})^{-1} $
\!\!\!\!\!\!\!
\!\!\!
\parbox[c]{5.1em}{\includegraphics[scale=0.22]{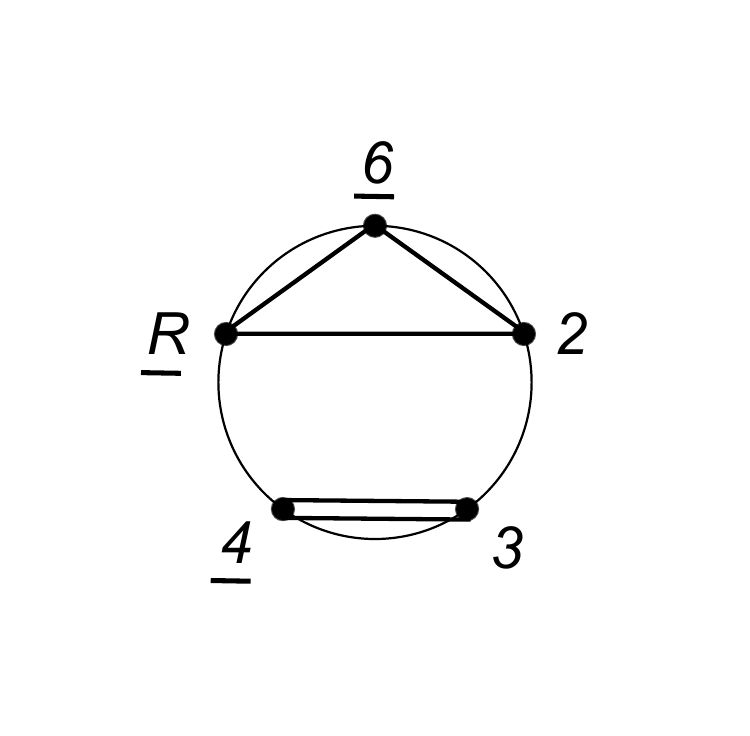}}\\
\vspace{0.65cm}
(c)\\
\parbox[c]{5.4em}{\includegraphics[scale=0.19]{graph-d.pdf}} 
$ \rightarrow$
\!\!\!\!\!
\parbox[c]{6.3em}{\includegraphics[scale=0.19]{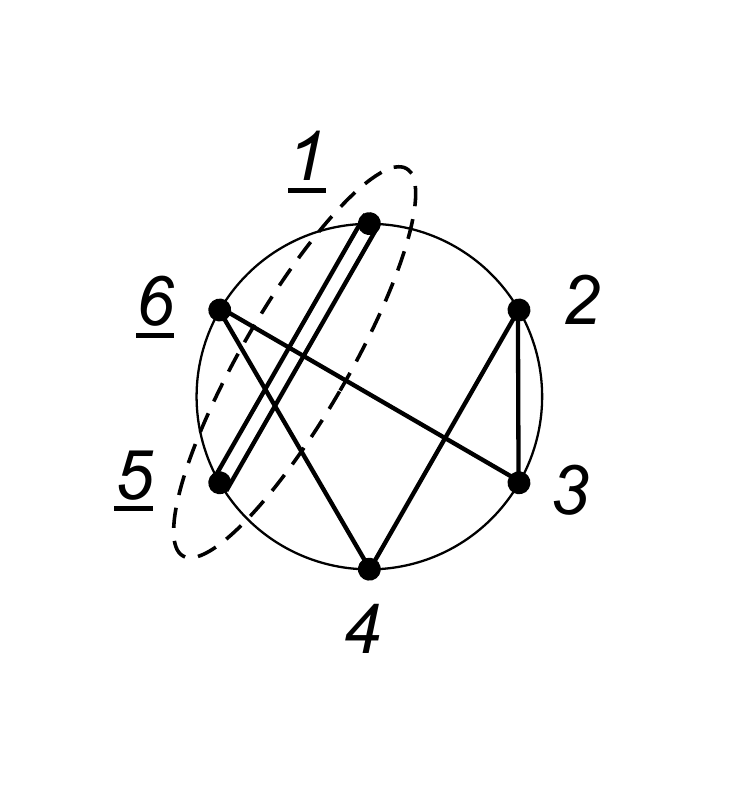}} 
\!\!\!\!\! \!\!\!\!\!\!
\!\!\!
=$ (\a_{15})^{-1} $
\!\!\!\!\!\!\!
\!\!\!
\parbox[c]{5.0em}{\includegraphics[scale=0.19]{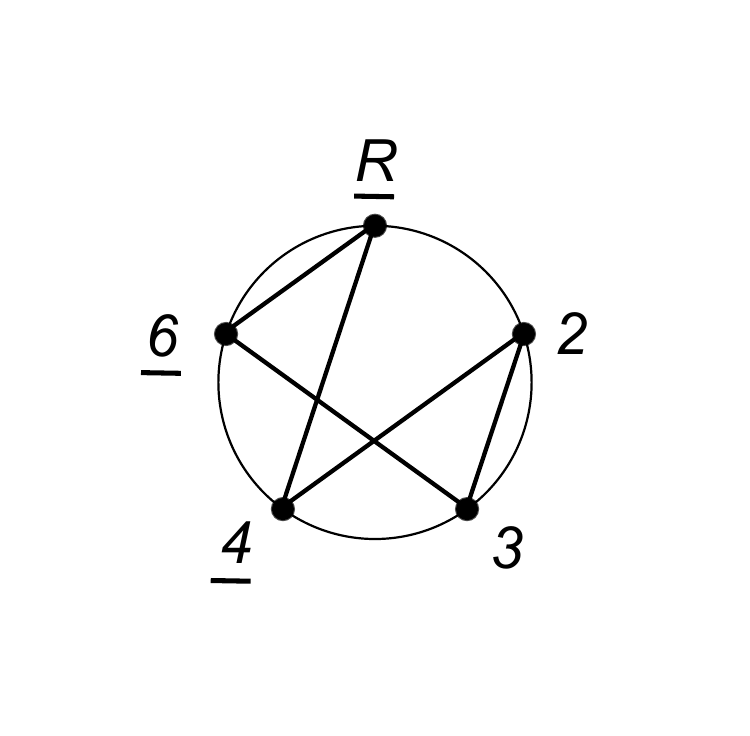}} 
\parbox[c]{5.0em}{\includegraphics[scale=0.18]{graph-e.pdf}} 
$ \rightarrow$
\!\!\!\!\!\!
\parbox[c]{6.3em}{\includegraphics[scale=0.18]{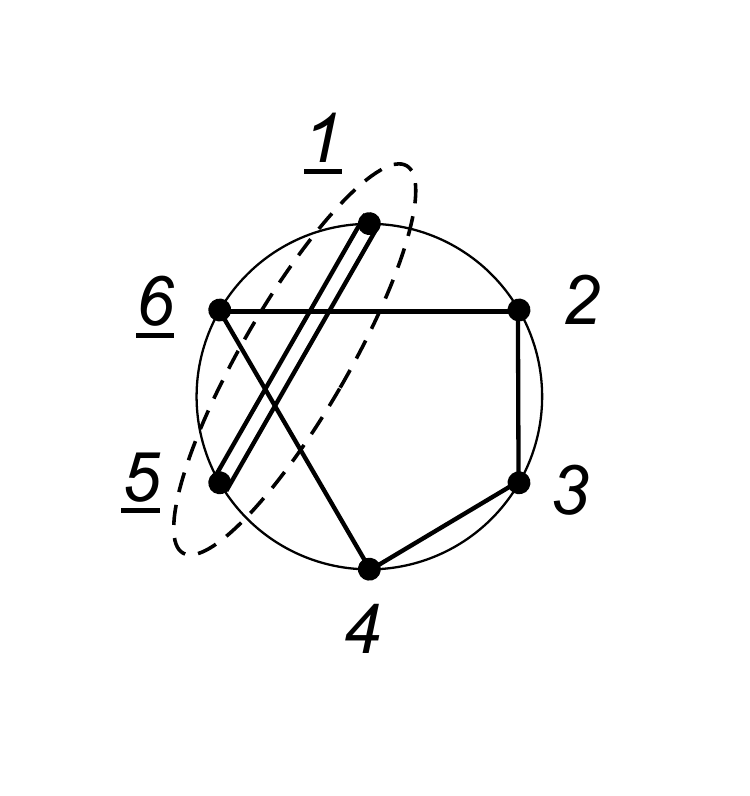}} 
\!\!\!\!\!\!\!\!\!\!
= $ (\a_{15})^{-1} $
\!\!\!\!\!\!\!
\!\!\!
\parbox[c]{5.1em}{\includegraphics[scale=0.18]{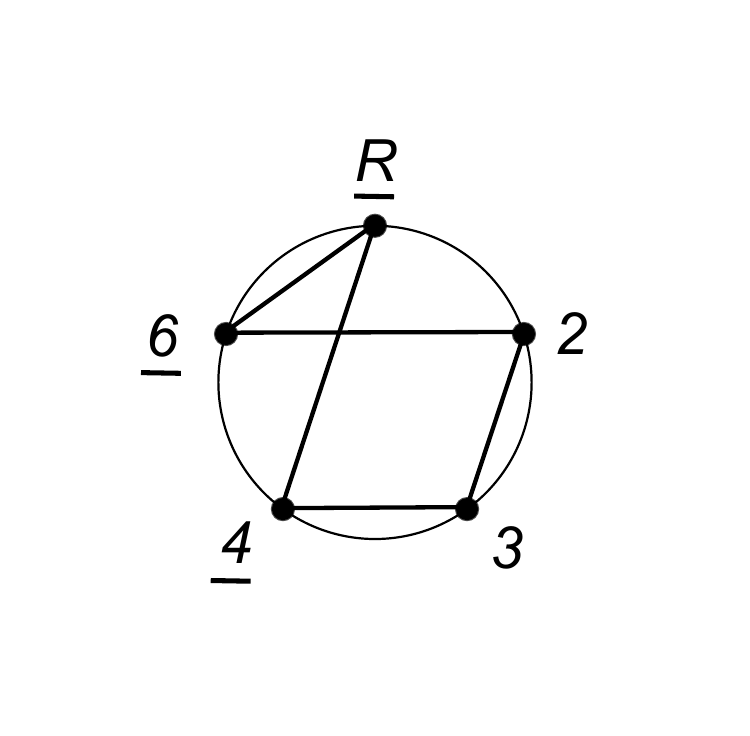}} 
\parbox[c]{5.3em}{\includegraphics[scale=0.19]{graph-f.pdf}} 
$ \rightarrow$
\!\!\!\!\!
\parbox[c]{5.5em}{\includegraphics[scale=0.19]{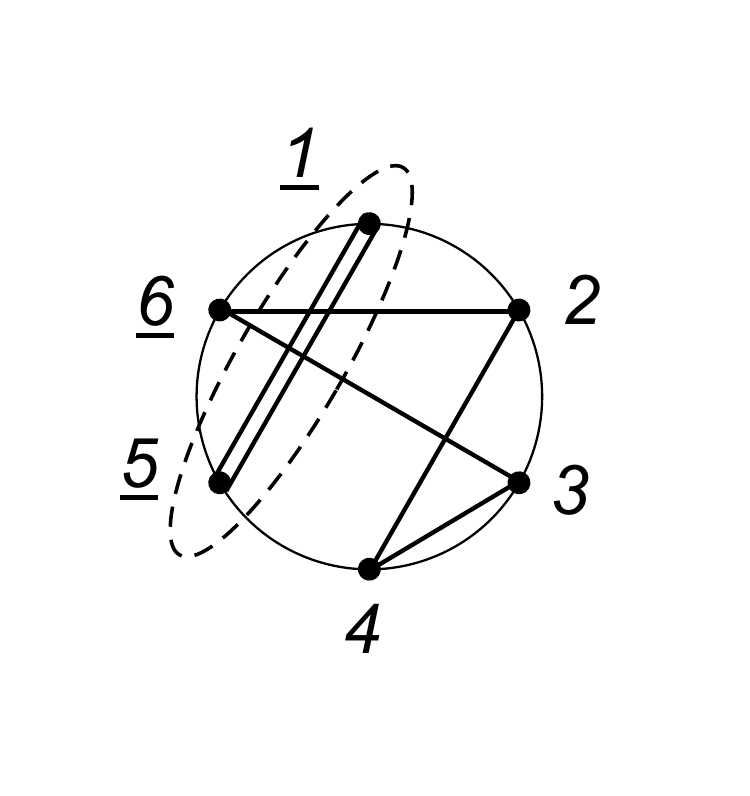}} 
\!\!\!\!
= $ (\a_{15})^{-1} $
\!\!\!\!\!\!\!
\!\!\!
\parbox[c]{5.1em}{\includegraphics[scale=0.19]{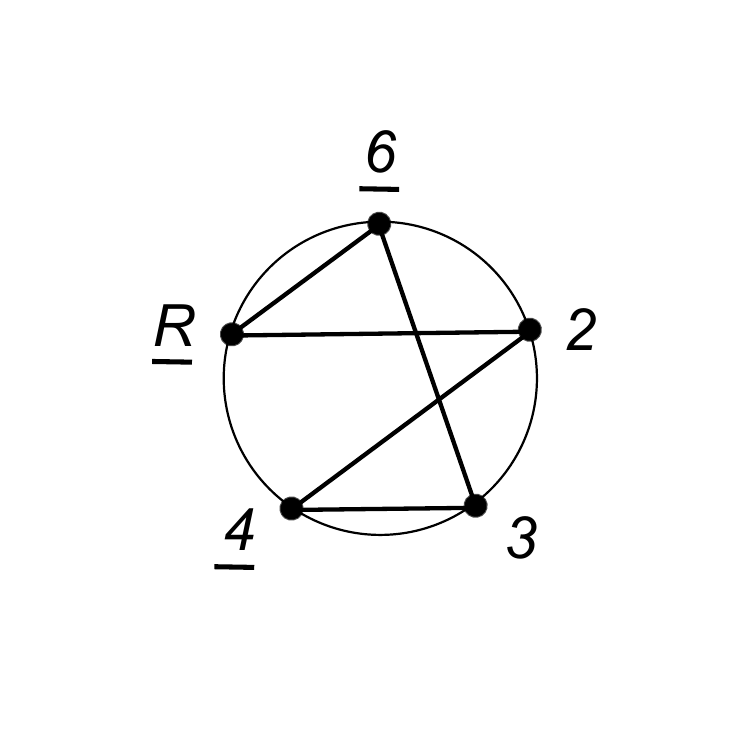}} 
\vspace{-0.15cm}
\caption{(a) Computing the ${\cal A}^{(b)}$ diagram.
(b) Computing the ${\cal A}^{(c)}$ diagram.
(c) Computing the ${\cal A}^{(d)}$, ${\cal A}^{(e)}$, and ${\cal A}^{(f)}$ diagrams.}\label{Adef-contributions} 
\end{figure}

The other diagrams in Figure \ref{Fig6-point} can be computed using similar methods, so we just state the final results:
\begin{align}\label{otherfigs}
{\cal A}^{(b)} 
 &= 
 -
 \left[ ( {\cal D}_2+ {\cal D}_3+ {\cal D}_4)^2 + m^2 \right]^{-1} (\a_{15})^{-1} (\a_{24})^{-1} - (\a_{15})^{-1} (\a_{24})^{-1}(\a_{36})^{-1}, \nonumber \\
{\cal A}^{(c)}
 &= 
 \left[ ( {\cal D}_3+{\cal D}_4+ {\cal D}_5 )^2 + m^2 \right]^{-1} (\a_{26})^{-1} \left(\a_{34} \right)^{-1} \left( - \a_{15} \right) \left(\a_{34} \right)^{-1}
 \nonumber\\
 &+  \left[ ( {\cal D}_2+{\cal D}_3+ {\cal D}_4 )^2 + m^2 \right]^{-1} (\a_{15})^{-1} \left(\a_{34} \right)^{-1} \left(  \a_{23} + \a_{34} \right) \left(\a_{34} \right)^{-1}
  \nonumber\\
 &+  (\a_{15})^{-1} (\a_{26})^{-1} \left(\a_{34} \right)^{-1} \left(  \a_{23} + \a_{34}+\a_{36} \right) \left(\a_{34} \right)^{-1}, \nonumber \\
{\cal A}^{(d)} 
 &= 
 \left[ ( {\cal D}_2+{\cal D}_3+ {\cal D}_4 )^2 + m^2 \right]^{-1} (\a_{15})^{-1} \left(\a_{23} \right)^{-1} , \nonumber \\
{\cal A}^{(e)} 
 &= -
 \left[ ( {\cal D}_2+{\cal D}_3+ {\cal D}_4 )^2 + m^2 \right]^{-1} (\a_{15})^{-1} \left\{ \left(\a_{23} \right)^{-1}+ \left(\a_{34} \right)^{-1}\right\}, \nonumber\\
{\cal A}^{(f)} 
 &= 
 \left[ ( {\cal D}_2+{\cal D}_3+ {\cal D}_4 )^2 + m^2 \right]^{-1} (\a_{15})^{-1} \left(\a_{34} \right)^{-1} \, .
\end{align}
The factorisation cuts from which these were derived are illustrated in Figure \ref{Adef-contributions}. The complete result is then obtained by combining \eqref{figa} with \eqref{otherfigs}. Neglecting commutators, it is not difficult to see that this agrees with \eqref{nlsmaction6pt}.

\section{Four-point diagram with double pole}\label{4point-DP}

In this Appendix, we will illustrate how to evaluate a worldsheet integral with a double pole. This integral arises in the 6-point NLSM computation as explained in the previous Appendix. Non-simple poles can also appear in four-point worldsheet formulae of higher-derivative theories like DBI and sGal, as explained in section \ref{worldsheet}, although they can be avoided using an appropriate choice of Pfaffians. Let us consider the integral
\begin{equation}
{\cal A}_4=\int_{\gamma_{S_d}} \dif \s_d \, (S_d)^{-1} (\s_{ab} \s_{bc} \s_{ca})^2\, {\rm PT}(a,b,c,d)\, \frac{1}{\s_{ab}^2 \s_{cd}^2},
 \end{equation}
where
\begin{equation}
S_{d}=\frac{\a_{da}}{\s_{da}} + \frac{\a_{db}}{\s_{db}} + \frac{\a_{dc}}{\s_{dc}}, 
\label{cwi4}
\end{equation}
and $\a_{da}+\a_{db}+\a_{dc}=0$ when acting on a four-point contact diagram with ${\cal D}_a^2\neq {\cal D}_b^2 \neq {\cal D}_c^2 \neq -m^2$ (see section 6.2 of \cite{Gomez:2021ujt} for a discussion of how CWI and ${\rm SL}(2,\mathbb{C})$ symmetry are realised in this context).

\begin{figure}[h]
\centering
\parbox[c]{6.3em}{\includegraphics[scale=0.22]{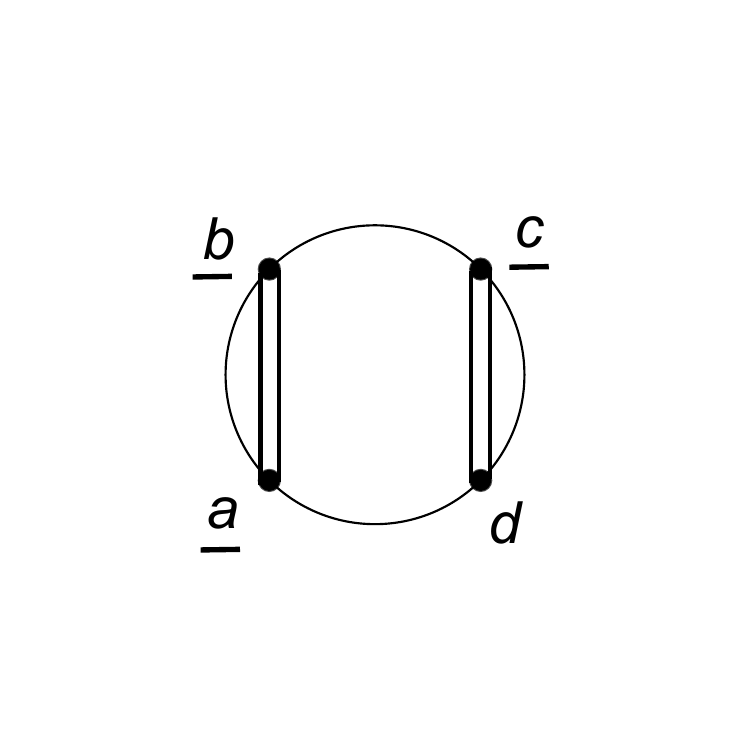}} 
$ \rightarrow$
\!\!\!\!\!
\parbox[c]{6.3em}{\includegraphics[scale=0.22]{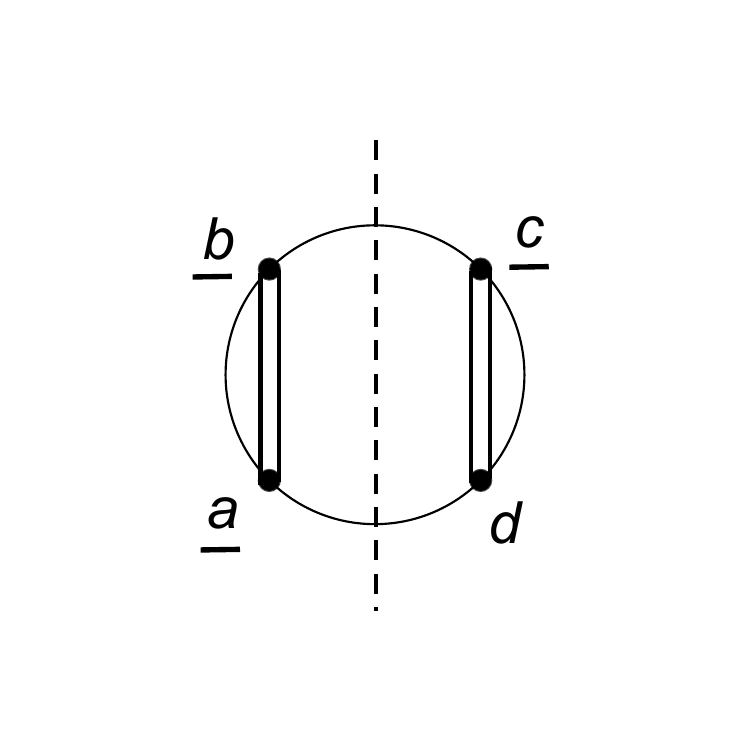}} 
\vspace{-0.45cm}
\caption{Four-point diagrams with double pole and its factorisation contribution.}\label{A4-dp} 
\end{figure}

In the following, it will be understood that $\mathcal{A}_4$ acts on a contact diagram and that momentum in conserved along the boundary. In Fig. \ref{A4-dp}, we give the graph representation for ${\cal A}_4$ and its factorisation contribution after applying the integration rules reviewed in the previous Appendix. 
This factorisation contribution is given when $\s_d\rightarrow \s_c$. To carry out this computation we use the parametrisation $\sigma_i=\epsilon x_i+\sigma_c$,  
with $i=c,d$, $x_d=\text{constant}$, $x_c=0$, $\sigma_c=\sigma_L$,
and expand around $\epsilon=0$: 
\begin{align}\label{eq:measure4pdp}
&
\dif\sigma_d=  x_{dc}\, \dif\epsilon
\nonumber \\
&
(\sigma_{ab} \sigma_{bc} \sigma_{ca})^2 \, 
 {\rm PT}(a,b,c,d) \, \frac{1}{\s_{ab}^2 \,\s_{cd}^2 }
=\frac{1}{\epsilon^3}\, \frac{(\sigma_{ab} \sigma_{bL} \sigma_{La})^2 }{ ( \sigma_{ab}^3\, \sigma_{b L }\, \sigma_{La} \, x^3_{cd}) }\, \left(     
1+\epsilon \frac{x_{cd}}{\s_{La}}
+ {\cal O}(\epsilon^{2})
\right), 
\end{align}
and
\begin{align}\label{eq:SEdp}
S_d =  \frac{1}{\epsilon}    \left[
  \hat S_d(\epsilon)+ {\cal O}(\epsilon^2) \right] ,\qquad
 \hat S_d(\epsilon)\equiv
 \frac{\a_{dc}}{x_{dc}} +\epsilon \left(
\frac{\a_{da}}{\s_{La}} + \frac{\a_{db}}{\s_{Lb}}
\right).
\end{align}
Using the global residue theorem, the integral becomes an integration around $\epsilon=0$, {\it i.e.}
\begin{equation}\label{eq:A4DP}
{\cal A}_4=
\int_{\gamma_{\epsilon}} \frac{\dif \epsilon}{\epsilon} \, \left[
  \hat S_d(\epsilon)+ {\cal O}(\epsilon^2) \right]^{-1} \frac{\sigma_{bL}  }{ ( \sigma_{ab} \, x_{cd}) }
 + 
\int_{\gamma_{\epsilon}} \frac{\dif \epsilon}{\epsilon^2} \, \left[
  \hat S_d(\epsilon)+ {\cal O}(\epsilon^2) \right]^{-1} \frac{(\sigma_{bL} \sigma_{La}) }{ ( \sigma_{ab}\ \, x^2_{cd}) } .
 \end{equation}
The first term has a simple pole and is simple to evaluate
\begin{equation}
\int_{\gamma_{\epsilon}} \frac{\dif \epsilon}{\epsilon} \, \left[
  \hat S_d(\epsilon)+ {\cal O}(\epsilon^2) \right]^{-1} \frac{\sigma_{bL}  }{ ( \sigma_{ab} \, x_{cd}) }
=-
\frac{\sigma_{bL}  }{  \sigma_{ab} } \, (\a_{dc})^{-1}.
 \end{equation}

Since the second term in \eqref{eq:A4DP} has a double pole, we must compute the derivative $\frac{\dif [( \hat S_d(\epsilon))^{-1}] }{\dif \epsilon}$, which must be done with care since $\hat{S}_d$ is a differential operator.
Using the identity
\begin{equation}
\hat S_d(\epsilon) \, [\hat S_d(\epsilon)]^{-1} = \mathbb{I},
\end{equation}
it is straightforward to arrive at the following result:
\begin{equation}
\frac{\dif [( \hat S_d(\epsilon))^{-1}] }{\dif \epsilon}\Big|_{\epsilon=0}=-
[\hat S_d(0)]^{-1}  \hat S^\prime_d(0) \, [\hat S_d(0)]^{-1} 
=-x_{dc}^2 \, (\a_{dc})^{-1} \left(
\frac{\a_{da}}{\s_{La}} + \frac{\a_{db}}{\s_{Lb}}
\right) (\a_{dc})^{-1}.
\end{equation}
Therefore, the second integral in \eqref{eq:A4DP} is given by
\begin{equation}
\int_{\gamma_{\epsilon}} \frac{\dif \epsilon}{\epsilon^2} \, \left[
  \hat S_d(\epsilon)+ {\cal O}(\epsilon^2) \right]^{-1} \frac{(\sigma_{bL} \sigma_{La}) }{ ( \sigma_{ab}\ \, x^2_{cd}) }
 =\frac{1}{\s_{ab}} (\a_{dc})^{-1} \left(
\s_{Lb} \, \a_{da} + \s_{La}\, \a_{db}
\right) (\a_{dc})^{-1}.
\end{equation}
Finally, we obtain
\begin{align}
{\cal A}_4
&
=\frac{1}{\s_{ab}} (\a_{dc})^{-1} \left(
\s_{Lb} \, \a_{dc}+ 
\s_{Lb} \, \a_{da} + \s_{La}\, \a_{db}
\right) (\a_{dc})^{-1}\nonumber \\
&
=-(\a_{cd})^{-1} ( \a_{db}) (\a_{cd})^{-1}\nonumber\\
&
=(\a_{cd})^{-1} ( \a_{ad}+\a_{cd}) (\a_{cd})^{-1},
\end{align}
where we have used the CWI in \eqref{cwi4} in the last two lines.

\section{Six-derivative results}
\label{app:sGalMin}

For minimally coupled scalars we find that
\eqs{
(\hat{s}^{3}+\hat{t}^{3}+\hat{u}^{3})\mathcal{C}_{4}^{\Delta=3} &= \frac{360}{E^7}k_1k_2k_3k_4 k_1^\mu k_{2, \mu} k_3^\nu k_{4, \nu}\left(k_1^\sigma k_{2, \sigma} + k_3^\sigma k_{4, \sigma}\right)\left(1 + \frac{E}{6}\sum_{i=1}^{4}\frac{1}{k_i} + \frac{E^2}{30k_1k_2k_3k_4}\sum_{i<j}k_ik_j\right)\\
&\qquad + \frac{60}{E^6}k_1k_2k_3k_4\left((k_3+k_4)(k_1^\mu k_{2, \mu})^2 + (k_1+k_2)(k_3^\mu k_{4, \mu})^2\right)\\
&\qquad + \frac{24}{E^5}\bigg(10 k_1k_2k_3k_4 k_1^\mu k_{2, \mu} k_3^\nu k_{4, \nu}\\
&\qquad\qquad  + \frac{1}{2}\sum_{i<j}k_ik_j\Big(2k_1k_2k_3k_4\left(k_1^\mu k_{2, \mu}+k_3^\mu k_{4, \mu}-k_1k_2-k_3k_4\right) -2k_3^2k_4^2k_1^\mu k_{2, \mu}\\
&\qquad\qquad\qquad-2k_1^2k_2^2k_3^\mu k_{4, \mu}+2(k_1k_2+k_3k_4)k_1^\mu k_{2, \mu} k_3^\nu k_{4, \nu}+k_3k_4(k_1^\mu k_{2, \mu})^2\\
&\qquad\qquad\qquad+k_1k_2(k_3^\mu k_{4, \mu})^2\Big) + \left(1+\sum_{i\neq j}\frac{k_i}{k_j}\right)k_1k_2k_3k_4\vec{k}_1\cdot \vec{k}_2 \vec{k}_3\cdot \vec{k}_4\\
&\qquad\qquad -(k_1^2k_2^2+k_3^2k_4^2)\vec{k}_1\cdot \vec{k}_2 \vec{k}_3\cdot \vec{k}_4 + k_3^2k_4^2\vec{k}_1\cdot \vec{k}_2\left(-k_1^2-k_2^2+2(k_1k_2+k_3k_4)\right)\\
&\qquad\qquad + k_1^2k_2^2\vec{k}_3\cdot \vec{k}_4\left(-k_3^2-k_4^2+2(k_1k_2+k_3k_4)\right)\\
&\qquad\qquad-\frac{1}{2}\left(4k_1^2k_2^2k_3^2k_4^2+k_1^2k_2^2 + (\vec{k}_3\cdot \vec{k}_4)^2 + (\vec{k}_1\cdot \vec{k}_2)^2\right)\bigg)\\
&\qquad + \frac{6}{E^4}\Bigg(\left(k_1k_2k_3k_4 \sum_{i=1}^4\frac{1}{k_i} + \sum_{i\neq j}k_ik_j^2\right)\vec{k}_1\cdot\vec{k}_2 \vec{k}_3\cdot\vec{k}_4\\
&\qquad\qquad -2k_3^2k_4^2(k_3+k_4)\vec{k}_1\cdot\vec{k}_2 -2k_1^2k_2^2(k_1+k_2)\vec{k}_3\cdot\vec{k}_4\Bigg)\\
&\qquad + \frac{2}{E^3}\Big(2\vec{k}_1\cdot\vec{k}_2\vec{k}_3\cdot\vec{k}_4(\vec{k}_1\cdot\vec{k}_2+\vec{k}_3\cdot\vec{k}_4)-\sum_{i<j}k_ik_j \vec{k}_1\cdot\vec{k}_2\vec{k}_3\cdot\vec{k}_4 \\
&\qquad\qquad + 2k_3^2k_4^2\vec{k}_1\cdot\vec{k}_2 + 2k_1^2k_2^2\vec{k}_3\cdot\vec{k}_4\Big)\\
&\qquad + \frac{4}{E}\vec{k}_1\cdot\vec{k}_2\vec{k}_3\cdot\vec{k}_4 + \mathrm{Cyc.}[234],
}
where $k_i^\mu k_{j, \mu} = \vec{k}_i\cdot\vec{k}_j - k_ik_j$. This expression mixes different types of dot products in order to obtain a more compact form.

\bibliography{EffDoubleCopy}

\providecommand{\href}[2]{#2}\begingroup\raggedright\begin{thebibliography}{10}

\bibitem{Guth:1980zm}
A.~H. Guth, \emph{{The Inflationary Universe: A Possible Solution to the
  Horizon and Flatness Problems}},
  \href{http://dx.doi.org/10.1103/PhysRevD.23.347}{\emph{Phys. Rev. D}
  {\bfseries 23} (1981) 347--356}.

\bibitem{Linde:1981mu}
A.~D. Linde, \emph{{A New Inflationary Universe Scenario: A Possible Solution
  of the Horizon, Flatness, Homogeneity, Isotropy and Primordial Monopole
  Problems}}, \href{http://dx.doi.org/10.1016/0370-2693(82)91219-9}{\emph{Phys.
  Lett. B} {\bfseries 108} (1982) 389--393}.

\bibitem{Albrecht:1982wi}
A.~Albrecht and P.~J. Steinhardt, \emph{{Cosmology for Grand Unified Theories
  with Radiatively Induced Symmetry Breaking}},
  \href{http://dx.doi.org/10.1103/PhysRevLett.48.1220}{\emph{Phys. Rev. Lett.}
  {\bfseries 48} (1982) 1220--1223}.

\bibitem{Mukhanov:1981xt}
V.~F. Mukhanov and G.~V. Chibisov, \emph{{Quantum Fluctuations and a
  Nonsingular Universe}}, {\emph{JETP Lett.} {\bfseries 33} (1981) 532--535}.

\bibitem{Hartle:1983ai}
J.~B. Hartle and S.~W. Hawking, \emph{{Wave Function of the Universe}},
  \href{http://dx.doi.org/10.1103/PhysRevD.28.2960}{\emph{Phys. Rev. D}
  {\bfseries 28} (1983) 2960--2975}.

\bibitem{Strominger:2001gp}
A.~Strominger, \emph{{Inflation and the dS / CFT correspondence}},
  \href{http://dx.doi.org/10.1088/1126-6708/2001/11/049}{\emph{JHEP} {\bfseries
  11} (2001) 049}, [\href{https://arxiv.org/abs/hep-th/0110087}{{\ttfamily
  hep-th/0110087}}].

\bibitem{Maldacena:2002vr}
J.~M. Maldacena, \emph{{Non-Gaussian features of primordial fluctuations in
  single field inflationary models}},
  \href{http://dx.doi.org/10.1088/1126-6708/2003/05/013}{\emph{JHEP} {\bfseries
  05} (2003) 013}, [\href{https://arxiv.org/abs/astro-ph/0210603}{{\ttfamily
  astro-ph/0210603}}].

\bibitem{McFadden:2009fg}
P.~McFadden and K.~Skenderis, \emph{{Holography for Cosmology}},
  \href{http://dx.doi.org/10.1103/PhysRevD.81.021301}{\emph{Phys. Rev. D}
  {\bfseries 81} (2010) 021301},
  [\href{https://arxiv.org/abs/0907.5542}{{\ttfamily 0907.5542}}].

\bibitem{McFadden:2010vh}
P.~McFadden and K.~Skenderis, \emph{{Holographic Non-Gaussianity}},
  \href{http://dx.doi.org/10.1088/1475-7516/2011/05/013}{\emph{JCAP} {\bfseries
  05} (2011) 013}, [\href{https://arxiv.org/abs/1011.0452}{{\ttfamily
  1011.0452}}].

\bibitem{Maldacena:2011nz}
J.~M. Maldacena and G.~L. Pimentel, \emph{{On graviton non-Gaussianities during
  inflation}}, \href{http://dx.doi.org/10.1007/JHEP09(2011)045}{\emph{JHEP}
  {\bfseries 09} (2011) 045},
  [\href{https://arxiv.org/abs/1104.2846}{{\ttfamily 1104.2846}}].

\bibitem{Raju:2011mp}
S.~Raju, \emph{{Recursion Relations for AdS/CFT Correlators}},
  \href{http://dx.doi.org/10.1103/PhysRevD.83.126002}{\emph{Phys. Rev. D}
  {\bfseries 83} (2011) 126002},
  [\href{https://arxiv.org/abs/1102.4724}{{\ttfamily 1102.4724}}].

\bibitem{Ghosh:2014kba}
A.~Ghosh, N.~Kundu, S.~Raju and S.~P. Trivedi, \emph{{Conformal Invariance and
  the Four Point Scalar Correlator in Slow-Roll Inflation}},
  \href{http://dx.doi.org/10.1007/JHEP07(2014)011}{\emph{JHEP} {\bfseries 07}
  (2014) 011}, [\href{https://arxiv.org/abs/1401.1426}{{\ttfamily 1401.1426}}].

\bibitem{Weinberg:2005vy}
S.~Weinberg, \emph{{Quantum contributions to cosmological correlations}},
  \href{http://dx.doi.org/10.1103/PhysRevD.72.043514}{\emph{Phys. Rev. D}
  {\bfseries 72} (2005) 043514},
  [\href{https://arxiv.org/abs/hep-th/0506236}{{\ttfamily hep-th/0506236}}].

\bibitem{Arkani-Hamed:2017fdk}
N.~Arkani-Hamed, P.~Benincasa and A.~Postnikov, \emph{{Cosmological Polytopes
  and the Wavefunction of the Universe}},
  \href{https://arxiv.org/abs/1709.02813}{{\ttfamily 1709.02813}}.

\bibitem{Bzowski:2020kfw}
A.~Bzowski, P.~McFadden and K.~Skenderis, \emph{{Conformal correlators as
  simplex integrals in momentum space}},
  \href{http://dx.doi.org/10.1007/JHEP01(2021)192}{\emph{JHEP} {\bfseries 01}
  (2021) 192}, [\href{https://arxiv.org/abs/2008.07543}{{\ttfamily
  2008.07543}}].

\bibitem{Arkani-Hamed:2015bza}
N.~Arkani-Hamed and J.~Maldacena, \emph{{Cosmological Collider Physics}},
  \href{https://arxiv.org/abs/1503.08043}{{\ttfamily 1503.08043}}.

\bibitem{Arkani-Hamed:2018kmz}
N.~Arkani-Hamed, D.~Baumann, H.~Lee and G.~L. Pimentel, \emph{{The Cosmological
  Bootstrap: Inflationary Correlators from Symmetries and Singularities}},
  \href{http://dx.doi.org/10.1007/JHEP04(2020)105}{\emph{JHEP} {\bfseries 04}
  (2020) 105}, [\href{https://arxiv.org/abs/1811.00024}{{\ttfamily
  1811.00024}}].

\bibitem{Baumann:2019oyu}
D.~Baumann, C.~Duaso~Pueyo, A.~Joyce, H.~Lee and G.~L. Pimentel, \emph{{The
  cosmological bootstrap: weight-shifting operators and scalar seeds}},
  \href{http://dx.doi.org/10.1007/JHEP12(2020)204}{\emph{JHEP} {\bfseries 12}
  (2020) 204}, [\href{https://arxiv.org/abs/1910.14051}{{\ttfamily
  1910.14051}}].

\bibitem{Baumann:2020dch}
D.~Baumann, C.~Duaso~Pueyo, A.~Joyce, H.~Lee and G.~L. Pimentel, \emph{{The
  Cosmological Bootstrap: Spinning Correlators from Symmetries and
  Factorization}},
  \href{http://dx.doi.org/10.21468/SciPostPhys.11.3.071}{\emph{SciPost Phys.}
  {\bfseries 11} (2021) 071},
  [\href{https://arxiv.org/abs/2005.04234}{{\ttfamily 2005.04234}}].

\bibitem{Baumann:2021fxj}
D.~Baumann, W.-M. Chen, C.~Duaso~Pueyo, A.~Joyce, H.~Lee and G.~L. Pimentel,
  \emph{{Linking the Singularities of Cosmological Correlators}},
  \href{https://arxiv.org/abs/2106.05294}{{\ttfamily 2106.05294}}.

\bibitem{Meltzer:2021zin}
D.~Meltzer, \emph{{The inflationary wavefunction from analyticity and
  factorization}},
  \href{http://dx.doi.org/10.1088/1475-7516/2021/12/018}{\emph{JCAP} {\bfseries
  12} (2021) 018}, [\href{https://arxiv.org/abs/2107.10266}{{\ttfamily
  2107.10266}}].

\bibitem{Hillman:2021bnk}
A.~Hillman and E.~Pajer, \emph{{A differential representation of cosmological
  wavefunctions}}, \href{http://dx.doi.org/10.1007/JHEP04(2022)012}{\emph{JHEP}
  {\bfseries 04} (2022) 012},
  [\href{https://arxiv.org/abs/2112.01619}{{\ttfamily 2112.01619}}].

\bibitem{Alday:2017vkk}
L.~F. Alday and S.~Caron-Huot, \emph{{Gravitational S-matrix from CFT
  dispersion relations}},
  \href{http://dx.doi.org/10.1007/JHEP12(2018)017}{\emph{JHEP} {\bfseries 12}
  (2018) 017}, [\href{https://arxiv.org/abs/1711.02031}{{\ttfamily
  1711.02031}}].

\bibitem{Meltzer:2020qbr}
D.~Meltzer and A.~Sivaramakrishnan, \emph{{CFT unitarity and the AdS Cutkosky
  rules}}, \href{http://dx.doi.org/10.1007/JHEP11(2020)073}{\emph{JHEP}
  {\bfseries 11} (2020) 073},
  [\href{https://arxiv.org/abs/2008.11730}{{\ttfamily 2008.11730}}].

\bibitem{Goodhew:2020hob}
H.~Goodhew, S.~Jazayeri and E.~Pajer, \emph{{The Cosmological Optical
  Theorem}}, \href{http://dx.doi.org/10.1088/1475-7516/2021/04/021}{\emph{JCAP}
  {\bfseries 04} (2021) 021},
  [\href{https://arxiv.org/abs/2009.02898}{{\ttfamily 2009.02898}}].

\bibitem{Melville:2021lst}
S.~Melville and E.~Pajer, \emph{{Cosmological Cutting Rules}},
  \href{http://dx.doi.org/10.1007/JHEP05(2021)249}{\emph{JHEP} {\bfseries 05}
  (2021) 249}, [\href{https://arxiv.org/abs/2103.09832}{{\ttfamily
  2103.09832}}].

\bibitem{Jazayeri:2021fvk}
S.~Jazayeri, E.~Pajer and D.~Stefanyszyn, \emph{{From locality and unitarity to
  cosmological correlators}},
  \href{http://dx.doi.org/10.1007/JHEP10(2021)065}{\emph{JHEP} {\bfseries 10}
  (2021) 065}, [\href{https://arxiv.org/abs/2103.08649}{{\ttfamily
  2103.08649}}].

\bibitem{Goodhew:2021oqg}
H.~Goodhew, S.~Jazayeri, M.~H. Gordon~Lee and E.~Pajer, \emph{{Cutting
  cosmological correlators}},
  \href{http://dx.doi.org/10.1088/1475-7516/2021/08/003}{\emph{JCAP} {\bfseries
  08} (2021) 003}, [\href{https://arxiv.org/abs/2104.06587}{{\ttfamily
  2104.06587}}].

\bibitem{Sleight:2019hfp}
C.~Sleight and M.~Taronna, \emph{{Bootstrapping Inflationary Correlators in
  Mellin Space}}, \href{http://dx.doi.org/10.1007/JHEP02(2020)098}{\emph{JHEP}
  {\bfseries 02} (2020) 098},
  [\href{https://arxiv.org/abs/1907.01143}{{\ttfamily 1907.01143}}].

\bibitem{Sleight:2021plv}
C.~Sleight and M.~Taronna, \emph{{From dS to AdS and back}},
  \href{http://dx.doi.org/10.1007/JHEP12(2021)074}{\emph{JHEP} {\bfseries 12}
  (2021) 074}, [\href{https://arxiv.org/abs/2109.02725}{{\ttfamily
  2109.02725}}].

\bibitem{Armstrong:2020woi}
C.~Armstrong, A.~E. Lipstein and J.~Mei, \emph{{Color/kinematics duality in
  AdS$_{4}$}}, \href{http://dx.doi.org/10.1007/JHEP02(2021)194}{\emph{JHEP}
  {\bfseries 02} (2021) 194},
  [\href{https://arxiv.org/abs/2012.02059}{{\ttfamily 2012.02059}}].

\bibitem{Albayrak:2020fyp}
S.~Albayrak, S.~Kharel and D.~Meltzer, \emph{{On duality of color and
  kinematics in (A)dS momentum space}},
  \href{http://dx.doi.org/10.1007/JHEP03(2021)249}{\emph{JHEP} {\bfseries 03}
  (2021) 249}, [\href{https://arxiv.org/abs/2012.10460}{{\ttfamily
  2012.10460}}].

\bibitem{Alday:2021odx}
L.~F. Alday, C.~Behan, P.~Ferrero and X.~Zhou, \emph{{Gluon Scattering in AdS
  from CFT}}, \href{http://dx.doi.org/10.1007/JHEP06(2021)020}{\emph{JHEP}
  {\bfseries 06} (2021) 020},
  [\href{https://arxiv.org/abs/2103.15830}{{\ttfamily 2103.15830}}].

\bibitem{Diwakar:2021juk}
P.~Diwakar, A.~Herderschee, R.~Roiban and F.~Teng, \emph{{BCJ amplitude
  relations for Anti-de Sitter boundary correlators in embedding space}},
  \href{http://dx.doi.org/10.1007/JHEP10(2021)141}{\emph{JHEP} {\bfseries 10}
  (2021) 141}, [\href{https://arxiv.org/abs/2106.10822}{{\ttfamily
  2106.10822}}].

\bibitem{Sivaramakrishnan:2021srm}
A.~Sivaramakrishnan, \emph{{Towards Color-Kinematics Duality in Generic
  Spacetimes}},  \href{https://arxiv.org/abs/2110.15356}{{\ttfamily
  2110.15356}}.

\bibitem{Cheung:2022pdk}
C.~Cheung, J.~Parra-Martinez and A.~Sivaramakrishnan, \emph{{On-shell
  Correlators and Color-Kinematics Duality in Curved Symmetric Spacetimes}},
  \href{https://arxiv.org/abs/2201.05147}{{\ttfamily 2201.05147}}.

\bibitem{Herderschee:2022ntr}
A.~Herderschee, R.~Roiban and F.~Teng, \emph{{On the Differential
  Representation and Color-Kinematics Duality of AdS Boundary Correlators}},
  \href{https://arxiv.org/abs/2201.05067}{{\ttfamily 2201.05067}}.

\bibitem{Drummond:2022dxd}
J.~M. Drummond, R.~Glew and M.~Santagata, \emph{{BCJ relations in ${AdS}_5
  \times S^3$ and the double-trace spectrum of super gluons}},
  \href{https://arxiv.org/abs/2202.09837}{{\ttfamily 2202.09837}}.

\bibitem{Farrow:2018yni}
J.~A. Farrow, A.~E. Lipstein and P.~McFadden, \emph{{Double copy structure of
  CFT correlators}},
  \href{http://dx.doi.org/10.1007/JHEP02(2019)130}{\emph{JHEP} {\bfseries 02}
  (2019) 130}, [\href{https://arxiv.org/abs/1812.11129}{{\ttfamily
  1812.11129}}].

\bibitem{Lipstein:2019mpu}
A.~E. Lipstein and P.~McFadden, \emph{{Double copy structure and the flat space
  limit of conformal correlators in even dimensions}},
  \href{http://dx.doi.org/10.1103/PhysRevD.101.125006}{\emph{Phys. Rev. D}
  {\bfseries 101} (2020) 125006},
  [\href{https://arxiv.org/abs/1912.10046}{{\ttfamily 1912.10046}}].

\bibitem{Jain:2021qcl}
S.~Jain, R.~R. John, A.~Mehta, A.~A. Nizami and A.~Suresh, \emph{{Double copy
  structure of parity-violating CFT correlators}},
  \href{http://dx.doi.org/10.1007/JHEP07(2021)033}{\emph{JHEP} {\bfseries 07}
  (2021) 033}, [\href{https://arxiv.org/abs/2104.12803}{{\ttfamily
  2104.12803}}].

\bibitem{Zhou:2021gnu}
X.~Zhou, \emph{{Double Copy Relation in AdS Space}},
  \href{http://dx.doi.org/10.1103/PhysRevLett.127.141601}{\emph{Phys. Rev.
  Lett.} {\bfseries 127} (2021) 141601},
  [\href{https://arxiv.org/abs/2106.07651}{{\ttfamily 2106.07651}}].

\bibitem{Gomez:2021qfd}
H.~Gomez, R.~L. Jusinskas and A.~Lipstein, \emph{{Cosmological Scattering
  Equations}},
  \href{http://dx.doi.org/10.1103/PhysRevLett.127.251604}{\emph{Phys. Rev.
  Lett.} {\bfseries 127} (2021) 251604},
  [\href{https://arxiv.org/abs/2106.11903}{{\ttfamily 2106.11903}}].

\bibitem{Gomez:2021ujt}
H.~Gomez, R.~L. Jusinskas and A.~Lipstein, \emph{{Cosmological Scattering
  Equations at Tree-level and One-loop}},
  \href{https://arxiv.org/abs/2112.12695}{{\ttfamily 2112.12695}}.

\bibitem{Fichet:2021xfn}
S.~Fichet, \emph{{Field Holography in General Background and Boundary Effective
  Action from AdS to dS}},  \href{https://arxiv.org/abs/2112.00746}{{\ttfamily
  2112.00746}}.

\bibitem{Herderschee:2021jbi}
A.~Herderschee, \emph{{A New Framework for Higher Loop Witten Diagrams}},
  \href{https://arxiv.org/abs/2112.08226}{{\ttfamily 2112.08226}}.

\bibitem{Gadde:2022ghy}
A.~Gadde and T.~Sharma, \emph{{A Scattering Amplitude for Massive Particles in
  AdS}},  \href{https://arxiv.org/abs/2204.06462}{{\ttfamily 2204.06462}}.

\bibitem{Heckelbacher:2022hbq}
T.~Heckelbacher, I.~Sachs, E.~Skvortsov and P.~Vanhove, \emph{{Analytical
  evaluation of cosmological correlation functions}},
  \href{https://arxiv.org/abs/2204.07217}{{\ttfamily 2204.07217}}.

\bibitem{Bern:2008qj}
Z.~Bern, J.~J.~M. Carrasco and H.~Johansson, \emph{{New Relations for
  Gauge-Theory Amplitudes}},
  \href{http://dx.doi.org/10.1103/PhysRevD.78.085011}{\emph{Phys. Rev. D}
  {\bfseries 78} (2008) 085011},
  [\href{https://arxiv.org/abs/0805.3993}{{\ttfamily 0805.3993}}].

\bibitem{Bern:2010ue}
Z.~Bern, J.~J.~M. Carrasco and H.~Johansson, \emph{{Perturbative Quantum
  Gravity as a Double Copy of Gauge Theory}},
  \href{http://dx.doi.org/10.1103/PhysRevLett.105.061602}{\emph{Phys. Rev.
  Lett.} {\bfseries 105} (2010) 061602},
  [\href{https://arxiv.org/abs/1004.0476}{{\ttfamily 1004.0476}}].

\bibitem{fairlie}
D.~B. Fairlie and D.~E. Roberts, \emph{{DUAL MODELS WITHOUT TACHYONS - A NEW
  APPROACH}}, .

\bibitem{Gross:1987kza}
D.~J. Gross and P.~F. Mende, \emph{{The High-Energy Behavior of String
  Scattering Amplitudes}},
  \href{http://dx.doi.org/10.1016/0370-2693(87)90355-8}{\emph{Phys. Lett. B}
  {\bfseries 197} (1987) 129--134}.

\bibitem{Cachazo:2013hca}
F.~Cachazo, S.~He and E.~Y. Yuan, \emph{{Scattering of Massless Particles in
  Arbitrary Dimensions}},
  \href{http://dx.doi.org/10.1103/PhysRevLett.113.171601}{\emph{Phys. Rev.
  Lett.} {\bfseries 113} (2014) 171601},
  [\href{https://arxiv.org/abs/1307.2199}{{\ttfamily 1307.2199}}].

\bibitem{Cachazo:2013iea}
F.~Cachazo, S.~He and E.~Y. Yuan, \emph{{Scattering of Massless Particles:
  Scalars, Gluons and Gravitons}},
  \href{http://dx.doi.org/10.1007/JHEP07(2014)033}{\emph{JHEP} {\bfseries 07}
  (2014) 033}, [\href{https://arxiv.org/abs/1309.0885}{{\ttfamily 1309.0885}}].

\bibitem{Mason:2013sva}
L.~Mason and D.~Skinner, \emph{{Ambitwistor strings and the scattering
  equations}}, \href{http://dx.doi.org/10.1007/JHEP07(2014)048}{\emph{JHEP}
  {\bfseries 07} (2014) 048},
  [\href{https://arxiv.org/abs/1311.2564}{{\ttfamily 1311.2564}}].

\bibitem{Cachazo:2014xea}
F.~Cachazo, S.~He and E.~Y. Yuan, \emph{{Scattering Equations and Matrices:
  From Einstein To Yang-Mills, DBI and NLSM}},
  \href{http://dx.doi.org/10.1007/JHEP07(2015)149}{\emph{JHEP} {\bfseries 07}
  (2015) 149}, [\href{https://arxiv.org/abs/1412.3479}{{\ttfamily 1412.3479}}].

\bibitem{Casali:2015vta}
E.~Casali, Y.~Geyer, L.~Mason, R.~Monteiro and K.~A. Roehrig, \emph{{New
  Ambitwistor String Theories}},
  \href{http://dx.doi.org/10.1007/JHEP11(2015)038}{\emph{JHEP} {\bfseries 11}
  (2015) 038}, [\href{https://arxiv.org/abs/1506.08771}{{\ttfamily
  1506.08771}}].

\bibitem{Cheung:2017ems}
C.~Cheung, C.-H. Shen and C.~Wen, \emph{{Unifying Relations for Scattering
  Amplitudes}}, \href{http://dx.doi.org/10.1007/JHEP02(2018)095}{\emph{JHEP}
  {\bfseries 02} (2018) 095},
  [\href{https://arxiv.org/abs/1705.03025}{{\ttfamily 1705.03025}}].

\bibitem{Eberhardt:2020ewh}
L.~Eberhardt, S.~Komatsu and S.~Mizera, \emph{{Scattering equations in AdS:
  scalar correlators in arbitrary dimensions}},
  \href{http://dx.doi.org/10.1007/JHEP11(2020)158}{\emph{JHEP} {\bfseries 11}
  (2020) 158}, [\href{https://arxiv.org/abs/2007.06574}{{\ttfamily
  2007.06574}}].

\bibitem{Roehrig:2020kck}
K.~Roehrig and D.~Skinner, \emph{{Ambitwistor strings and the scattering
  equations on AdS$_{3}\times$S$^{3}$}},
  \href{http://dx.doi.org/10.1007/JHEP02(2022)073}{\emph{JHEP} {\bfseries 02}
  (2022) 073}, [\href{https://arxiv.org/abs/2007.07234}{{\ttfamily
  2007.07234}}].

\bibitem{Adler:1964um}
S.~L. Adler, \emph{{Consistency conditions on the strong interactions implied
  by a partially conserved axial vector current}},
  \href{http://dx.doi.org/10.1103/PhysRev.137.B1022}{\emph{Phys. Rev.}
  {\bfseries 137} (1965) B1022--B1033}.

\bibitem{Cheung:2014dqa}
C.~Cheung, K.~Kampf, J.~Novotny and J.~Trnka, \emph{{Effective Field Theories
  from Soft Limits of Scattering Amplitudes}},
  \href{http://dx.doi.org/10.1103/PhysRevLett.114.221602}{\emph{Phys. Rev.
  Lett.} {\bfseries 114} (2015) 221602},
  [\href{https://arxiv.org/abs/1412.4095}{{\ttfamily 1412.4095}}].

\bibitem{Low:2014nga}
I.~Low, \emph{{Adler\textquoteright{}s zero and effective Lagrangians for
  nonlinearly realized symmetry}},
  \href{http://dx.doi.org/10.1103/PhysRevD.91.105017}{\emph{Phys. Rev. D}
  {\bfseries 91} (2015) 105017},
  [\href{https://arxiv.org/abs/1412.2145}{{\ttfamily 1412.2145}}].

\bibitem{Hinterbichler:2015pqa}
K.~Hinterbichler and A.~Joyce, \emph{{Hidden symmetry of the Galileon}},
  \href{http://dx.doi.org/10.1103/PhysRevD.92.023503}{\emph{Phys. Rev. D}
  {\bfseries 92} (2015) 023503},
  [\href{https://arxiv.org/abs/1501.07600}{{\ttfamily 1501.07600}}].

\bibitem{Padilla:2016mno}
A.~Padilla, D.~Stefanyszyn and T.~Wilson, \emph{{Probing Scalar Effective Field
  Theories with the Soft Limits of Scattering Amplitudes}},
  \href{http://dx.doi.org/10.1007/JHEP04(2017)015}{\emph{JHEP} {\bfseries 04}
  (2017) 015}, [\href{https://arxiv.org/abs/1612.04283}{{\ttfamily
  1612.04283}}].

\bibitem{Bonifacio:2021mrf}
J.~Bonifacio, K.~Hinterbichler, A.~Joyce and D.~Roest, \emph{{Exceptional
  scalar theories in de Sitter space}},
  \href{https://arxiv.org/abs/2112.12151}{{\ttfamily 2112.12151}}.

\bibitem{Bittermann:2022nfh}
N.~Bittermann and A.~Joyce, \emph{{Soft limits of the wavefunction in
  exceptional scalar theories}},
  \href{https://arxiv.org/abs/2203.05576}{{\ttfamily 2203.05576}}.

\bibitem{Bjerrum-Bohr:2016axv}
N.~E.~J. Bjerrum-Bohr, J.~L. Bourjaily, P.~H. Damgaard and B.~Feng,
  \emph{{Manifesting Color-Kinematics Duality in the Scattering Equation
  Formalism}}, \href{http://dx.doi.org/10.1007/JHEP09(2016)094}{\emph{JHEP}
  {\bfseries 09} (2016) 094},
  [\href{https://arxiv.org/abs/1608.00006}{{\ttfamily 1608.00006}}].

\bibitem{Fu:2017uzt}
C.-H. Fu, Y.-J. Du, R.~Huang and B.~Feng, \emph{{Expansion of
  Einstein-Yang-Mills Amplitude}},
  \href{http://dx.doi.org/10.1007/JHEP09(2017)021}{\emph{JHEP} {\bfseries 09}
  (2017) 021}, [\href{https://arxiv.org/abs/1702.08158}{{\ttfamily
  1702.08158}}].

\bibitem{Edison:2020ehu}
A.~Edison and F.~Teng, \emph{{Efficient Calculation of Crossing Symmetric BCJ
  Tree Numerators}},
  \href{http://dx.doi.org/10.1007/JHEP12(2020)138}{\emph{JHEP} {\bfseries 12}
  (2020) 138}, [\href{https://arxiv.org/abs/2005.03638}{{\ttfamily
  2005.03638}}].

\bibitem{Bjerrum-Bohr:2020syg}
N.~E.~J. Bjerrum-Bohr, T.~V. Brown and H.~Gomez, \emph{{Scattering of Gravitons
  and Spinning Massive States from Compact Numerators}},
  \href{http://dx.doi.org/10.1007/JHEP04(2021)234}{\emph{JHEP} {\bfseries 04}
  (2021) 234}, [\href{https://arxiv.org/abs/2011.10556}{{\ttfamily
  2011.10556}}].

\bibitem{Low:2020ubn}
I.~Low, L.~Rodina and Z.~Yin, \emph{{Double Copy in Higher Derivative Operators
  of Nambu-Goldstone Bosons}},
  \href{http://dx.doi.org/10.1103/PhysRevD.103.025004}{\emph{Phys. Rev. D}
  {\bfseries 103} (2021) 025004},
  [\href{https://arxiv.org/abs/2009.00008}{{\ttfamily 2009.00008}}].

\bibitem{Heemskerk:2009pn}
I.~Heemskerk, J.~Penedones, J.~Polchinski and J.~Sully, \emph{{Holography from
  Conformal Field Theory}},
  \href{http://dx.doi.org/10.1088/1126-6708/2009/10/079}{\emph{JHEP} {\bfseries
  10} (2009) 079}, [\href{https://arxiv.org/abs/0907.0151}{{\ttfamily
  0907.0151}}].

\bibitem{Dolan:2013isa}
L.~Dolan and P.~Goddard, \emph{{Proof of the Formula of Cachazo, He and Yuan
  for Yang-Mills Tree Amplitudes in Arbitrary Dimension}},
  \href{http://dx.doi.org/10.1007/JHEP05(2014)010}{\emph{JHEP} {\bfseries 05}
  (2014) 010}, [\href{https://arxiv.org/abs/1311.5200}{{\ttfamily 1311.5200}}].

\bibitem{Creminelli:2012ed}
P.~Creminelli, J.~Nore\~na and M.~Simonovi\'c, \emph{{Conformal consistency
  relations for single-field inflation}},
  \href{http://dx.doi.org/10.1088/1475-7516/2012/07/052}{\emph{JCAP} {\bfseries
  07} (2012) 052}, [\href{https://arxiv.org/abs/1203.4595}{{\ttfamily
  1203.4595}}].

\bibitem{Hinterbichler:2013dpa}
K.~Hinterbichler, L.~Hui and J.~Khoury, \emph{{An Infinite Set of Ward
  Identities for Adiabatic Modes in Cosmology}},
  \href{http://dx.doi.org/10.1088/1475-7516/2014/01/039}{\emph{JCAP} {\bfseries
  01} (2014) 039}, [\href{https://arxiv.org/abs/1304.5527}{{\ttfamily
  1304.5527}}].

\bibitem{Kundu:2014gxa}
N.~Kundu, A.~Shukla and S.~P. Trivedi, \emph{{Constraints from Conformal
  Symmetry on the Three Point Scalar Correlator in Inflation}},
  \href{http://dx.doi.org/10.1007/JHEP04(2015)061}{\emph{JHEP} {\bfseries 04}
  (2015) 061}, [\href{https://arxiv.org/abs/1410.2606}{{\ttfamily 1410.2606}}].

\bibitem{Creminelli:2003iq}
P.~Creminelli, \emph{{On non-Gaussianities in single-field inflation}},
  \href{http://dx.doi.org/10.1088/1475-7516/2003/10/003}{\emph{JCAP} {\bfseries
  10} (2003) 003}, [\href{https://arxiv.org/abs/astro-ph/0306122}{{\ttfamily
  astro-ph/0306122}}].

\bibitem{Assassi:2012zq}
V.~Assassi, D.~Baumann and D.~Green, \emph{{On Soft Limits of Inflationary
  Correlation Functions}},
  \href{http://dx.doi.org/10.1088/1475-7516/2012/11/047}{\emph{JCAP} {\bfseries
  11} (2012) 047}, [\href{https://arxiv.org/abs/1204.4207}{{\ttfamily
  1204.4207}}].

\bibitem{Kundu:2015xta}
N.~Kundu, A.~Shukla and S.~P. Trivedi, \emph{{Ward Identities for Scale and
  Special Conformal Transformations in Inflation}},
  \href{http://dx.doi.org/10.1007/JHEP01(2016)046}{\emph{JHEP} {\bfseries 01}
  (2016) 046}, [\href{https://arxiv.org/abs/1507.06017}{{\ttfamily
  1507.06017}}].

\bibitem{Shukla:2016bnu}
A.~Shukla, S.~P. Trivedi and V.~Vishal, \emph{{Symmetry constraints in
  inflation, $\alpha$-vacua, and the three point function}},
  \href{http://dx.doi.org/10.1007/JHEP12(2016)102}{\emph{JHEP} {\bfseries 12}
  (2016) 102}, [\href{https://arxiv.org/abs/1607.08636}{{\ttfamily
  1607.08636}}].

\bibitem{Bzowski:2015pba}
A.~Bzowski, P.~McFadden and K.~Skenderis, \emph{{Scalar 3-point functions in
  CFT: renormalisation, beta functions and anomalies}},
  \href{http://dx.doi.org/10.1007/JHEP03(2016)066}{\emph{JHEP} {\bfseries 03}
  (2016) 066}, [\href{https://arxiv.org/abs/1510.08442}{{\ttfamily
  1510.08442}}].

\bibitem{Bzowski:2017poo}
A.~Bzowski, P.~McFadden and K.~Skenderis, \emph{{Renormalised 3-point functions
  of stress tensors and conserved currents in CFT}},
  \href{http://dx.doi.org/10.1007/JHEP11(2018)153}{\emph{JHEP} {\bfseries 11}
  (2018) 153}, [\href{https://arxiv.org/abs/1711.09105}{{\ttfamily
  1711.09105}}].

\bibitem{Bzowski:2018fql}
A.~Bzowski, P.~McFadden and K.~Skenderis, \emph{{Renormalised CFT 3-point
  functions of scalars, currents and stress tensors}},
  \href{http://dx.doi.org/10.1007/JHEP11(2018)159}{\emph{JHEP} {\bfseries 11}
  (2018) 159}, [\href{https://arxiv.org/abs/1805.12100}{{\ttfamily
  1805.12100}}].

\bibitem{paul}
A.~Bzowski, P.~McFadden and K.~Skenderis, \emph{{Handbook of holographic
  4-point functions, to appear}}, .

\bibitem{Cheung:2007st}
C.~Cheung, P.~Creminelli, A.~L. Fitzpatrick, J.~Kaplan and L.~Senatore,
  \emph{{The Effective Field Theory of Inflation}},
  \href{http://dx.doi.org/10.1088/1126-6708/2008/03/014}{\emph{JHEP} {\bfseries
  03} (2008) 014}, [\href{https://arxiv.org/abs/0709.0293}{{\ttfamily
  0709.0293}}].

\bibitem{Green:2020ebl}
D.~Green and E.~Pajer, \emph{{On the Symmetries of Cosmological
  Perturbations}},
  \href{http://dx.doi.org/10.1088/1475-7516/2020/09/032}{\emph{JCAP} {\bfseries
  09} (2020) 032}, [\href{https://arxiv.org/abs/2004.09587}{{\ttfamily
  2004.09587}}].

\bibitem{Pajer:2020wxk}
E.~Pajer, \emph{{Building a Boostless Bootstrap for the Bispectrum}},
  \href{http://dx.doi.org/10.1088/1475-7516/2021/01/023}{\emph{JCAP} {\bfseries
  01} (2021) 023}, [\href{https://arxiv.org/abs/2010.12818}{{\ttfamily
  2010.12818}}].

\bibitem{Kampf:2013vha}
K.~Kampf, J.~Novotny and J.~Trnka, \emph{{Tree-level Amplitudes in the
  Nonlinear Sigma Model}},
  \href{http://dx.doi.org/10.1007/JHEP05(2013)032}{\emph{JHEP} {\bfseries 05}
  (2013) 032}, [\href{https://arxiv.org/abs/1304.3048}{{\ttfamily 1304.3048}}].

\end{thebibliography}\endgroup
\bibliographystyle{JHEP}
  
\end{document}